\documentclass[12pt]{article}
\usepackage{lineno}
\usepackage{graphicx,graphbox}
\usepackage[export]{adjustbox}
\usepackage{amsmath,amsfonts,amssymb}
\usepackage{hhline}
\usepackage{amssymb}
\usepackage{cite}
\usepackage{array}
\usepackage{multirow} 

\usepackage{xspace}
\usepackage{bm}
\usepackage{sectsty}
\allsectionsfont{\boldmath} 
\usepackage{pifont}
\usepackage{float}
\usepackage[small, bf]{caption}
\usepackage{subcaption}
\usepackage{placeins} 
\usepackage{booktabs,ltablex}
\usepackage{pdflscape}
\usepackage{geometry}
\usepackage{dcolumn} 
\newcolumntype{d}[1]{D{.}{.}{#1}}

\usepackage[T1]{fontenc}
\usepackage[british]{babel}
\usepackage[language=british]{hyphsubst}

\usepackage{color}
\definecolor{rltred}{rgb}{0.75,0,0}
\definecolor{rltgreen}{rgb}{0,0.5,0}
\definecolor{rltblue}{rgb}{0,0,0.75}

\newif\ifpdf\pdftrue

\newif\ifpdf
\ifx\pdfoutput\undefined
    \pdffalse          
\else
    \pdfoutput=1       
    \pdftrue
\fi

\ifpdf
\usepackage{thumbpdf}
\usepackage[pdftex,
        colorlinks=true,
        urlcolor=rltblue,       
        filecolor=rltgreen,     
        linkcolor=rltred,       
        pdftitle={H1Draft Rho Photoproduction},
        pdfauthor={H1},
        pdfsubject={Template file},
        pdfkeywords={High-Energy Physics, Particle Physics},
        bookmarksopen=true]{hyperref}

\fi

\newlength{\dinwidth}
\newlength{\dinmargin}
\DeclareGraphicsExtensions{%
    .eps,
    .pdf,
    .png,
    }

\newcommand{\dd}{\ensuremath{{\mathrm{d}}}}
\newcommand{\ee}{\ensuremath{{\mathrm{e}}}}
\newcommand{\ii}{\ensuremath{{\mathrm{i}}}}

\newcommand{\ie}{i.e.,\ }
\newcommand{\eg}{e.g.,\ }

\newcommand{\mpipi}{{\ensuremath{{m_{\pi\pi}}}}\xspace}
\newcommand{\mpipiSq}{{\ensuremath{{m_{\pi\pi}^2}}}\xspace}
\newcommand{\pipi}{{\ensuremath{{\pi^+\pi^-}}}\xspace}

\newcommand{\ptsqrec}{{\ensuremath{{\left(p_{T,\pi\pi}^{\mathrm{rec}}\right)^2}}}\xspace}
\newcommand{\Epipirec}{{\ensuremath{{E_{\pi\pi}^{\mathrm{rec}}}}}\xspace}
\newcommand{\pzpipirec}{{\ensuremath{{p_{z,\pi\pi}^{\mathrm{rec}}}}}\xspace}
\newcommand{\ptpipirec}{{\ensuremath{{p_{T,\pi\pi}^{\mathrm{rec}}}}}\xspace}
\newcommand{\mpipirec}{{\ensuremath{{m_{\pi\pi}^\mathrm{rec}}}}\xspace}

\newcommand{\wgp}{{\ensuremath{{W_{\gamma p}}}}\xspace}

\newcommand{\gp}{{\ensuremath{{\gamma p}}}\xspace}

\newcommand{\My}{{\ensuremath{{m_Y}}}\xspace}
\newcommand{\MySq}{{\ensuremath{{m_Y^2}}}\xspace}
\newcommand{\dEdx}{{\ensuremath{{\dd E/\dd x}}}\xspace}

\newcommand{\wgprec}{{\ensuremath{{W_{\gamma p}^{\mathrm{rec}}}}}\xspace}
\newcommand{\trec}{\ensuremath{{t^\mathrm{rec}}}\xspace} 

\newcommand{\myrho}{{\ensuremath{{\rho^0}}}\xspace}
\newcommand{\rprim}{{\ensuremath{{\rho^\prime}}}\xspace}
\newcommand{\rhomega}{{\ensuremath{{\rho^0\text{-}\omega}}}\xspace}

\newcommand{\BR}{{\ensuremath{{\mathcal{BR}}}}\xspace}

\newcommand{\stat}{\ensuremath{{\mathrm{stat.}}}\xspace}
\newcommand{\syst}{\ensuremath{{\mathrm{syst.}}}\xspace}
\newcommand{\tot}{\ensuremath{{\mathrm{tot.}}}\xspace}

\newcommand{\reg}{\ensuremath{{I\!\!R}}\xspace}
\newcommand{\pom}{\ensuremath{{I\!\!P}}\xspace}
\newcommand{\phiWW}{\ensuremath{{\Phi_{\gamma/e}}}\xspace}
\newcommand{\chirStat}{\ensuremath{{\chi_\mathrm{stat}^2/\mathrm{n}_\mathrm{dof}}}\xspace}

\newcommand{\sigmaPiPiY}{\ensuremath{{\sigma(\gamma p \rightarrow \pipi Y)}}\xspace}
\newcommand{\sigmaPiPip}{\ensuremath{{\sigma(\gamma p \rightarrow \pipi p)}}\xspace}
\newcommand{\sigmaRhoY}{\ensuremath{{\sigma(\gamma p \rightarrow \rho^0 Y)}}\xspace}
\newcommand{\sigmaRhop}{\ensuremath{{\sigma(\gamma p \rightarrow \rho^0 p)}}\xspace}
\newcommand{\ddSigmaPiPiYdmdt}{{\ensuremath{{\dd^2 \sigma(\gamma p \rightarrow \pipi Y)/\dd \mpipi \dd t}}}\xspace}
\newcommand{\ddSigmaPiPipdmdt}{{\ensuremath{{\dd^2 \sigma(\gamma p \rightarrow \pipi p)/\dd \mpipi \dd t}}}\xspace}
\newcommand{\dSigmaPiPiYdm}{{\ensuremath{{\dd \sigma(\gamma p \rightarrow \pipi Y)/\dd \mpipi}}}\xspace}
\newcommand{\dSigmaPiPipdm}{{\ensuremath{{\dd \sigma(\gamma p \rightarrow \pipi p)/\dd \mpipi}}}\xspace}

\newcommand{\dSigmaRhoYdt}{{\ensuremath{{\dd \sigma(\gamma p \rightarrow \myrho Y)/\dd t}}}\xspace}
\newcommand{\dSigmaRhopdt}{{\ensuremath{{\dd \sigma(\gamma p \rightarrow \myrho p)/\dd t}}}\xspace}
\newcommand{\sigmarho}{{\ensuremath{{\sigma_{\rho}}}}\xspace}

\newcommand{\gev}{{\ensuremath{{\mathrm{GeV}}}}\xspace}
\newcommand{\mev}{{\ensuremath{{\mathrm{MeV}}}}\xspace}
\newcommand{\gevSq}{{\ensuremath{{\mathrm{GeV}^2}}}\xspace}
\newcommand{\gevSqInv}{{\ensuremath{{\mathrm{GeV}^{-2}}}}\xspace}
\newcommand{\cm}{{\ensuremath{{\mathrm{cm}}}}\xspace}

\newcommand{\m}{{\ensuremath{{\mathrm{m}}}}\xspace}

\newcommand{\mub}{\ensuremath{{\mu\mathrm{b}}}\xspace}
\newcommand{\invpb}{{\ensuremath{{\mathrm{pb}^{-1}}}}\xspace}

\newcommand{\eqnref}[1]{Equation~(\ref{#1})}
\newcommand{\tabref}[1]{Table~\ref{#1}}
\newcommand{\secref}[1]{Section~\ref{#1}}

\newcommand{\apxref}[1]{Appendix~\ref{#1}}
\newcommand{\figref}[1]{Figure~\ref{#1}}

\newcommand{\el}{\ensuremath{{\mathrm{el}}}\xspace}
\newcommand{\pd}{\ensuremath{{\mathrm{pd}}}\xspace}
\newcommand{\nr}{\ensuremath{{\mathrm{nr}}}\xspace}
\newcommand{\VM}{\ensuremath{{\mathrm{VM}}}\xspace}

\begin{document}  

\def\Journal#1#2#3#4{{#1} {\bf #2} (#3) #4}
\def\NCA{\em Nuovo Cimento}
\def\NIM{\em Nucl. Instrum. Methods}
\def\NIMA{{\em Nucl. Instrum. Methods} {\bf A}}
\def\NPB{{\em Nucl. Phys.}   {\bf B}}
\def\PLB{{\em Phys. Lett.}   {\bf B}}
\def\PRL{\em Phys. Rev. Lett.}
\def\PRD{{\em Phys. Rev.}    {\bf D}}
\def\ZPC{{\em Z. Phys.}      {\bf C}}
\def\EJC{{\em Eur. Phys. J.} {\bf C}}
\def\CPC{\em Comp. Phys. Commun.}

\begin{titlepage}

\noindent
\begin{flushleft}
{\tt DESY 20-080    \hfill    ISSN 0418-9833} \\
{\tt May 2020}                  \\
\end{flushleft}

\vspace{1.0cm}

\begin{center}
\begin{Large}

  {\bf Measurement of Exclusive $\bm{{\pipi}}$ and $\bm{{\myrho}}$ Meson Photoproduction at HERA}

\vspace{1.0cm}

H1 Collaboration

\end{Large}
\end{center}

\vspace{0.5cm}

\begin{abstract}
\noindent
Exclusive photoproduction of $\myrho(770)$ mesons is studied using the H1 detector at the $ep$ collider HERA. A sample of about 900000 events is used to measure single- and double-differential cross sections for the reaction $\gamma p \to \pi^{+}\pi^{-}Y$. Reactions where the proton stays intact (${\My{=}m_p}$) are statistically separated from those where the proton dissociates to a low-mass hadronic system ($m_p{<}\My{<}10~\gev$). The double-differential cross sections are measured as a function of the invariant mass $\mpipi$ of the decay pions and the squared $4$-momentum transfer $t$ at the proton vertex. The measurements are presented in various bins of the photon-proton collision energy $\wgp$. The phase space restrictions are ${0.5 < m_{\pi\pi} < 2.2~\gev}$, ${\vert t\vert < 1.5~\gevSq}$, and ${20 < W_{\gamma p} < 80~\gev}$. Cross section measurements are presented for both elastic and proton-dissociative scattering. The observed cross section dependencies are described by analytic functions.
Parametrising the \mpipi dependence with resonant and non-resonant contributions added at the amplitude level leads to a measurement of the $\myrho(770)$ meson mass and width at $m_\rho = 770.8\ {}^{+2.6}_{-2.7}~(\tot)~\mev$ and $\Gamma_\rho = 151.3\ {}^{+2.7}_{-3.6}~(\tot)~\mev$, respectively.
The model is used to extract the $\myrho(770)$ contribution to the $\pi^{+}\pi^{-}$ cross sections and measure it as a function of $t$ and \wgp. In a Regge asymptotic limit in which one Regge trajectory $\alpha(t)$ dominates, the intercept $\alpha(t{=}0) = 1.0654\ {}^{+0.0098}_{-0.0067}~(\tot)$ and the slope $\alpha^\prime(t{=}0) = 0.233\ {}^{+0.067 }_{-0.074 }~(\tot) ~\gevSqInv$ of the $t$ dependence are extracted for the case $m_Y{=}m_p$.
\end{abstract}

  \vspace{0.5cm}

\begin{center}
  This publication is dedicated to the memory of our dear colleague Peter Tru\"ol.
\end{center}

\vspace{0.5cm}
\begin{center}
To be submitted to \EJC 
\end{center}

\end{titlepage}

\begin{flushleft}

V.~Andreev$^{19}$,             
A.~Baghdasaryan$^{30}$,        
A.~Baty$^{45}$,                
K.~Begzsuren$^{27}$,           
A.~Belousov$^{19}$,            
A.~Bolz$^{10,12}$,             
V.~Boudry$^{22}$,              
G.~Brandt$^{40}$,              
D.~Britzger$^{20}$,            
A.~Buniatyan$^{2}$
% ...


\bigskip{\it
% ...
$ ^{2}$ School of Physics and Astronomy, University of Birmingham,
          Birmingham, UK$^{ b}$ \\
% ...
$ ^{10}$ DESY, Hamburg, Germany \\
% ...
$ ^{12}$ Physikalisches Institut, Universit\"at Heidelberg,
          Heidelberg, Germany$^{ a}$ \\
% ...
$ ^{19}$ Lebedev Physical Institute, Moscow, Russia \\
 $ ^{20}$ Max-Planck-Institut f\"ur Physik, M\"unchen, Germany \\
% ...
$ ^{22}$ LLR, Ecole Polytechnique, CNRS/IN2P3, Palaiseau, France \\
% ...
$ ^{27}$ Institute of Physics and Technology of the Mongolian
          Academy of Sciences, Ulaanbaatar, Mongolia \\
% ...
$ ^{30}$ Yerevan Physics Institute, Yerevan, Armenia \\
% ...
$ ^{40}$ II. Physikalisches Institut, Universit\"at G\"ottingen,
          G\"ottingen, Germany \\
% ...
$ ^{45}$ Rice University, Houston, USA \\
% ...
$ ^a$ Supported by the Bundesministerium f\"ur Bildung und Forschung, FRG,
      under contract numbers 05H09GUF, 05H09VHC, 05H09VHF,  05H16PEA \\
 $ ^b$ Supported by the UK Science and Technology Facilities Council,
      and formerly by the UK Particle Physics and
      Astronomy Research Council \\
% ...
}

\end{flushleft}


\newpage

\section{Introduction}
\label{sec:Introduction}

Diffractive hadron interactions at high scattering energies are characterised by final states consisting of two systems well separated in rapidity, which carry the quantum numbers of the initial state hadrons. Most diffractive phenomena are governed by \text{soft}, large distance processes. Despite being dominated by the strong interaction, they remain largely inaccessible to the description by Quantum Chromodynamics (QCD) in terms of quark and gluon interactions. In many cases, perturbative QCD calculations are not applicable because the typical scales involved are too low. Instead, other models have to be employed, such as Regge theory~\cite{Collins:1977jy}, in which interactions are described in terms of the coherent exchange of reggeons and the pomeron.

\begin{figure}[!htb]\centering
  \includegraphics[width=0.33\textwidth,align=t]{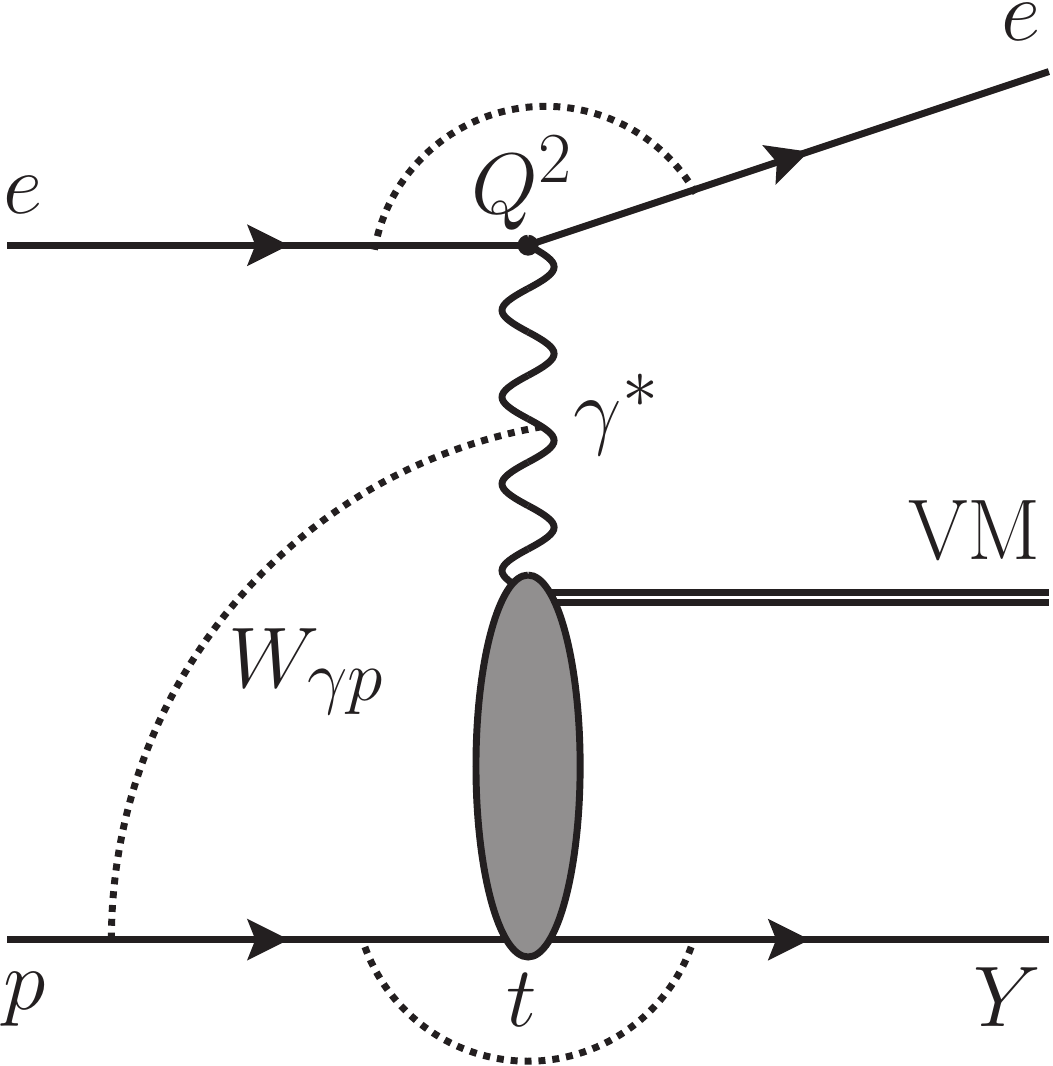
}
  \caption{Diffractive vector meson electroproduction.}
  \label{fig:intro_VMProd}
\end{figure}
 
Exclusive vector meson (VM) \text{electroproduction} $e+p \rightarrow e + \VM + Y$ is particularly suited to study diffractive phenomena. This process is illustrated in \figref{fig:intro_VMProd}. In leading order QED, the interaction of the electron with the proton is the exchange of a virtual photon which couples to the proton in a diffractive manner to produce a VM ($\myrho$, $\omega$, $\phi$, $J/\psi$, \dots) in the final state. The proton is scattered into a system $Y$, which can be a proton (\textit{elastic} scattering) or a diffractively excited system (\textit{diffractive proton dissociation}). Scales for the process are provided by the vector meson mass squared $m^2_{\VM}$, the photon virtuality $Q^2 = -q^2$, and the squared $4$-momentum transfer at the proton vertex $t$, the dependence on each of which can be studied independently. 

In this paper, elastic and proton-dissociative \textit{photoproduction} ($Q^2=0$) of $\myrho$ mesons is studied using electroproduction data at small $Q^2 < 2.5~\gevSq$. Electroproduction of \myrho mesons has been studied previously at HERA in both the photoproduction regime and for large $Q^2\gg \Lambda_\mathrm{QCD}^2$ (the perturbative cut-off in QCD), as well as for elastic and proton-dissociative scattering~\cite{Aaron:2009wg,  Aaron:2009xp, Aktas:2006qs, Adloff:2002tb, Adloff:1999kg, Adloff:1997jd, Aid:1996ee, Aid:1996bs, Chekanov:2007zr, Chekanov:2002rm, Breitweg:1999fm, Breitweg:1999jy, Breitweg:1998nh, Breitweg:1997ed, Derrick:1996vw, Derrick:1995vq, Derrick:1995yd,Abramowicz:2011pk}. Measurements at lower photon-proton centre-of-mass energy \wgp have been published in fixed-target interactions~\cite{Ballam:1971wq,Park:1971ts,Ballam:1972eq,Struczinski:1975ik,Egloff:1979mg,Aston:1982hr}. Most recently, a measurement of exclusive \myrho photoproduction has been performed at the CERN LHC in ultra-peripheral lead-proton collisions~\cite{Sirunyan:2019nog}. 

The present measurement is based on a data set collected during the 2006/2007 HERA running period by the H1 experiment. Since \myrho mesons decay almost exclusively into a pair of charged pions, the analysis is based on a sample of \pipi photoproduction events. Compared with previous HERA results, the size of the sample makes possible a much more precise measurement with a statistical precision at the percent level. It is then possible to extract up to three-dimensional kinematic distributions as a function of the invariant mass of the \pipi system $\mpipi$, of $\wgp$, and of $t$. However, the size of the dataset is such that the systematic uncertainties of the modelling of the H1 experiment are important.

The structure of the paper is as follows: Theoretical details of \myrho meson photoproduction are discussed in \secref{sec:theory} with a focus on the description of \pipi photoproduction in terms of Regge theory (\secref{sec:theo_pipiRegge}), cross section definitions (\secref{sec:theo_xSecDef}), and Monte Carlo modelling of relevant processes (\secref{sec:theo_MCModel}). In \secref{sec:experimental}, the experimental method is detailed, including a description of the H1 detector (\secref{sec:exp_detector}), the dataset underlying the analysis (\secref{sec:exp_dataSample}), the unfolding procedure applied to correct detector level distributions (\secref{sec:exp_unfolding}), and systematic uncertainties of the measurement (\secref{sec:systematics}). Results are presented in \secref{sec:results}. They encompass a measurement of the integrated \pipi production cross section in the fiducial phase space (\secref{sec:res_sigmaPiPiFid}), a study of the invariant \mpipi distributions (\secref{sec:res_sigmaRhoFid}), measurements of the scattering energy (\secref{sec:res_sigmaRhoOfW}) and $t$ dependencies of the \myrho meson production cross sections (\secref{sec:res_dSigmaRhoOft}), as well as the extraction of the leading Regge trajectory from the two-dimensional $t$ and \wgp dependencies (\secref{sec:res_dSigmaRhoOfWt}).

\section{Theory}
\label{sec:theory}
\subsection{\pipi and \myrho meson photoproduction}
\label{sec:theo_pipiRegge}
In electron\footnote{In the following, the term ``electron'' is used indistinctly to refer to both positrons and electrons.}-proton collisions, \pipi and \myrho meson photoproduction is studied in the scattering process
\begin{equation}
  e(e) + p(p) \rightarrow e(e^\prime) + \pi^+(k_1) + \pi^-(k_2) + Y(p^\prime)\, \mathrm,
\end{equation}
where the 4-momenta of the participating particles are given in parentheses. The relevant kinematic variables are the electron-proton centre-of-mass energy squared
\begin{equation}
 s = (e+p)^2 \mathrm,     
\end{equation}
the photon virtuality, \ie the negative squared $4$-momentum transfer at the electron vertex
\begin{equation}
    Q^2 = -q^2 = (e-e^\prime)^2 \mathrm,
\end{equation}
the inelasticity 
\begin{equation}
   y= (q \cdot p)/(e\cdot p) \,\mathrm,
\end{equation}
the $\gamma p$ centre-of-mass energy squared
\begin{equation}
  W_{\gamma p}^2 = (q+p)^2 \mathrm, 
  \label{eqn:theo_Wgp}
\end{equation}
the invariant mass of the \pipi system squared
\begin{equation}
  \mpipiSq = (k_1+k_2)^2\mathrm,
 \end{equation}
the squared $4$-momentum transfer at the proton vertex 
\begin{equation}
  t = (p-p^\prime)^2\mathrm,
  \label{eqn:theo_t}
\end{equation}
and the invariant mass squared of the (possibly dissociated) final state proton system
\begin{equation}
  \MySq = (p^\prime)^2\mathrm.
\end{equation}

In general, diffractive photoproduction of (light) vector mesons shares the characteristics of soft hadron-hadron scattering: In the high energy limit, the total and elastic cross sections are observed to rise slowly with increasing centre-of-mass energy. Differential cross sections $\dd \sigma/\dd t$  are \textit{peripheral}, favouring low $|t|$, \textit{forward} scattering. With increasing scattering energy, the typical $|t|$ of elastic cross sections becomes smaller, i.e.~forward peaks appear to shrink.

In vector dominance models (VDM)~\cite{VDM}, the photon is modelled as a superposition of (light) vector mesons which can interact strongly with the proton to subsequently form a bound VM state. Like hadron-hadron interactions in general, VM production is dominated by colour singlet exchange in the $t$-channel. The lack of a sufficiently hard scale makes these exchanges inaccessible to perturbative QCD in a large portion of the phase space. Empirical and phenomenological models are used instead to describe the data. At low $|t| \ll 1~\gevSq$, differential cross sections are observed to follow exponential dependencies $\dd \sigma/ \dd t \propto \exp(bt)$. In the optical interpretation, the exponential slope $b$ is related to the transverse size of the scattered objects. Towards larger $|t|$, cross section dependencies change into a less steep power-law dependence $\dd \sigma / \dd t \propto |t|^a$. In Regge theory~\cite{Collins:1977jy}, the dependence of hadronic cross sections on the scattering energy $\wgp$ and the shrinkage of the forward peak are characterised by \textit{Regge trajectories} $\alpha(t)$. The contribution of a single \textit{Regge pole} to the differential elastic cross section is $\dd\sigma_{\el} / \dd t(\wgp) \propto W_{\gamma p}^{4 (\alpha(t)-1)}$~\cite{Collins:1977jy}. At low energies $\wgp \lesssim 10~\gev$, \textit{reggeon trajectories} $\alpha_{\reg}(t)$ dominate which are characterised by intercepts $\alpha_{\reg}(0) < 1$, \ie they result in cross sections that fall off with increasing energy~\cite{Donnachie:1994zb}. In the high energy limit, only the contribution of what is known as the \textit{pomeron} Regge pole remains because its trajectory $\alpha_{\pom}(t)$ has a large enough intercept $\alpha_{\pom}(0) \gtrsim 1$ for it not to have decreased to a negligible level. The shrinkage of the elastic forward peak with increasing energy is the result of a positive slope $\alpha_{\pom}^\prime > 0$ of the trajectory $\alpha(t)$ at $t=0$.

\begin{figure}[!htb]\centering
 \includegraphics[width=0.33\textwidth,align=c]{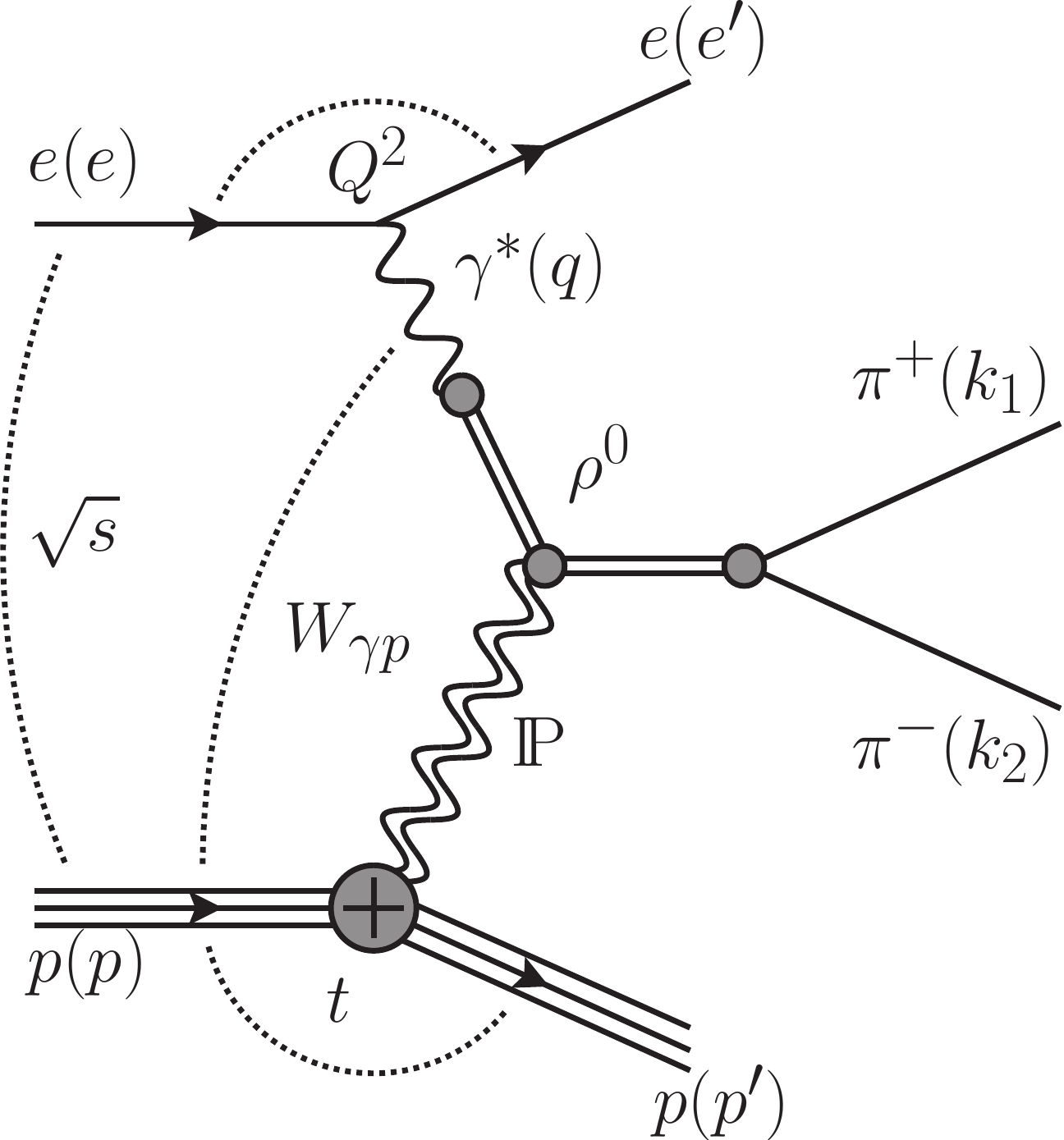}\hspace*{.1\textwidth}
 \includegraphics[width=0.33\textwidth,align=c]{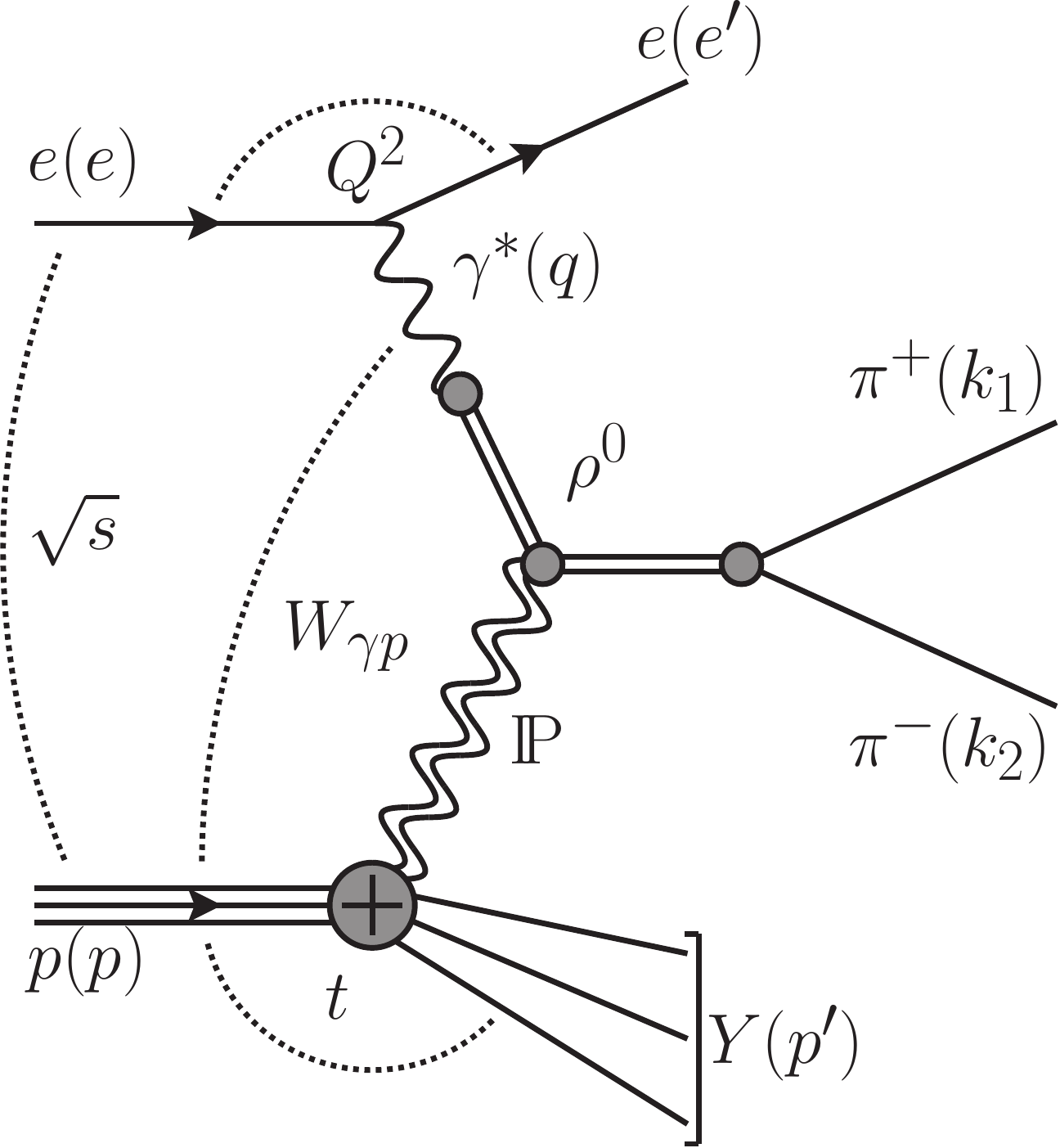}
  \caption{Diagram of \myrho meson production and decay in elastic (left) and proton-dissociative (right) $ep$ scattering in the VDM and Regge picture, where the interaction is governed by soft pomeron exchange in the high energy limit.}
  \label{fig:intro_eptorhopDiag}
\end{figure}

Feynman-like diagrams which illustrate the interpretation of \myrho meson photoproduction in the VDM/Regge picture are given in \figref{fig:intro_eptorhopDiag}. In the diagrams, the \myrho meson is shown to decay into a pair of charged pions. This is the dominant decay channel with a branching ratio $\BR(\myrho \rightarrow \pipi ) \simeq 99\%$~\cite{PDG}. 
The decay structure of the \myrho meson into \pipi is described by two decay angles~\cite{Schilling:1973ag}. These also give insight into the production mechanism of the \myrho meson. Contributions with $s$-channel helicity conservation are expected to dominate, such that the \myrho meson retains the helicity of the photon, \ie in photoproduction the \myrho meson is transversely polarised~\cite{Breitweg:1997ed}.

While vector mesons dominate photon-proton interactions, the VDM approach does not provide a complete picture. This is particularly evident for \myrho meson production where in the region of the \myrho meson resonance peak also non-resonant \pipi production plays an important role. The non-resonant \pipi amplitude interferes with the resonant \myrho meson production amplitude to produce a skewing of the Breit-Wigner resonance profile in the \pipi mass spectrum~\cite{Soding:1965nh}. Newer models thus aim to take a more general approach, \eg a recently developed model for \pipi photoproduction based on tensor pomeron exchange~\cite{Nachtmann:pipi}. That model seems to be in fair agreement with H1 data when certain model parameters are adjusted \cite{Phd:Abolz}. A more detailed investigation is beyond the scope of this paper.

\subsection{Cross section definition}
\label{sec:theo_xSecDef}
\subsubsection{Photon flux normalisation}
In this paper, photoproduction cross sections $\sigma_{\gp}$ are derived from the measured electron-proton cross sections $\sigma_{ep}$. At low $Q^2$, these can be expressed as a product of a flux of virtual photons $f^{T/L}_{\gamma/e}$ and a virtual photon-proton cross section $\sigma^{T/L}_{\gamma^* p}$:
\begin{equation}
  \dfrac{\dd^2 \sigma_{ep}}{\dd y \dd Q^2} =f^T_{\gamma/e}\left(y, Q^2\right) \, \sigma^T_{\gamma^* p}\left(\wgp(y,Q^2), Q^2\right) + f^L_{\gamma/e}\left(y, Q^2\right) \, \sigma^L_{\gamma^* p}\left(\wgp(y,Q^2), Q^2\right).
  \label{eqn:sigmaEpFactoriz}
\end{equation}
A distinction is made between transversely and longitudinally polarised photons, as indicated by the superscripts $T$ and $L$, respectively. In the low $Q^2$ regime studied here, the transverse component is dominant. The transverse and longitudinal photon fluxes are given in the Weizs\"acker-Williams approximation~\cite{equivPhoton} by
\begin{align}
  f^T_{\gamma/e}(y,Q^2) &= \dfrac{\alpha_\mathrm{em}}{2\pi} \dfrac{1}{yQ^2}\left( 1 + (1-y)^2 - 2(1-y) \dfrac{Q^2_\mathrm{min}}{Q^2} \right)  
\intertext{ and} 
  f^L_{\gamma/e}(y,Q^2) &= \dfrac{\alpha_\mathrm{em}}{\pi} \dfrac{1}{yQ^2}\left(1-y\right) \text,
\end{align}
respectively, where $\alpha_\mathrm{em}$ is the fine structure constant, $m_e$ the mass of the electron, and $Q^2_\mathrm{min} = m_e^2 y^2/(1-y)$ is the smallest kinematically accessible $Q^2$ value. 

In vector meson production in the VDM approach, the cross section for virtual photon interactions ($Q^2>0$) can be related to the corresponding photoproduction cross section $\sigma_{\gamma p}$ at $Q^2=0$:
\begin{align}
  \sigma^T_{\gamma^* p}\left(\wgp, Q^2\right) &= \sigma_{\gamma p}\left(\wgp\right) \, \left( 1 + \frac{Q^2}{m^2_{\VM}}  \right)^{-2}  \mathrm, \\
  \sigma^L_{\gamma^* p}\left(\wgp, Q^2\right) &= \sigma^T_{\gamma^* p}\left(\wgp, Q^2\right) \, \dfrac{Q^2}{m^2_{\VM}} \xi^2\mathrm.
\end{align}
In the equations, $m_{\VM}$ denotes the mass of the considered vector meson and $\xi^2$ is a proportionality constant, that in the following is set to unity. The real photoproduction cross section can then be factored out of \eqnref{eqn:sigmaEpFactoriz} according to
\begin{equation}
  \dfrac{\dd^2 \sigma_{ep}}{\dd y \dd Q^2} = \sigma_{\gamma p}\left( \wgp(y) \right)\, \varphi_\mathrm{eff}\left(y, Q^2\right) \mathrm,
\end{equation}
were the so-called \textit{effective photon flux} is given by
\begin{equation}
  \varphi_\mathrm{eff}(y, Q^2) = \dfrac{\alpha_\mathrm{em}}{2\pi} \dfrac{1}{yQ^2} \left[  1 + (1-y)^2 - 2(1-y) \left( \dfrac{Q^2_\mathrm{min}}{Q^2}-\dfrac{Q^2}{m^2_{\VM}} \right) \right]  \left( 1 + \frac{Q^2}{m^2_{\VM}}  \right)^{-2}\mathrm.
  \label{eqn:theo_effFlux}
\end{equation}

In practice, measurements of $\sigma_{ep}$ are evaluated as integrals over finite regions in $Q^2$ and $\wgp$. In order to extract corresponding photoproduction cross sections at $Q^2=0$ and for an appropriately chosen average energy $\langle \wgp \rangle$, the measured values are normalised by the effective photon flux integrated over the corresponding $Q^2$ and $\wgp$ ranges:
\begin{equation}
  \sigma_{\gamma p}\left(\langle \wgp \rangle\right) = \dfrac{ \sigma_{ep} }{\phiWW} \mathrm,
\end{equation}
with 
\begin{equation}
  \phiWW = \displaystyle \int_{\scriptsize\begin{array}{c}Q^2_{\min}<Q^2<Q^2_{\max} \\
      W_{\min}<\wgp<W_{\max}\end{array}}  \varphi_\mathrm{eff}(y, Q^2) \, \dd y\dd Q^2 \mathrm.
  \label{eqn:theo_intFlux}
\end{equation}

\subsubsection{\myrho meson photoproduction cross section}
\label{sec:theo_xSecRho}
With the H1 detector, \pipi photoproduction is measured. In the kinematic region studied, there are significant contributions from the \myrho meson resonance, non-resonant \pipi production, the $\omega$ meson resonance and excited $\rho$ meson states~\cite{PDG,Nachtmann:pipi}. Photoproduction of \myrho mesons has to be disentangled from these processes by analysis of the \pipi mass spectrum.
Here, a model is used which parametrises the spectrum in terms of \myrho and $\omega$ meson, as well as non-resonant amplitudes~\cite{Phd:Abolz}, similar to the original proposal made by S\"oding~\cite{Soding:1965nh}.

The contributions are added at the amplitude level, and interference effects are taken into account. The model is defined as
\begin{equation}
  \dfrac{\dd \sigma_{\pi\pi}}{\dd\mpipi }(\mpipi) = A\ \dfrac{q^3(\mpipi)}{q^3(m_\rho)} \ \bigg \vert A_{\rho,\omega}(\mpipi) + A_{\nr}(\mpipi) \bigg \vert^2\text,
 \label{eqn:theo_rhoSoedMass}
\end{equation}
where $A$ is a global normalisation factor and $q^3(\mpipi)$ describes the phase space, with $q(\mpipi) = \frac{1}{2} \sqrt{\mpipiSq - 4m_\pi^2}$ being the momentum of one of the pions in the \pipi centre-of-mass frame~\cite{Jackson:1964zd}. It is normalised to the value $q^3(m_\rho)$ at the \myrho mass. The amplitude $A_{\rho,\omega}$ takes into account the $\myrho$ and $\omega$ meson resonance contributions, whereas the non-resonant component is modelled by $A_{\nr}(\mpipi)$. The components are considered to be fully coherent. Since additional resonances with masses above 1~\gev are neglected, the model can only be applied to the region near the \myrho meson mass peak. Models of this form have been widely used in similar past analyses because they preserve the physical amplitude structure while being parametric and thus easily applicable.
More sophisticated models are designed to preserve unitarity \cite{Basdevant:1977ya} or are regularised by barrier factors at higher masses \cite{VonHippel:1972fg}.

The combined $\rhomega$ amplitude is modelled following a parametrisation given by \cite{Akhmetshin:2001ig} 
\begin{equation}
 A_{\rho,\omega}(\mpipi) =  \mathcal{BW}_{\rho}(\mpipi) \ \left( 1 + f_\omega  \ee^{\ii \phi_{\omega} } \dfrac{ \mpipiSq}{ m_\omega^2 } \mathcal{BW}_{\omega}(\mpipi) \right),
\end{equation}
where $f_\omega$ and $\phi_\omega$ are a normalisation factor and a mixing phase for the $\omega$ contribution, respectively. The $\omega$ meson is not expected to decay into a pair of charged pions directly because of the conservation of G-parity by the strong interaction. However, electromagnetic $\omega\rightarrow\myrho$-mixing with a subsequent $\myrho \rightarrow \pipi$ decay is possible~\cite{OConnell:1995nse}. Both resonances are modelled by a relativistic Breit-Wigner function~\cite{Breit:1936zzb}:
\begin{equation}
 \mathcal{BW}_{\VM}(\mpipi) =  \dfrac{ m_{\VM}  \Gamma_{{\VM}} }{ m^2_{\pi\pi} - m_{\VM}^2 + \ii \, m_{\VM} \Gamma(\mpipi) }.
 \label{eqn:effingBW}
\end{equation}
The parameters $m_{\VM}$ and $\Gamma_{{\VM}}$ are the respective vector meson's Breit-Wigner mass and width. The Breit-Wigner function is normalised to $|\mathcal{BW}_{\VM}(m_{\VM})|=1$. For the \myrho resonance a p-wave mass-dependent width~\cite{Jackson:1964zd} is used:
\begin{equation}
\Gamma(\mpipi) = \Gamma_{{\VM}} \dfrac{q^3(\mpipi)}{q^3(m_{\VM})} \dfrac{m_{\VM}}{\mpipi}\mathrm{,}
\label{eqn:gamma_rho}
\end{equation}
whereas a constant width is assumed for the very narrow $\omega$ resonance.

The unknown non-resonant amplitude is parametrised by the function
\begin{equation}
  A_{\nr}(\mpipi) =  \dfrac{f_{\nr}}{  \left( \mpipiSq - 4m_\pi^2 + \Lambda^2_{\nr} \right)^{\delta_{\nr}}} \mathrm{,}
  \label{eqn:theo_Anr}
\end{equation}
where the relative normalisation is given by $f_{\nr}$, while $\Lambda_{\nr}$ and $\delta_{\nr}$ are free model parameters. They can shape the amplitude for the modelling of a possible internal structure of the non-resonant $\gamma\pi\pi$-coupling. In similar past analyses, typically a purely real non-resonant amplitude has been assumed. Following that assumption, $f_{\nr}$ is set to be real. For $\delta_{\nr} > \frac{3}{4}$, the non-resonant contribution to the cross section (cf. \eqnref{eqn:theo_rhoSoedMass}) has a local maximum at
\begin{equation}
  m_{\nr}^{\text{max}} = \dfrac{ \sqrt{\Lambda^2_{\nr} + (\frac{4}{3}\delta_{\nr}-1)\, 4m_\pi^2} }{\sqrt{\frac{4}{3}\delta_{\nr}-1}}\mathrm,
\end{equation}
and falls proportionally to $\left( 1/\mpipiSq \right)^{2\delta_{\nr}-3}$ in the high mass region.

In order to extract the \myrho meson contribution to the \pipi photoproduction cross section, the measured \pipi mass distributions are fitted using \eqnref{eqn:theo_rhoSoedMass}. The \myrho meson Breit-Wigner contribution is then conventionally defined by the integral
\begin{equation}
  \sigma(\gamma p \rightarrow \myrho Y) =  \dfrac{A}{q^3(m_\rho)} \int_{2m_\pi}^{m_\rho + 5 \Gamma_{\rho} } \big\vert \mathcal{BW}_{\rho}(m) \big\vert^2 q^3(m)  \dd m.
  \label{eqn:theo_intRho}
\end{equation}
As the \myrho meson resonance decays almost exclusively into two charged pions, this is taken to be equal to the total \myrho meson photoproduction cross section without correcting for the $\myrho \rightarrow \pipi$ branching fraction.

Kinematic dependencies of the \myrho meson production cross section on the variables \wgp, $t$, and \My are measured by fitting the mass distributions in bins of the respective variables, such that all model parameters may have kinematic dependencies.
Physical considerations and statistical and technical limitations affect the assumed dependencies. Physical parameters, \ie $m_{\VM}$, $\Gamma_{{\VM}}$, and $\delta_{\nr}$ are assumed to be constants. The small width of the $\omega$ meson cannot be constrained by the present data and the PDG value $\Gamma_{\omega}=8.5~\mev$~\cite{PDG} is assumed and kept fixed in all fits. Dependencies of $f_\omega$, $\phi_\omega$, and $f_{\nr}$ on $t$ or \wgp cannot be constrained with the present data. However, these parameters are allowed to depend on \My, \ie to be different for elastic and proton-dissociative distributions. The non-resonant background is observed to change with $t$. This is modelled by a $t$ dependence of the parameter $\Lambda_{\nr}$, which also can be different for elastic and proton-dissociative distributions. The normalisation $A$ is a free parameter in each kinematic bin. All fit set-ups with the corresponding parameter assumptions are summarised in \tabref{tab:theo_SodingPars}.

\begin{table}\centering
  \small
\begin{tabular}{@{}l@{\hspace{0.5em}} l@{\hspace{1.0em}} c c c c@{}}
\toprule
  &       & \multicolumn{4}{c}{Number of free fit parameters} \\
                                         \cmidrule{3-6}
  Parameter               & Dependencies     & $\frac{\dd \sigma}{\dd m}(m;\, \My)$ & $\frac{\dd \sigma}{\dd m}(m;\, \My,W)$ & $\frac{\dd^2 \sigma}{\dd m \dd t}(m;\, \My, t)$ & $\frac{\dd^2 \sigma}{\dd m \dd t}(m;\, \My, W, t)$\\

\midrule 
$A$                & $\My,\ W,\ t$  & $1^{\el}+1^{\pd}$         & $9_{W}^{\el\vphantom{p}} + 6_{W}^{\pd}$   & $12_{t}^{\el\vphantom{p}} + 9_{t}^{\pd}$          & $4_{W}^{\el\vphantom{p}}\cdot7_{t}^{\el\vphantom{p}} + 4_{W}^{\pd}\cdot5_{t}^{\pd}$ \\
$m_\rho$           & -              & $1$                                 & $1$                                     & $1$                                             & $1$   \\
$\Gamma_\rho$      & -              & $1$                                 & $1$                                     & $1$                                             & $1$   \\
$f_\omega$         & $\My$          & $1^{\el}+1^{\pd}$                   & fixed                                   & fixed                                           & fixed \\
$\phi_\omega$      & $\My$          & $1^{\el}+1^{\pd}$                   & fixed                                   & fixed                                           & fixed \\
$m_\omega$         & -              & $1$                                 & fixed                                   & fixed                                           & fixed \\
$\Gamma_\omega$    & -              & PDG                                 & PDG                                     & PDG                                             & PDG   \\
$f_{\nr}$          & $\My$          & $1^{\el}+1^{\pd}$                   & $1^{\el}+1^{\pd}$                       & $1^{\el}+1^{\pd}$                               & $1^{\el}+1^{\pd}$ \\
$\delta_{\nr}$     & -              & $1$                                 & $1$                                     & $1$                                             & $1$   \\
$\Lambda_{\nr}$    & $\My,\ t$      & $1^{\el}+1^{\pd}$                   & $1^{\el}+1^{\pd}$                       & $12_{t}^{\el\vphantom{p}} + 9_{t}^{\pd}$      & $7_{t}^{\el\vphantom{p}} + 5_{t}^{\pd}$\\
\midrule
  \multicolumn{2}{@{}l}{Total}        & $14$                                & $22$                                    & $47$                                            & $65$  \\
\bottomrule
\end{tabular}

\caption{%
Parameter assumptions and resulting number of parameters used to fit \eqnref{eqn:theo_rhoSoedMass} to the invariant \pipi mass distributions.
The number of parameters depends on the number of bins in the extracted cross sections. The number of bins in which a parameter is fitted freely is given and the corresponding distribution indicated by sub- and superscripts. For the fits of the \mpipi distributions in multiple \wgp or $t$ bins, the $\omega$ meson model parameters are fixed to the values obtained from the fit to the one-dimensional distributions. The $\omega$ meson width is always fixed to the PDG value. }
  \label{tab:theo_SodingPars}
\end{table}

\subsection{Monte Carlo modelling}
\label{sec:theo_MCModel}
For the purpose of quantifying detector effects, the data are modelled using Monte Carlo (MC) simulations of elastic and proton-dissociative electroproduction of $\myrho$, $\omega(782)$, $\phi(1020)$, $\rho(1450)$, and $\rho(1700)$ vector mesons, as well as for non-resonant diffractive photon dissociation. The samples are all generated using the DIFFVM event generator\cite{List:1999} that models VM production on the principles of equivalent photon approximation\cite{equivPhoton}, VDM\cite{VDM}, and pomeron exchange \cite{ReggeTraj}. Proton dissociation is modelled by DIFFVM assuming the following dependence of the cross section on the mass of the dissociated system:
\begin{equation}
 \dfrac{\dd \sigma_\gp}{\dd \MySq} = \dfrac{f(\My)}{ (\MySq)^{1+\epsilon_Y} }\mathrm{.}
\end{equation}
Here, $\epsilon_Y = 0.0808$ and $f(\My)$ is a phenomenological function that is fitted to the experimental data~\cite{ProtonDissoc} to parametrise the low-mass resonance structure in the region $m_p < \My < 1.9~\gev$. For $\My>1.9~\gev$, the function $f(\My)=1$ becomes constant. In the low mass region the dissociative system is treated as an $N^*$ resonance and decays are modelled according to measured branching fractions\cite{PDG}. For higher masses the decay is simulated using the Lund fragmentation model as implemented in JETSET~\cite{Sjostrand:1986hx}. Non-resonant photon dissociation is modelled analogously by assuming a dissociative mass $m_X$ spectrum $\dd \sigma_{\gp}/\dd m_X^2 = 1/(m_X^2)^{1+\epsilon_X}$ with $\epsilon_X = \epsilon_Y$, and simulating the decay using the Lund model.

\begin{table}[htb] \centering
\small
\renewcommand{\arraystretch}{1.1}
\begin{tabular}{ @{}l@{\hspace{2.5em}} l   @{\hspace{2em}}r@{\hspace{2em}} c c@{}}
\toprule
          &             &               & \multicolumn{2}{c}{Number of events}   \\
  \cmidrule{4-5}
  Process & Decay modes & $\BR$ $[\%]$  & elastic & $p$-dissociative \\
   \midrule
    \multirow{2}{*}{$\myrho(770)$}                & $\pi^+\pi^-$       & $   99.0$   & \multirow{2}{*}{$10^7$} & \multirow{2}{*}{$10^7$}  \\
                                                  & $\pipi\gamma$      & $   1.0$    \\
   \multicolumn{3}{l}{$\quad \hookrightarrow$ reweighted to describe all \pipi final states} &\\
   
   \cmidrule(r){1-3} \cmidrule{4-5}
   \multirow{3}{*}{$\omega(782)$}    & $\pi^+\pi^-\pi^0$                           & $  89.2$  & \multirow{3}{*}{$10^6$} & \multirow{3}{*}{$10^6$} \\
                                     & $\pi^0\gamma$                               & $   8.6$    \\
                                     & $\pi^+\pi^-$ (removed, included in signal)  & $   2.2$  \\

   \cmidrule(r){1-3} \cmidrule{4-5}
   \multirow{5}{*}{$\phi(1020)$}    & $K^+ K^-$                                  & $   49.0$         & \multirow{5}{*}{$10^6$}  & \multirow{5}{*}{$10^6$}\\
                                    & $K_L K_S$                                  & $   34.4$  \\
                                    & $\pi^+\rho^-,\ \pi^-\rho^+,\ \pi^0\rho^0$  & $4.3$, $4.3$, $4.3$ \\
                                    & $\pi^+\pi^-\pi^0$                          & $    2.4 $\\
                                    & $\eta \gamma$                              & $    1.3 $\\
   
   \cmidrule(r){1-3} \cmidrule{4-5}
   \multirow{2}{*}{$\rho(1450)$}    & $\rho^0\pi^+\pi^-,\ \rho^+\pi^-\pi^0,\ \rho^-\pi^+\pi^0$ & $25.0$, $25.0$, $ 25.0 $ & \multirow{2}{*}{$10^6$} & \multirow{2}{*}{$10^6$}  \\
   \multirow{2}{*}{\& }             & $\pi^+\pi^-\pi^+\pi^-$                                   & $   15.0$       \\
   \multirow{2}{*}{$\rho(1700)$}    & $\pi^+\pi^-\pi^0\pi^0$                                   & $    8.0$      & \multirow{2}{*}{$10^6$} & \multirow{2}{*}{$10^6$}\\
                                    & $\pi^+\pi^-$ (removed, included in signal)               & $    2.0$ \\ 
   \multicolumn{3}{l}{$\quad \hookrightarrow$ merged $\rho(1450):\rho(1700) = 1:1$ } & \\

   \cmidrule(r){1-3} \cmidrule{4-5}
   \multirow{2}{*}{$\gamma$-dissoc.}  & \multicolumn{2}{l}{Lund fragmentation  model } & \multirow{2}{*}{$10^7$}& \multirow{2}{*}{$10^7$}         \\
                                      & \multicolumn{2}{l}{ (exclusive $\pipi$ removed, included in signal) } \\
\bottomrule
\end{tabular}

  \caption{DIFFVM MC samples used to model the \pipi photoproduction dataset. All decay modes with a branching fraction $\gtrsim1\%$ are simulated~\cite{List:1999}. For the $\rho(1450)$ and $\rho(1700)$ samples a ratio of 1:1 is assumed. The \pipi final states are removed from background samples and included in the signal definition.}
  \label{tab:theo_diffVMSamples}
\end{table}
In \tabref{tab:theo_diffVMSamples}, details on the samples and in particular on the simulated decay modes are listed\footnote{For the simulation of the $\rho(1450)$ and $\rho(1700)$ mesons, DIFFVM was modified to account for the finite width of intermediate $\rho(770)$ resonances, and decay modes involving the final state $\pipi \pi^0\pi^0$ were added.}.  Several of the considered processes result in an exclusive \pipi final state. They are simulated by DIFFVM independently, so that interference effects are not considered. However, these can be significant. For example, the interference between the \myrho meson resonance and non-resonant \pipi production causes a strong skewing of the resonance lineshape. To consider these interference effects, the $\myrho$ meson samples are reweighted to describe exclusive \pipi production including contributions from \myrho, $\omega$, and a single \rprim meson resonance and non-resonant production, that are all added at the amplitude level.
For the reweighting, a $t$ and \mpipi dependent lineshape is used, which is similar to the model introduced in \secref{sec:theo_xSecRho} but is extended by an additional \rprim Breit-Wigner amplitude~\cite{Phd:Abolz}. 

All generated events are passed through the full GEANT-3\cite{Brun:1994aa} based simulation of the H1 apparatus and are reconstructed using the same program chain as used for the data. Trigger scaling factors are applied to correct differences in the trigger performance between data and simulation. They are obtained from a \pipi electroproduction sample, that is triggered independently of the tracking devices~\cite{Phd:Abolz}. 

A template is constructed from all MC samples to describe the data. For a better description of the \wgp and $t$ distributions, all samples are tuned to data~\cite{Phd:Abolz}. An additional background contribution from beam-gas events is estimated in a data driven method. The MC normalisations are obtained from data control regions that are enriched with events from a given process through modified event selection requirements as described below. The $\rho(1450)$ and $\rho(1700)$ samples cannot be well distinguished experimentally in this analysis. They are thus combined at a $1{:}1$ ratio and treated as a single MC sample. In order to obtain normalisations for the elastic and proton-dissociative samples independently, information from the forward detector components is used as described below.

Neither initial and final state radiation of real photons from the electron, nor vacuum polarisation effects are simulated. Consequently, these effects are not corrected for in the present measurement. In a comparable phase space, their effect on the overall cross section has been estimated to be smaller than 2\%~\cite{Kurek:1996ez}. 

\section{Experimental Method and Data Analysis}
\label{sec:experimental}
\subsection{H1 detector}
\label{sec:exp_detector}
A detailed description of the H1 detector is given elsewhere~\cite{Abt:1996xvhi,Appuhn:1996na}. The components that are relevant for the present analysis are briefly described in the following. A right-handed Cartesian coordinate system is used with the origin at the nominal $ep$ interaction point. The proton beam direction defines the positive $z$-axis (\textit{forward direction}). Transverse momenta are measured in the $x$-$y$ plane. Polar ($\theta$) and azimuthal ($\phi$) angles are measured with respect to this frame of reference.

The interaction point is surrounded in the central region ($15^\circ < \theta < 165^\circ$) by the central tracking detector. Two large coaxial cylindrical jet chambers (CJC1 and CJC2) for precise track reconstruction form its core. They are supported by the central inner proportional chamber (CIP) used for the reconstruction of the primary vertex position on the trigger level, a $z$-drift chamber for an improved reconstruction of $z$ coordinates, and a silicon vertex detector for the reconstruction of secondary decay vertices~\cite{Pitzl:2000wz}. In the forward direction (${7^\circ < \theta < 30^\circ}$), additional coverage is provided by the forward tracking detector, a set of planar drift chambers. The tracking detectors are operated in a solenoidal magnetic field of 1.16~T. For offline track reconstruction, track helix parameters are fitted to the inner detector hits in a general broken lines fit~\cite{Blobel:2006yi}. The procedure considers multiple scattering and corrects for energy loss by ionisation in the detector material. The primary vertex position is calculated from all tracks and optimised as part of the fitting procedure. Transverse track momenta are measured with a resolution of ${\sigma(p_T)/p_T \simeq 0.002~p_T/\gev \oplus 0.015}$. The CJCs also provide a measurement of the specific energy loss of charged particles by ionisation $\dEdx$ with a relative resolution of 6.5\% for long tracks. 

The tracking detectors are surrounded by the liquid argon (LAr) sampling calorimeter~\cite{Andrieu:1993kh}. It provides coverage in the region ${4^\circ < \theta < 154^\circ}$ and over the full azimuthal angle. The inner electromagnetic section of the LAr is interlaced with lead, the outer hadronic section with steel absorbers. With the LAr, the energies of electromagnetic and hadronic showers are measured with a precision of ${\sigma(E)/E \simeq 12\%/\sqrt{E/\gev} \oplus 1\%}$ and ${\sigma(E)/E \simeq 50\%/\sqrt{E/\gev}\oplus 2\%}$, respectively~\cite{Andrieu:1994yn}. In the backward region \linebreak (${153^\circ < \theta < 178^\circ}$), energies are measured with a spaghetti calorimeter (SpaCal) of lead absorbers interlaced with scintillating fibres~\cite{Appuhn:1996na}. 

Detector components positioned in the very forward direction are used in this analysis to identify proton dissociation events. These are the forward muon detectors (FMD), the PLUG calorimeter and the forward tagging system (FTS). The lead-scintillator plug calorimeter is positioned around the beampipe at $z=4.9~\m$ to measure the energies of particles in the pseudorapidity region $3.5 < \eta < 5.5$. The FMD is a system of six drift chambers positioned outside of the LAr and covering the range $1.9 < \eta < 3.7$. Particles at larger pseudorapidity up to $\eta \lesssim 6.5$ can still induce spurious signals via secondary particles produced in interactions with the beam transport system and detector support structures~\cite{Phd:Dirkmann}. The very forward region, $6.0 < \eta < 7.5$, is covered by an FTS station of scintillation detectors positioned around the beampipe at $z=28~\m$.

The H1 trigger is operated in four stages. The first trigger level (L1) is implemented in dedicated hardware reading out fast signals of selected sub-detector components. Those signals are combined and refined at the second level (L2). A third, software-based level (L3) combines L1 and L2 information for partial event reconstruction. After full detector read-out and full event reconstruction, events are subject to a final software-based filtering (L4). The data used for the present analysis are recorded using mainly information from the fast track trigger (FTT)~\cite{Baird:2001xc}. The FTT makes it possible to measure transverse track parameters at the first trigger level and complete three-dimensional tracks at L2. This is achieved through applying pattern recognition and associative memory technology to identify predefined tracks in the hit-patterns produced by charged particles in a subset of the CJC signal wires.

The instantaneous luminosity is measured by H1 with a dedicated photon detector located close to the beampipe at $z=-103~\m$. With it, the rate of the Bethe-Heitler process $ep \rightarrow ep\gamma$ is monitored. The integrated luminosity is measured more precisely with the main H1 detector using the elastic QED Compton process. In this process, the electron and photon in the $ep\gamma$ final state have large transverse momenta and can be reconstructed in a back-to-back topology in the SpaCal. The integrated luminosity has been measured with a total uncertainty of $2.7\%$~\cite{Aaron:2012kn} that is dominated by systematic effects.

\subsection{Data sample}
\label{sec:exp_dataSample}
The present analysis is based on data collected by the H1 experiment during the 2006/2007 HERA running period. In that period, the accelerator was operated with positrons having an energy of $E_e = 27.6~\gev$ and protons with an energy of $E_p = 920~\gev$. Due to bandwidth limitations, only a subset of the H1 dataset is available for the trigger conditions relevant for this analysis, corresponding to an effective integrated luminosity of $\mathcal{L}_\mathrm{int} = 1.3~\invpb$. In the kinematic range considered in this analysis, the pions from $\myrho \rightarrow \pipi$ photoproduction are produced within the acceptance of the CJC and with low transverse momenta $p_T \lesssim 0.5~\gev$. In the diffractive photoproduction regime, both the outgoing proton and electron avoid detection by escaping through the beampipe\footnote{In the studied energy region, the elastically scattered protons are mostly outside of the acceptance region of the H1 forward proton spectrometer (FPS) and the very forward proton spectrometer (VFPS)~\cite{FPSVFPS}.}.

\subsubsection{Trigger}
A dedicated, track-based \pipi photoproduction trigger condition was used for online event selection. Track information within the \mbox{2.3~$\mu$s} decision time of the L1 trigger was available through the FTT.  For a positive trigger decision, at least two FTT tracks above a transverse momentum threshold of 160~\mev and at most three tracks above a threshold of 100~\mev were required. The sum of the charges of these tracks was restricted to $\pm 1$ elementary electric charge. In addition, trigger information from the CIP was used to ensure a low multiplicity interaction within the nominal interaction region along the \mbox{$z$-axis}. Vetoes on the inner forward part of the LAr calorimeter and on a scintillator wall in the forward direction were applied to suppress non-diffractive inelastic interactions. Further SpaCal and timing vetoes rejected events from beam-gas and beam-machine interactions. To keep under control the expected rate from the large \myrho meson production cross section, the trigger was scaled down by an average factor of $\sim100$.

\subsubsection{Event reconstruction and selection}
In order to select a sample of \pipi photoproduction events, a set of offline selection cuts is applied on top of the trigger requirements:
\begin{itemize}
  \item The \pipi topology is ensured by requiring exactly two primary-vertex fitted, central tracks to be reconstructed. They need to satisfy some additional quality requirements, have opposite charge, and be within the acceptance region\footnote{The polar acceptance is reduced with respect to the CJC geometry to improve the performance of the \pipi photoproduction trigger and its MC simulation.} defined as $25^\circ < \theta < 155^\circ$ and \mbox{$p_T>0.16~\gev$}. Low-momentum kaons, protons, and deuterons are suppressed using the difference between the measured energy loss $\dEdx$ of the tracks in the CJC and the expected loss for the respective particle hypothesis in a likelihood-based approach. The two tracks are then taken to be the pion candidates, and their 4-momentum vectors are calculated with the corresponding mass hypothesis. 
 \item The photoproduction kinematic regime is ensured by vetoing events with a scattered electron candidate in the SpaCal or LAr. The SpaCal acceptance then limits the photon virtuality to $Q^2\lesssim 2.5~\gevSq$.
 \item The diffractive topology is ensured by requiring a large rapidity gap between the central tracks and any forward detector activity. Events with LAr clusters above a noise level of 0.6~\gev in the forward region $\theta < 20^\circ$ are rejected. Information from the FTD is used to reconstruct forward tracks, and events with more than one forward track that cannot be matched to one of the central tracks are also rejected. The presence of a single unmatched track is permitted to reduce the sensitivity on the modelling of the forward energy flow in the forward detectors. This rapidity gap selection in particular also limits the mass of the proton-dissociative system to approximately $\My \lesssim 10~\gev$.
 \item Background processes with additional neutral particles or charged particles outside of the central tracker acceptance are suppressed by cuts on the LAr and SpaCal energy. LAr and SpaCal clusters above respective noise levels of 0.6~\gev and 0.3~\gev are geometrically matched to the two central tracks: A cluster is associated to a track if it is within a cylinder of a 60~\cm radius in the direction of the track upon calorimeter entry. The energies from clusters not associated to either track are summed up. Events are rejected if the total unassociated LAr or SpaCal energies exceed thresholds of 0.8~\gev or 0.4~\gev, respectively. This allows for a small amount of unassociated energy to account for residual noise or secondary particles produced in interactions of the pion candidates with the detector material. A further suppression of background events with additional final state particles is achieved by requiring a transverse opening angle between the two pion tracks $\Delta \phi > 50^\circ$.
 \item For a reliable trigger performance and MC modelling thereof, the difference in the FTT track angles\footnote{The FTT $\phi$ angle is determined at a radial distance of $r=22~\cm$ from the $z$ axis.} must exceed $\Delta \phi_\mathrm{FTT} > 20^\circ$.
 \item The background is further reduced by rejecting out-of-time events via cuts on the LAr and CJC event timing information. Background events from beam-gas and beam-wall interactions are suppressed by restricting the $z$ coordinate of the primary vertex to be within 25~\cm of the nominal interaction point.
\end{itemize}

The reaction $ep \rightarrow e \pipi Y$ is kinematically underconstrained since only the two pions in the final state are reconstructed. The mass of the \pipi system \mpipirec is reconstructed from the 4-momenta of the two tracks under pion hypothesis. The momentum transfer at the proton vertex $t$ and the scattering energy $\wgp$ are reconstructed from the two pion 4-momenta:
\begin{align}
  \trec   &= -\ptsqrec \label{eqn:exp_trec}\\
  \intertext{and}
  \wgprec &=  \sqrt{2E_p \left( \Epipirec - \pzpipirec \right)}\,\text. \label{eqn:exp_wrec}
\end{align}
Here, $E_p$ denotes the proton-beam energy and $\Epipirec$, $\ptpipirec$, and $\pzpipirec$ are the measured energy, transverse, and longitudinal 4-momentum components of the \pipi system. These two equations are approximations to \eqnref{eqn:theo_Wgp} and \eqnref{eqn:theo_t}, respectively. In some regions of the probed phase space, $Q^2$ may be similar in size to $t$, or $\My$ may be similar in size to \wgp, such that these approximations are poor. Such effects are corrected for in the unfolding procedure discussed later in the text (cf. \secref{sec:exp_unfolding}). 

The analysis phase space probed by this measurement is explicitly defined by detector-level cuts $15 < \wgprec <90~\gev$, $\trec < 3~\gevSq$, and $0.3 <\mpipirec < 2.3~\gev$. The exclusivity requirements, which veto events with detector activity not related to the \pipi pair, further restrict the phase space to $Q^2 \lesssim 2.5~\gevSq$ and $\My\lesssim 10~\gev$. The mean and median $Q^2$ in that phase space are approximately $0.02~\gevSq$ and $8\cdot10^{-6}~\gevSq$, respectively, as evaluated in the MC simulation. 

A total of $943\,962$ \pipi photoproduction event candidates pass all selection requirements. In \figref{fig:exp_ctrl}, the selected number of events is shown as a function of \mpipirec, \wgprec, and \trec. The distributions are compared to the MC model introduced in \secref{sec:theo_MCModel}. The \myrho meson resonance at a mass of $\sim 770~\mev$ clearly dominates the sample. Background contamination amounts to about 11\% and is investigated in the next section.

\subsubsection{Background processes}
The \pipi photoproduction sample includes various background contributions, even after the full event selection. The dominant background processes are the decays of diffractively produced $\omega \rightarrow \pipi \pi^0$, $\phi\rightarrow K^+K^-$, or $\rho^\prime \rightarrow 4\pi$, as well as diffractive photon dissociation. Another source of background originates from interactions of the electron and proton beams with residual gas. Such \textit{reducible} background events are wrongly selected when charged kaons or protons are misidentified as pions or additional charged or neutral particles escape detection, \eg by being outside the central tracker acceptance or having energies below the calorimeter noise threshold. In addition to the \myrho meson, some other vector mesons also decay to an exclusive \pipi state (cf. \tabref{tab:theo_diffVMSamples}). Rather than being treated as \textit{irreducible} background, these are included in the signal for the analysis of the \pipi production cross section.

To study the reducible background contributions in more detail, multiple dedicated control regions are introduced:
\begin{itemize}
  \item $\omega(782)$ mesons predominantly decay into the $\pipi \pi^0$ final state. The $\pi^0$ meson can be identified via energy deposits in the calorimeters that are not associated with either of the two pion tracks. An $\omega$ control region is defined by replacing the empty calorimeter selection by a cut $E_\mathrm{LAr}^\mathrm{!assoc} > 0.8~\gev$ on the unassociated energy deposited in the LAr. Events with an $\omega$ meson are distinguished from those with a \rprim meson by cuts on $\mpipi < 0.55~\gev$ and on the invariant mass of both tracks (assumed to be pions) and all unassociated clusters $m_\mathrm{evt} < 1.2~\gev$. The $\omega$ meson purity achieved in this region is roughly 54\%.
  \item $\phi(1020)$ mesons predominantly decay into pairs of charged kaons. A $\phi$ control region is defined by replacing the $\dd E/\dd x$ pion identification selection cuts by a kaon selection and requiring the invariant $K^+K^-$ mass to be within 15~\mev of the $\phi$ meson mass. Also the cut on the opening angle between the two tracks at the vertex is removed. The $\phi$ meson purity achieved in this region is roughly 89\%. 
  \item The excited $\rprim$ meson dominantly decay into 4 pions in various charge configurations. Due to the track veto in the trigger, additional tracks cannot be used to identify $\rprim \rightarrow 4\pi$ events. Instead, unassociated energy deposits in the LAr  $E_\mathrm{LAr}^\mathrm{!assoc} > 0.8~\gev$ are required in place of the empty LAr signal selection. Events with a $\rprim$ meson are distinguished from those with an $\omega$ meson by requiring $m_\mathrm{evt}>1.2~\gev$. The $\rprim$ meson purity achieved in this region is roughly 48\%.
  \item Particles from photon dissociation emerge primarily in the backwards direction. A photon dissociation control region is thus defined by replacing the empty SpaCal signal requirement by a cut $ 4 < E_\mathrm{SpaCal}^\mathrm{!assoc} < 10~\gev$ on the unassociated energy deposit in the SpaCal. The lower cut removes $\omega$ and $\rprim$ events, the upper cut is retained as a veto against the scattered electron. The photon dissociation purity achieved in this region is roughly 78\%.
\end{itemize}

For the \myrho meson photoproduction cross section measurement, reducible background processes are subtracted in the unfolding procedure, where the templates for the respective diffractive background processes are taken from the DIFFVM MC samples, as described in \secref{sec:theo_MCModel}. The respective normalisation factors are determined by making use of the control regions in the unfolding process. The residual beam-gas induced background is modelled in a data driven approach, using events from electron and proton {\em pilot} bunches. For electron (proton) pilot bunches, there is a corresponding gap in the proton (electron) beam bunch structure, such that only interactions with rest-gas atoms may occur. The beam gas background shape predictions estimated from pilot bunch events are scaled to match the integrated beam current of the colliding bunches. The beam-gas induced background amounts to about 2\%.

\subsubsection{Proton dissociation tagging}
\label{sec:forwardTagging}
\noindent
In proton-dissociative events, the proton remnants are produced in the very forward direction where the H1 detector is only sparsely instrumented. Consequently, the remnants cannot be fully reconstructed, and elastic and proton-dissociative scattering cannot be uniquely identified on an event-by-event basis. However, in many cases some of the remnants do induce signals in the forward instruments, either directly or via secondary particles that are produced in interactions with the detector or machine infrastructure. These signals are used to tag proton-dissociative events (\textit{tagging fraction}). At a much lower level, such signals can also be present in elastic scattering events (\textit{mistagging fraction}), \eg in the presence of detector noise or when the elastically scattered proton produces secondaries in interactions with the beam transport system. By simulating the respective tagging and mistagging fractions of proton-dissociative and elastic scattering, the respective proton-dissociative and elastic contributions to the cross section can be determined. 

The forward detectors used for tagging in this analysis are the FMD, the PLUG, and the FTS. An event is considered to be tagged by the FMD if there is at least one hit in any of the first three FMD layers, by the PLUG if there is more than one cluster above a noise level of 1.2~\gev, or by the FTS if it produces at least one hit. A small contribution to the FTS signal, induced by secondary particles produced by the elastically scattered proton hitting a beam collimator, is discarded by applying acceptance cuts depending on $\trec$ and the location of hits in the FTS~\cite{Phd:Abolz}. 

The tagging information from the three detectors is combined by summing the total number of tags in an event: $ 0 \leq N_\mathrm{tags} \leq 3$. In turn, three tagging categories are defined: a zero-tag ($N_\mathrm{tag} = 0$), single-tag ($N_\mathrm{tag} = 1$), and multi-tag ($N_\mathrm{tag} \geq 2$) category. The respective tagging fractions for events passing the standard selection cuts are shown in \figref{fig:exp_ctrl_tagEff_ptsq} as a function of \trec. The tagging categories are used to split the dataset into 3 tagging control regions. The zero-tag region is dominated by $\sim 90\%$ elastic events, in the single-tag region their fraction is reduced to $\sim 64\%$, and in the multi-tag region proton dissociation dominates at $\sim 91\%$.

\subsection{Unfolding}
\label{sec:exp_unfolding}
An unfolding procedure is applied to correct binned reconstructed detector-level distributions for various detector effects. The unfolding corrects the data for reducible background contributions, the finite resolution of reconstructed variables, and efficiency and acceptance losses. Furthermore, it is set up to separate elastic and proton-dissociative scattering events. This makes possible the determination of elastic and proton-dissociative particle level distributions from which the corresponding differential \pipi photoproduction cross sections are derived. The cross sections are measured in a fiducial phase space that is slightly smaller than the analysis phase space defined by the event selection. This makes it possible to account for contributions migrating into and out of the fiducial phase space. The fiducial and analysis phase space cuts are summarised in \tabref{tab:exp_fiducialPS}. 

\begin{table}\centering
\small
\renewcommand{\arraystretch}{1.2}
\setlength{\tabcolsep}{3pt}
\begin{tabular}{@{}c@{\hspace{5em}}  c@{}}
\toprule
\multicolumn{1}{@{}l}{Analysis phase space} & \multicolumn{1}{@{}l}{Fiducial measurement phase space} \\
  \midrule

 \begin{tabular}[t]{ @{}r c c c  r l@{} }
             15.0     &$<$ & $\wgp$                 & $<$ & 90.0  & \gev \\
                      &    & $|t|$                  & $<$ & 3.0 & \gevSq \\
             0.3      &$<$ & $\mpipi$               & $<$ & 2.3 & \gev\\
                      &    & $Q^2$                  & $<$ & 2.5 & \gevSq  \\
                      &    & \multirow{2}{*}{$\My$} & \multirow{2}{*}{$<$} & \multirow{2}{*}{10.0} & \multirow{2}{*}{\gev}\\
  \end{tabular}
  & 
  \begin{tabular}[t]{ @{}l   r c c c  r l@{}}
             &20.0       &$<$ & $\wgp$    & $<$ & 80.0  & \gev \\
             &         &    & $|t|$     & $<$ & 1.5 & \gevSq \\
             &0.5      &$<$ & $\mpipi$  & $<$ & 2.2 & \gev\\
             &         &    & $Q^2$     & $<$ & 2.5 & \gevSq\\
    \multicolumn{1}{@{}l}{elastic:} & & & $\My$ & $=$ & $m_p$ \\ 
    \multicolumn{1}{@{}l}{$p$-dissociative:} & $m_p$ & $<$ & $\My$ & $<$ & 10.0 & \gev \\
  \end{tabular}\\
\bottomrule
\end{tabular}

  \caption{Analysis and fiducial measurement phase space. At detector level, the respective cuts are applied to \wgprec, \trec, and \mpipirec, the $Q^2$ cut is replaced by the veto on the reconstruction of the scattered electron, and the $\My$ cut is replaced by the rapidity gap requirement as detailed in the text. }
  \label{tab:exp_fiducialPS}
\end{table}

\subsubsection{Regularised template fit}
Differential cross section measurements are performed for various distributions of the variables $\mpipi$, \wgp, and $t$, and combinations thereof. The beam-gas background template is subtracted from the considered data distribution and the result is used as input to the unfolding. The unfolding is performed by means of a regularised template fit within the TUnfold framework~\cite{Schmitt:2012kp}. A response matrix $\bm A$ is introduced, with elements $A_{ij}$ describing the probability that an event generated in bin $j$ of a truth-level distribution $\vec{x}$ is reconstructed in bin $i$ of a detector-level distribution $\vec{y}$. 
In the definition of the response matrix, elastic and proton-dissociative signal and background MC processes are included as dedicated sub-matrices. This makes it possible to separate the elastic and proton-dissociative signal components in the unfolding. Also, background subtraction is implicitly performed during the unfolding where the normalisation of the backgrounds is determined in the fit. At detector level, the response matrix is split into signal and background control regions as defined above. The signal region is further split into three and the background regions into two orthogonal forward tagging categories. This constrains the respective MC contributions in the template fit. Migrations into and out of the fiducial phase space are considered by including side bins in each sub-matrix both at detector and at truth level~\cite{Phd:Abolz}. These contain events passing the analysis phase space cuts but failing the fiducial cuts (cf. \tabref{tab:exp_fiducialPS}).
As an illustration, the response matrix used for unfolding the one-dimensional \mpipirec distributions is given in \figref{fig:exp_respMatrix_mpipi}. For all response matrices, most truth bins are found to have good constraints by at least one reconstruction level bin with a purity above 50\%. As a result, the statistical correlations between bins of the unfolded one-dimensional distributions are small. The somewhat larger statistical correlations in the unfolding of multi-dimensional distributions are reduced by introducing a Tikhonov regularisation term on the curvature into the template fit~\cite{Schmitt:2012kp,tikonov}. The truth-level MC distributions, that are adjusted to data as explained in \secref{sec:theo_MCModel}, are used as a bias in the regularisation. However, regularisation is not used in the unfolding of the one-dimensional \mpipi distributions. This makes possible the measurement of the $\omega$ meson mass and production cross section, which were found to be very sensitive to any sort of regularisation.

\subsubsection{\pipi cross section definition}
\label{sec:exp_xSecPiPi}
Differential \pipi photoproduction cross sections are calculated from the number of unfolded \pipi events. The double-differential cross section in \mpipi and $t$ is defined as
\begin{equation}
  \left[\dfrac{\dd^{2} \sigma(\gamma p \rightarrow \pipi Y) (\mpipi, t; \wgp)}{ \dd t \, \dd \mpipi} \right]_j = 
      \dfrac{ \hat{x}_j }{ \Delta t \, \Delta \mpipi} \dfrac{1}{ \mathcal{L}\,  \phiWW\left(\wgp\right)  }
      \label{eqn:exp_xSecDef}
\end{equation}
in bin $j$ of an unfolded distribution and with $\hat x_j$ being the number of unfolded events. Differentiation is approximated by dividing the event yields by the \mpipi and $t$ bin widths, $\Delta t$ and $\Delta \mpipi$, respectively. No bin centre correction is performed for the variables \mpipi and $t$. Instead, predictions or fit functions are integrated over the respective bin width when comparing to data. The event yields are normalised by the integrated $ep$ luminosity $\mathcal{L}$. Photoproduction cross sections are calculated by normalising the $ep$ cross sections by means of the integrated effective photon flux $\phiWW$, that is calculated for the relevant \wgp ranges according to \eqnref{eqn:theo_intFlux}.

\subsection{Cross section uncertainties}
\label{sec:systematics}
The measured cross sections are subject to statistical and systematic uncertainties. Statistical uncertainties originate from the limited size of the available data and MC samples. Systematic uncertainties originate from the MC modelling of the signal and background processes and their respective kinematic distributions, as well as from the simulation of the detector response. Detector uncertainties are estimated by either varying the simulated detector response to MC events or by modifying the event selection procedure simultaneously in both the MC samples and data.  Model uncertainties are estimated by varying the generation and reweighting parameters of the MC samples.

\subsubsection{Statistical uncertainties}
Two sources of statistical uncertainties are considered: uncertainties of the data distributions and uncertainties of the unfolding factors obtained from the MC samples. The data uncertainties include contributions from the subtraction of the beam-gas background. Statistical uncertainties are propagated through the unfolding as described in the TUnfold documentation~\cite{Schmitt:2012kp}. The finite detector resolution imposes correlations on the statistical uncertainties after the unfolding.

\subsubsection{Experimental systematic uncertainties}
In order to estimate uncertainties related to the knowledge of the H1 detector response, the unfolding is repeated with systematically varied response matrices. Two types of variations are considered for different sources of uncertainty. Variations of the MC samples on detector level are introduced and are propagated to the cross sections by repeating the unfolding with varied response matrices. Other uncertainties are estimated by modifying the selection procedure. Such variations affect both the migration matrices and the input data distributions simultaneously. For two-sided variations, the full absolute shifts between the nominal and varied unfolded distributions are used to define up and down uncertainties. For one-sided variations, the full shift is taken as a two-sided uncertainty~\cite{Phd:Abolz}. 

Concerning the trigger, the following uncertainties are considered:
\begin{itemize}
  \item The trigger scale factors for the CIP and FTT correction are varied to account for statistical and shape uncertainties. A further uncertainty is estimated to cover the propagation of the correction factors from the electroproduction regime, where they are derived, to photoproduction~\cite{Phd:Abolz}.
  \item The trigger scale factors for the correction of the forward vetoes are varied to account for statistical and shape uncertainties~\cite{Phd:Abolz}. 
  \item Pions in the very backward direction may potentially interfere with the SpaCal trigger veto via secondary particles produced in {\em nuclear interactions}. Nuclear interactions of the pions with the beam transport system and detector material are not modelled perfectly by the simulation. To estimate a potential impact, the cut on the upper track polar angle acceptance is varied between $150^\circ$ and $160^\circ$.
\end{itemize}

Concerning the track reconstruction and simulation, the following uncertainties are considered: 
\begin{itemize}
  \item The uncertainty of the modelling of the geometric CJC acceptance in $p_T$ is estimated by varying the acceptance cut to $p_T >0.18~\gev$.
  \item The uncertainty for potential mismodelling of the forward tracking detectors is estimated by varying the veto on the number of forward tracks to allow either none or up to two tracks in the selection.
  \item The uncertainty of the MC $z$-vertex distribution is estimated by varying the MC distribution to cover small discrepancies in the mean and tails relative to the observed $z$-vertex distribution in data.
  \item In the particle identification, the cuts on the kaon, proton, and deuteron \dEdx rejection likelihoods are varied up and down while simultaneously varying the pion \dEdx selection likelihood down and up.
  \item Inhomogeneities in the magnetic field may not be fully modelled in the simulation. Nor may the modelling of nuclear interactions of the pions with the detector material be completely accurate. These effects are estimated to result in an uncertainty on the modelled track $p_T$ resolution of 20\%.  Their impact is evaluated by smearing the reconstructed track $p_{T,\mathrm{rec}} \rightarrow p_{T,\mathrm{rec}} \pm 0.2\, \left(p_{T,\mathrm{rec}}-p_{T,\mathrm{gen}} \right)$ with respect to the generated true $p_{T,\mathrm{gen}}$ in the signal \pipi MC samples. The result is propagated to all kinematic variables reconstructed from the two pion 4-momenta.
  \item Similarly, a 20\% uncertainty is assumed on the resolution of the track $\theta$ measurement to cover potential mismodellings in the simulation.
\end{itemize}

Concerning the track momentum scale, the following uncertainties are considered:
\begin{itemize}
  \item The field strength of the H1 solenoid is known to a level of 0.3\%~\cite{Abt:1996xvhi} or better. To estimate a potential impact on the absolute track $p_T$ scale, the reconstructed MC $p_T$ values are varied up and down by $\pm 0.3\%$. The variation is performed simultaneously for both tracks.
  \item The energy loss correction depends on a precise knowledge of the material budget, which is known at a level of 7\%. Through the energy loss correction in the track reconstruction, this uncertainty results in an average track $p_T$ uncertainty of $0.4~\mev$.
\end{itemize}

Concerning the calorimeters, the following uncertainties are considered:
\begin{itemize}
  \item The noise cuts on LAr and SpaCal energy clusters are independently varied up and down by $\pm 0.2~\gev$ and $\pm 0.1~\gev$, respectively.
  \item The energy scales of the LAr and SpaCal clusters are independently varied by $\pm 10\%$. These variations are applied prior to the respective cluster noise cut.
  \item A mismodelling of nuclear interactions may also affect the calorimeter response. A potential impact is estimated by varying the cut applied for associating calorimeter clusters to tracks by $\pm 10~\cm$. 
\end{itemize}

Concerning the forward detectors, the following tagging uncertainties are considered:
\begin{itemize}
  \item The tagging and mistagging fractions of the forward detectors are varied independently. The FTS tagging fraction is varied by $\pm 5\%$ in the proton-dissociative MC samples and the mistagging fraction by $\pm 50\%$ in the elastic samples. Similarly, these fractions in the FMD are varied by $\pm 5\%$ and $\pm 5\%$ and in the PLUG by $\pm 5\%$ and $\pm 100\%$, respectively. The sizes of the variations are estimated to cover observed mismodellings in the tagging fractions of the respective detectors.
\end{itemize}

\subsubsection{Model uncertainties}

In order to estimate model uncertainties, the MC samples are modified on generator-level. Model uncertainties are propagated to the cross sections by repeating the unfolding with the varied migration matrices and MC bias distributions. Cross section uncertainties are then calculated from the shifted unfolded spectra in the same manner as for the experimental uncertainties. Generally, the unfolding approach is rather insensitive to many aspects of the MC modelling. The following model variations are considered:
\begin{itemize}
  \item Uncertainties on the $Q^2$ and $\My$ dependencies of the MC samples are estimated~\cite{Alexa:2013xxa}. The $Q^2$ dependence of the MC samples is varied by applying an event weight ${(1+Q^2/m_{\VM}^2)^{\pm 0.07}}$ in agreement with experimental uncertainties~\cite{Aaron:2009xp}. The $\My$ dependence of the proton-dissociative sample is varied by applying a weight ${(1/\MySq)^{\pm 0.15}}$.
  \item The tuning parameters of the MC samples derived to describe the kinematic data distributions (\secref{sec:theo_MCModel}) are independently varied up and down~\cite{Phd:Abolz}. 
  \item For the $\rprim$ background MC samples, an uncertainty on the relative ratio of $\rho(1450)$ and $\rho(1700)$ is estimated by varying it from $1{:}1$ up and down to $2{:}1$ and $1{:}2$, \linebreak respectively. An uncertainty on the $\rprim$ decay modes is estimated by varying \linebreak ${\BR(\rprim \rightarrow \rho^\pm \pi^\mp \pi^0) = (50\pm25)\%}$ while simultaneously scaling all other decay modes proportionally. 
  \item A shape uncertainty on the mass distribution of the photon-dissociative mass is estimated by reweighting the distribution by $(1/m_{X}^2)^{\pm 0.15}$. 
\end{itemize}

\subsubsection{Normalisation uncertainties}

Normalisation uncertainties are directly applied to the unfolded cross sections. The following uncertainties are considered:
\begin{itemize}
  \item The uncertainty of the track reconstruction efficiency due to modelling of the detector material is 1\% per track, leading to a 2\% normalisation uncertainty.
  \item The integrated luminosity of the used dataset is known at a precision of 2.7\%.
  \item Higher order QED effects have been estimated to be smaller than 2\%~\cite{Kurek:1996ez} which is taken as a normalisation uncertainty.
\end{itemize}
This results in a combined 3.9\% normalisation uncertainty on the measured cross sections. 

\subsection{Cross section fits}
\label{sec:exp_fits}
For the analysis, various parametric models are fitted to the measured cross section distributions. Fits are performed by varying model parameters to minimise a $\chi^2$ function. Parametrisations of differential distributions are integrated over the respective kinematic bins and divided by their bin widths in order to match the differential cross section definition (cf. \eqnref{eqn:exp_xSecDef}). The $\chi^2$ function takes into account only the statistical uncertainties, represented by the covariance matrix. Nominal fit parameters are obtained by fitting the nominal measured cross section distributions. The corresponding $\chi^2$ value relative to the number of degrees of freedom $\chirStat$ is used as a measure of the quality of a fit. Systematic uncertainties are then propagated through a fit via an \textit{offset method}, where each systematic cross section variation is fitted independently. The resulting shifts in the fit parameters relative to the nominal parameters are used to define the systematic parameter uncertainties.

\section{Results}
\label{sec:results}
\noindent
Elastic and proton-dissociative \pipi photoproduction cross sections are measured. The measurement is presented integrated over the fiducial phase space and as one-, two-, and three-dimensional cross section distributions as functions of \mpipi, \wgp, and $t$.
Subsequently, \myrho meson photoproduction cross sections as functions of \wgp and $t$ are extracted via fits to the invariant \pipi mass distributions (cf.~\eqnref{eqn:theo_rhoSoedMass}).
The procedure is exemplified with the differential \pipi photoproduction cross sections \dSigmaPiPiYdm as functions of \mpipi. One- and two-dimensional \myrho meson photoproduction cross section distributions are then parametrised and interpreted using fits. In particular, a Regge fit of the elastic \myrho meson production cross section as a function of \wgp and $t$ makes possible the extraction of the effective leading Regge trajectory $\alpha(t)$.

\subsection{Fiducial \pipi photoproduction cross sections}
\label{sec:res_sigmaPiPiFid}
\noindent
The fiducial \pipi electroproduction cross section is measured in the phase space defined in \tabref{tab:exp_fiducialPS} by unfolding one-dimensional \mpipirec distributions, integrating the event yield over all bins, and normalising the result by the integrated luminosity. It is subsequently turned into a photoproduction cross section by normalisation to the integrated effective photon flux. In the phase space considered, the integrated effective flux calculated with \eqnref{eqn:theo_intFlux} is \phiWW = 0.1368\text.
This yields a \pipi photoproduction cross section of
\begin{equation}
  \sigmaPiPiY = 16.20 \pm 0.05\ (\stat)\ {}^{+1.11}_{-1.15}\ (\syst)\ \mub \text, \ \text{ for }\ m_p \leq \My < 10~\gev \text. 
\end{equation}
The cross section is measured at an average energy $\langle \wgp \rangle \simeq 43~\gev$ as estimated from the MC simulation. The elastic and proton-dissociative components of the cross section are separated through the unfolding which yields
\begin{alignat}{3}
  \sigmaPiPip =&&\   11.&52 &&\pm 0.06\ (\stat)\ {}^{+0.76}_{-0.78}\ (\syst)\ \mub \text{ and}      \\
  \sigmaPiPiY =&&\    4.&68 &&\pm 0.06\ (\stat)\ {}^{+0.62}_{-0.64}\ (\syst)\ \mub \text, \ \text{ for }\ m_p < \My < 10~\gev \text, 
\end{alignat}
respectively\footnote{The fiducial cross sections can be obtained by alternatively unfolding the $\trec$, \wgprec, or multi-dimensional distributions. The variations of the results are similar in size to the statistical uncertainties. Comparable statistical and systematic uncertainties are observed.}. The uncertainties of the two components are correlated with a statistical and symmetrised total Pearson correlation coefficient of $\rho_\mathrm{stat} = -0.59$, and $\rho_\mathrm{tot} = +0.30$, respectively. The measurements are statistically very precise at a level of 0.5\% (1.2\%) but have large systematic uncertainties of  6.3\% (13.2\%) for the elastic (proton-dissociative) component. The uncertainty of the elastic component is dominated by the trigger and the normalisation uncertainties of 4.1\% and 3.9\%, respectively, whereas the uncertainties associated with the tagging and the calorimeter dominate the proton-dissociative component at 8.4\% and 7.3\%, respectively.  A more detailed breakdown of the cross section uncertainties is given in \tabref{tab:res_fidUncGroups}. 

\begin{table}[!ttt]\centering
\small
\begin{tabular}{@{}l@{\hspace{4em}} c@{\hspace{5em}} c@{}} \\ 
\toprule
  & \multicolumn{2}{c}{Relative cross section uncertainty [\%]} \\ 
\cmidrule{2-3}
  Source of uncertainty  & ${{\My} = m_p}$ & {${m_p < {\My} < 10~{\gev}}$}\\ 
  \midrule
    Statistical                                        & 0.5  & 1.2     \\[6pt]
    Trigger                                            & 4.1  & 5.3     \\ 
    Tracking                                           & 1.4  & 1.3     \\ 
    Momentum scale                                     & 0.1  & 0.1     \\
    Calorimeter                                        & 1.5  & 7.3     \\ 
    Tagging                                            & 2.0  & 8.4     \\[6pt] 
    Normalisation                                      & 3.9  & 3.9     \\[6pt] 
    MC model ($\My,Q^2,$ bgr.)                         & 2.0  & 2.7     \\ 
    MC model ($\mpipi,\wgp,t$)                         & 0.1  & 0.4     \\ 
    \midrule
    Total                                              & 6.6  & 13.3 $\  $     \\ 
  \bottomrule 
\end{tabular} 
    \caption{Summary of the combined impact of systematic uncertainties on the fiducial \pipi photoproduction cross sections. The numbers are obtained by unfolding the one-dimensional \mpipi distribution and with symmetrised systematic uncertainties.}
    \label{tab:res_fidUncGroups}
\end{table}

Photoproduction of \pipi mesons has a fairly large cross section and forms a sizeable contribution to the total $\gamma p$ cross section, which is known for similar \wgp from cosmic ray experiments~\cite{Vereshkov:2003cp}. Within the restrictions of the fiducial phase space in \mpipi, $t$, and \My, the elastic and proton-dissociative processes contribute to the total cross section by about 9\% and 4\%, respectively. The elastic \pipi photoproduction cross section has been measured before in $ep$ collisions at slightly higher \wgp \cite{Aid:1996bs,Breitweg:1997ed}. Considering the differences in \wgp and the energy dependence of the elastic cross section (see below), the measurements are found to be consistent. There has not been a previous measurement of the proton-dissociative \pipi photoproduction cross section in a comparable phase space.

\subsection{Differential \pipi photoproduction cross sections}
\label{sec:res_sigmaRhoFid}
\noindent
The elastic and proton-dissociative differential \pipi photoproduction cross sections \linebreak \dSigmaPiPiYdm are measured as functions of \mpipi by unfolding one-dimensional \mpipirec distributions. Numerical results are given in \apxref{apx:sec_pipiXSecTables} and the cross section distributions are displayed in \figref{fig:results_dSigma_dm}. The photoproduction of \pipi mesons is dominated by the \myrho meson resonance peaking at a mass of around 770~\mev. The differential cross sections fall off steeply towards higher masses with a second broad excited $\rho^\prime$ meson resonance appearing at around 1600~\mev. At the \myrho meson resonance peak, the proton-dissociative cross section is around 40\% of the elastic cross section with the difference becoming smaller towards higher masses. The uncertainties of the measurement vary slowly with \mpipi. At the \myrho meson peak the statistical and systematic uncertainty of the elastic differential cross section are roughly 1\% and 7\%, respectively. At the position of the $\rho^\prime$ peak they have grown to 7\% and 13\%, respectively. The corresponding uncertainties of the differential proton-dissociative cross section are about twice as large.

\subsubsection{Fit of the \mpipi dependence}

In order to extract the \myrho meson contributions to the \pipi photoproduction cross sections, the \mpipi lineshape defined in \eqnref{eqn:theo_rhoSoedMass} is fitted to the data.
As this paper focusses on the analysis of \myrho meson production, the fit is only performed in the analysis range $0.6~\gev \leq \mpipi \leq 1.0~\gev$ where contributions from excited $\rho^\prime$ meson states can be neglected. The parametrisation is fitted simultaneously to the elastic and proton-dissociative distributions with the parameter settings described in \tabref{tab:theo_SodingPars}. The model describes the data well with the fit yielding $\chirStat= 24.6/24$.
The resulting model parameters are presented in \tabref{tab:results_dSigma_dm_pars}.
The fitted curves are shown in \figref{fig:results_dSigma_dm_fit}, where they are compared to the measured data. 
\begin{table}[!thb]\centering
\small
\renewcommand{\arraystretch}{1.3}
\begin{tabular}{@{}l@{\hspace{4em}} d{3.4}  d{1.4}  l c@{\hspace{4em}}  d{3.4}  d{1.4}  l@{}} 
\toprule
Shared parameter & \multicolumn{1}{l}{Value} & \multicolumn{1}{l}{$\Delta_{\stat}$} & \multicolumn{1}{l}{$\Delta_{\syst}$} &\\ 
\midrule
$               m_{\rho}~[\mev]$ & 770.8 &  1.3  &$ ^{+2.3}_{-2.4}$ \\ 
$          \Gamma_{\rho}~[\mev]$ & 151.3 &  2.2  &$ ^{+1.6}_{-2.8}$ \\ 
$             m_{\omega}~[\mev]$ & 777.9 &  2.2  &$ ^{+4.3}_{-2.2}$ \\ 
$        \Gamma_{\omega}~[\mev]$ & 8.5   & \multicolumn{2}{l}{PDG fixed}  \\ 
$                   \delta_{nr}$ & 0.76  &  0.35 &$ ^{+0.14}_{-0.08}$ \\
\midrule
& \multicolumn{3}{c}{$\My = m_p$} && \multicolumn{3}{c}{$m_p < \My < 10~{\gev}$} \\ 
\cmidrule{2-4} \cmidrule{6-8}
Independent parameter & \multicolumn{1}{l}{Value} & \multicolumn{1}{l}{$\Delta_{\stat}$} & \multicolumn{1}{l}{$\Delta_{\syst}$} && \multicolumn{1}{l}{Value} & \multicolumn{1}{l}{$\Delta_{\stat}$} & \multicolumn{1}{l}{$\Delta_{\syst}$}\\ 
\midrule
$             A~[\mub/\gev^{2}]$ &   48.4  &   0.8   & $^{+3.3}_{-3.2}$      &&   20.8   &   0.4    & $^{+2.7}_{-2.6}$\\ 
$                    f_{\omega}$ &   0.166 &   0.017  & $^{+0.008}_{-0.023}$ &&   0.135  &   0.042 & $^{+0.042}_{-0.036}$\\ 
$                 \phi_{\omega}$ &   -0.53 &   0.22  & $^{+0.21}_{-0.17}$    &&   -0.02  &   0.34   & $^{+0.31}_{-0.19}$\\ 
$                        f_{nr}$ &   0.189 &   0.026  & $^{+0.025}_{-0.016}$ &&   0.145  &   0.025  & $^{+0.029}_{-0.014}$\\ 
$       \Lambda_{nr}~[\gev]$     &   0.18  &   0.59  & $^{+0.20}_{-0.10}$    &&   0.1    &   1.3   & $^{+0.4}_{-0.3}$\\
\bottomrule
\end{tabular}\\
\caption{%
Free parameters for the fit of the single-differential elastic and proton-dissociative cross section \dSigmaPiPiYdm as a function of \mpipi.
The nominal fit values and statistical and systematic uncertainties are given. A full uncertainty breakdown and statistical correlations are provided~\cite{H1_data}. The corresponding fit is shown in \figref{fig:results_dSigma_dm_fit}.}
  \label{tab:results_dSigma_dm_pars}
\end{table}
Within the fit model, the differential \pipi photoproduction cross sections obtain significant contributions from the \myrho meson resonance and non-resonant \pipi production.
The two corresponding amplitudes give rise to large interference terms that result in a skewing of the resonance lineshapes. At the measured \myrho meson mass, the relative contributions of the \myrho resonance and non-resonant \pipi production to the elastic differential \pipi photoproduction cross section are $[79.1 \pm 3.2~(\stat)~{}^{+4.4}_{-1.4}~(\syst)]\%$  and $[7.1 \pm 0.6~(\stat)~{}^{+0.7}_{-0.5}~(\syst)]\%$, respectively. The corresponding contributions to the proton-dissociative cross section are $[85.9 \pm 3.8~(\stat)~{}^{+3.8}_{-3.2}~(\syst)]\%$  and $[4.8 \pm 0.6~(\stat)~{}^{+0.6}_{-0.5}~(\syst)]\%$, respectively. Most noticeably, the relative non-resonant contribution is smaller for proton-dissociative than for elastic scattering resulting in a weaker skewing. In the analysed mass range, the non-resonant differential cross section exhibits little dependence on \mpipi. For this reason the two fit parameters $\delta_{\nr}$ and $\Lambda_{\nr}$ are strongly correlated and cannot be constrained very well individually.
Qualitatively the result of the fit is similar to past HERA \pipi photoproduction analyses~\cite{Breitweg:1997ed}.
Quantitative comparisons of fit parameters are not possible because different parametrisations were used.
Variants of the fit were also explored but did not significantly improve the compatibility with data. When including a possible admixture of incoherent background, the respective contribution is found to be compatible with zero. This conclusion is largely independent of the assumed incoherent background shape. Models including barrier factors \cite{VonHippel:1972fg} also can be fitted well by the data. Obviously, all extracted particle masses and widths are model dependent. A detailed discussion of these effects is beyond the scope of this paper.

The rich structure of the interference term near 770~\mev, visible in \figref{fig:results_dSigma_dm_fit}, originates from the $\omega$ resonance.
In the fit model, the $\omega$ resonance amplitude prevents the net interference contribution from vanishing at the \myrho meson mass and shifts the position of the sign change of the interference term towards the $\omega$ meson mass.
In the \pipi production cross section this is visible as a characteristic step.
The impact of the $\omega$ meson on the measured differential cross sections is similar in size to observations made in $e^+e^- \rightarrow \pipi$ production~\cite{ee_mrho}. In the present analysis, this effect is measured for the first time at HERA.

The masses of the $\myrho$ and $\omega$ mesons and the width of the $\myrho$ meson obtained from the fit are compared to the world average PDG values~\cite{PDG} in \tabref{tab:res_mgRhoComp}.
The PDG lists two sets of \myrho parameters from \myrho meson photoproduction and production in $e^+e^-$ scattering, respectively. The value measured for the \myrho mass is compatible with previous photoproduction measurements~\cite{Breitweg:1997ed,Abramowicz:2011pk,Bartalucci:1977cp}, confirming a slightly lower \myrho mass parameter as compared to typical $e^+e^-$ measurements~\cite{ee_mrho}. The mass uncertainty is dominated by the uncertainties in the magnetic field and the material in the detector affecting the track $p_T$ scale.  The precision of the presented value is similar to the most precise previous photoproduction measurements. The \myrho width is consistent with both PDG values and offers an uncertainty as good as previous photoproduction measurements. The $\omega$ mass is also compatible with the PDG value but has a sizeable uncertainty. The observed difference between the measured and the PDG $\omega$ mass is similar to that of the $\myrho$ mass relative to the $e^+e^-$ measurement.

\begin{table}[!ttt] \centering
\small
\renewcommand{\arraystretch}{1.3}
\setlength{\tabcolsep}{1em}
 \begin{tabular}{@{}l@{\hspace{3em}} c   c c c@{}}
 \toprule
 Parameter                    & This measurement        & PDG $\gamma p$ & PDG $e^+e^-$             & $\Delta$($e^+e^-$, H1)\\
\midrule
    $ m_{\myrho}[\mev]$       & $770.8~{}^{+2.6}_{-2.7}$   & $769.0\pm 1.0$          & $775.26 \pm 0.25$     & $4.5 ^{+2.7}_{-2.6}$  \\
    $ \Gamma_{\myrho}~[\mev]$ & $151.3~{}^{+2.7}_{-3.6}$   & $151.7\pm 2.6$          & $147.8\ \, \pm 0.9\ \,$  &  \\
    $ m_\omega~[\mev]$        & $777.9~{}^{+4.8}_{-3.1}$   &                         & $782.65\pm 0.12$           & $4.8 ^{+3.1}_{-4.8}$ \\
\bottomrule
\end{tabular}
  \caption{Comparison of measured $\myrho$ and $\omega$ meson properties to the PDG values~\cite{PDG}. Only the total uncertainties are given. The last column gives the mass differences between the PDG $e^+e^-$ and the present measurement.}
  \label{tab:res_mgRhoComp}
\end{table}

\subsubsection{Fiducial \myrho and $\omega$ meson photoproduction cross sections}

The result of the fit is used to calculate the \myrho meson contributions to the fiducial \pipi photoproduction cross sections.
This is achieved by integrating the \myrho meson resonance amplitude in the range $2m_\pi < \mpipi < 1.53~\gev$ as discussed in \secref{sec:theo_xSecRho}. The integration yields 
\begin{alignat}{3}
  \sigmaRhop =&&\   10.&97 &&\pm 0.18\ (\stat)\ {}^{+0.72}_{-0.73}\ (\syst)\ \mub \text{ and}      \\
  \sigmaRhoY =&&\    4.&71 &&\pm 0.09\ (\stat)\ {}^{+0.60}_{-0.59}\ (\syst)\ \mub \text, \ \text{ for }\ m_p < \My < 10~\gev \text, 
\end{alignat}
for the elastic and proton-dissociative $\myrho$ meson photoproduction cross sections, respectively. The ratio of the elastic and proton-dissociative \myrho meson photoproduction cross sections in the measurement phase space is
\begin{equation}
  \dfrac{\sigmaRhoY}{\sigmaRhop} = 0.429 \pm 0.009\ (\stat)\ {}^{+0.056}_{-0.053}\ (\syst) \text, \ \text{ for }\ m_p < \My < 10~\gev \text,
\end{equation}
taking into account correlations. The measured elastic \myrho meson photoproduction cross section is consistent with previous HERA measurements~\cite{Aid:1996bs,Derrick:1995vq,Breitweg:1997ed}, when accounting for \wgp differences between the measurements. Suitable reference measurements for the proton-dissociative cross section are not available. However, the ratio of the proton-dissociative to the elastic \myrho meson photoproduction cross section is consistent with a previous measurement~\cite{Breitweg:1997ed}, assuming that phase space differences are covered by the large uncertainties.

With the fit model \eqnref{eqn:theo_rhoSoedMass}, also the indirect measurement of the fiducial $\omega$ meson photoproduction cross sections is possible by integrating the $\omega$-\myrho mixing amplitude over the range $2m_\pi \leq \mpipi \leq 0.82~\gev$.
The result is corrected for the $\omega \rightarrow \pipi$ branching fraction $\mathcal{BR}(\omega \rightarrow \pipi) = 0.0153 \pm 0.0006\ (\tot)$~\cite{PDG} to yield the cross sections
\begin{alignat}{3}
  \sigma(\gamma p \rightarrow \omega  p) 
            =&&\   1.&06 &&\pm 0.26\ (\stat)\ {}^{+0.13}_{-0.30}\ (\syst)\ \mub \ \text{ and}      \\
  \sigma(\gamma p \rightarrow \omega Y)
            =&&\   0.&31 &&\pm 0.23\ (\stat)\ {}^{+0.19}_{-0.16}\ (\syst)\ \mub \text, \ \text{ for }\ m_p < \My < 10~\gev \text. 
\end{alignat}
The uncertainty of the branching fraction is included in the systematic uncertainty. The ratios of the $\omega$ meson to the $\myrho$ meson photoproduction cross sections are then determined to be
\begin{alignat}{2}
  \dfrac{ \sigma(\gamma p \rightarrow \omega p)}{\sigmaRhop} =&\ 0.097 &&\pm 0.020\ (\stat)\ {}^{+0.011}_{-0.026}\ (\syst) \ \text{ and} \\
  \dfrac{ \sigma(\gamma p \rightarrow \omega Y)}{\sigmaRhoY} =&\ 0.065 &&\pm 0.041\ (\stat)\ {}^{+0.042}_{-0.032}\ (\syst) \text, \ \text{ for }\ m_p < \My < 10~\gev \text.
\end{alignat} 
The elastic $\omega$ photoproduction cross section has been directly measured at HERA in the $\omega \rightarrow \pipi \pi^0$ channel by the ZEUS collaboration. In a similar phase space, a compatible value for the cross section was observed~\cite{Derrick:1996yt}. The present value is also consistent with the expectation that the \myrho and $\omega$ meson photoproduction cross sections differ by a factor $c_{\omega}/c_{\rho} \simeq 1/9$ that originates from SU(2) flavour symmetry and the quark electric charges.


\subsection{Energy dependence of \myrho meson photoproduction}
\label{sec:res_sigmaRhoOfW}
\noindent
The energy dependencies of the elastic and proton-dissociative \myrho meson photoproduction cross sections are obtained by unfolding two-dimensional distributions $\wgprec \otimes \mpipirec$ and calculating the corresponding two-dimensional \pipi cross sections. Numerical results are presented in \apxref{apx:sec_pipiXSecTables} and the cross section distributions are displayed in \figref{fig:results_dSigma_dmW_wmFit_el} and \figref{fig:results_dSigma_dmW_wmFit_pd} for the elastic and proton-dissociative components, respectively.
To extract the \myrho meson contributions, \eqnref{eqn:theo_rhoSoedMass} is fitted simultaneously to the \mpipi distributions in every unfolded \wgp bin with the parameter settings described in \tabref{tab:theo_SodingPars}.
In particular, the $\omega$ meson model parameters cannot be constrained well by the fit and are fixed to the values obtained in the fit to the one-dimensional mass distributions. The fit gives $\chirStat = 222.0/188$. The fitted curves are also shown figures \ref{fig:results_dSigma_dmW_wmFit_el} and \ref{fig:results_dSigma_dmW_wmFit_pd}.

The fit results are used to calculate the elastic and proton-dissociative \myrho meson photoproduction cross sections \sigmaRhoY as a function of \wgp.
The results are given in \tabref{tab:tab_xSec_wRho} and displayed in \figref{fig:results_sigmaRho_w_fit}. The cross sections only have weak dependencies on \wgp, as is expected for diffractive processes at such energies. Many sources of uncertainty considered here are observed to vary with \wgp. For the elastic cross section, the full systematic uncertainty is about 6\% in the centre of the considered phase space and increases to 10\% and 8\% at the lowest and highest considered \wgp, respectively. The corresponding proton-dissociative uncertainties are about twice as large. 

\subsubsection{Fit of the energy dependence} 
The energy dependencies of the elastic and proton-dissociative \myrho meson photoproduction cross sections in the range ${20 < \wgp < 80~\gev}$ are parametrised by a simple power law:
\begin{equation}
 \sigmarho(\wgp) = \sigmarho(W_0)\ \left(\dfrac{\wgp}{W_0} \right)^\delta\,.
 \label{eqn:results_sigma_w_fit_fcn}
\end{equation}
The parametrisation is fitted simultaneously to the measured elastic and proton-disso\-cia\-tive distributions assuming independent elastic and proton-dissociative fit parameters. The function is evaluated at the geometric bin centres, and the reference energy is set to $W_0 = 40~\gev$. The fit yields a very poor $\chirStat = 33.0 / 11 $, where however systematic uncertainties are not considered. The resulting fit parameters are presented in \tabref{tab:results_sigmaRho_w_pars}. The fitted curves are compared to the data in \figref{fig:results_sigmaRho_w_fit}.
The fitted power parameter ${\delta_{\el}=0.171\pm0.009~(\stat)\ {}^{+0.039}_{-0.026}~(\syst)}$ characterises the slow rise of the elastic cross section with increasing scattering energy. The measured value is compatible with previous measurements, \eg by the ZEUS collaboration~\cite{Breitweg:1997ed} or the CMS Collaboration ($pPb$)~\cite{Sirunyan:2019nog}, but is more precise. The fit results in a proton-dissociative parameter ${\delta_{\pd}=-0.156\pm0.026~(\stat)\ {}^{+0.081}_{-0.079}~(\syst)}$ that is significantly different from the elastic parameter and suggests a decrease of the proton-dissociative cross section with increasing energy. However, the proton-dissociative cross section is expected to be strongly shaped by the restriction of the fiducial phase space; most notably by the requirement $\My < 10~\gev$. Since energy is required to excite high masses \My, the cut suppresses the cross section more strongly for high than for low \wgp. The phase space is not accounted for in \eqnref{eqn:results_sigma_w_fit_fcn} so that the proton-dissociative result $\delta_{\pd}$ cannot be interpreted directly in terms of a scattering amplitude. 

In \figref{fig:results_sigmaRho_w_hist}, the measured elastic \myrho meson photoproduction cross section data are compared to selected measurements by fixed-target~\cite{Ballam:1971wq,Park:1971ts,Ballam:1972eq,Struczinski:1975ik,Egloff:1979mg,Aston:1982hr}, HERA~\cite{Aid:1996bs,Derrick:1995vq,Breitweg:1997ed}, and LHC~\cite{Sirunyan:2019nog} experiments\footnote{The phase space differences between the measurements with respect to the considered $t$ ranges are found to be negligible ($\lesssim 2\%$) relative to much larger overall normalisation uncertainties.}. The present elastic data are in agreement with other measurements and connect the fixed-target data with the high energy regime. The combination of data makes possible the study of the region of reggeon exchange at small scattering energies and of pomeron exchange at large energies simultaneously. For describing the energy range $2 < \wgp < 200~\gev$, the energy dependence of the elastic cross section is parametrised by the sum of two power-law functions:
\begin{equation}
  \sigmarho(\wgp) = \sigmarho(W_0)\ \left( \left(\dfrac{\wgp}{W_0} \right)^{\delta_{\pom}} + f_{\reg} \left(\dfrac{\wgp}{W_0} \right)^{\delta_{\reg}} \right). 
  \label{eqn:results_sigma_w_fit_fcn_DL}
\end{equation}
The two parameters $\delta_{\pom}$ and $\delta_{\reg}$ are associated with pomeron exchange at high and reggeon exchange at low \wgp, respectively. The parametrisation is fitted to the HERA data~\cite{Aid:1996bs,Derrick:1995vq,Breitweg:1997ed} together with data from fixed-target experiments~\cite{Ballam:1971wq,Park:1971ts,Ballam:1972eq,Struczinski:1975ik,Egloff:1979mg,Aston:1982hr} as shown in \figref{fig:results_sigmaRho_w_hist}. In the fit, previous measurements enter with their respective uncorrelated and normalisation uncertainties. For the H1~\cite{Aid:1996bs} and ZEUS measurements~\cite{Breitweg:1997ed}, normalisation uncertainties of 8\% and 9\%, respectively, are extracted from the total uncertainties. Correlations between experiments are not considered. The uncorrelated uncertainties are accounted for in the data covariance, together with the statistical uncertainties of the present measurement. The normalisation uncertainties are included by offsetting the data and repeating the fit, as is also done for 
systematic and normalisation uncertainties of the present data.

The fit yields $\chirStat = 84.3/43$ and results in $\delta_{\pom} = 0.207 \pm 0.015~(\stat) ^{+0.053}_{-0.033}~(\syst)$ and $\delta_{\reg} = -1.45 \pm 0.12~(\stat) ^{+0.35}_{-0.21} ~(\syst)$. All fit parameters are given in table \ref{tab:results_world_w_pars}.
The fitted curve is compared to the data in \figref{fig:results_sigmaRho_w_hist}. The extracted parameters are consistent with the broader picture of vector meson electroproduction, in particular with the observed dependence of $\delta_{\pom}$ on the vector meson mass~\cite{Favart:2015umi}. The fit further suggests a small reggeon contribution of ${f_{\reg} = [2.0 \pm 0.7~(\stat)\ {}^{+2.9}_{-1.3}~(\syst)]\%}$ at $W_0=40~\gev$. For this reason, the parameter $\delta_{\el}$ from the fit with a single power-law function (\tabref{tab:results_sigmaRho_w_pars}, $\My=m_p$) is interpreted as an effective parameter describing both pomeron and reggeon exchange contributions. 

\subsection{$t$ dependence of \myrho meson photoproduction}
\label{sec:res_dSigmaRhoOft}
\noindent
The $t$ dependencies of the elastic and proton-dissociative \myrho meson photoproduction cross sections are obtained by unfolding two-dimensional distributions $\trec \otimes \mpipirec$ and calculating the corresponding two-dimensional \pipi cross sections. Numerical results are presented in \apxref{apx:sec_pipiXSecTables} and the cross section distributions are displayed in \figref{fig:results_dSigma_dmdt_mtFit_el} and \figref{fig:results_dSigma_dmdt_mtFit_pd}  for the elastic and proton-dissociative component, respectively.
To extract the \myrho meson contributions, \eqnref{eqn:theo_rhoSoedMass} is fitted simultaneously to the \mpipi distributions in every unfolded $t$ bin with the parameter settings described in \tabref{tab:theo_SodingPars}.
The fit yields ${\chirStat = 353.4/247}$. The fitted curves are also shown figures \ref{fig:results_dSigma_dmdt_mtFit_el} and \ref{fig:results_dSigma_dmdt_mtFit_pd}. 

The fit results are used to calculate the elastic and proton-dissociative differential \myrho meson photoproduction cross sections $\dSigmaRhoYdt$ as a function of $t$.
The results are given in \tabref{tab:tab_xSec_tRho} and displayed in \figref{fig:results_dSigmaRho_dt_fit}. The cross sections fall of exponentially with $t$ as is expected for diffractive processes. The relative systematic uncertainties of roughly 6\% for the elastic distribution vary little with $t$ and increase moderately to 11\% in the highest $|t|$ bin. For the proton-dissociative distribution, the uncertainties increase from roughly 13\% at intermediate $|t|$ to 17\% in both the highest and lowest considered $|t|$ bins.

\subsubsection{Fit of the $t$ dependence}

The $t$ dependencies of the elastic and proton-dissociative \myrho meson photoproduction cross sections are parametrised by the function
\begin{align}
  \dfrac{\dd \sigmarho}{\dd t}(t) =  \dfrac{\dd \sigmarho}{\dd t}(t=0)\ \left(1 - \dfrac{ b\, t}{a} \right)^{-a} \text.
 \label{eqn:results_dSigma_dt_fit_fcn}
\end{align}
It interpolates between an exponential $\dd \sigmarho/\dd t \propto \exp\left(b\, t \right) $ at low $|t|$ and a power-law $ \dd \sigmarho / \dd t \propto |t|^{-a}$ at large $|t|$. The function is fitted simultaneously to the elastic and proton-dissociative distributions assuming independent elastic and proton-dissociative model parameters. To match the cross section definition, the function is integrated over each bin and divided by the respective bin width in the fit. The proton-dissociative distribution is affected by phase space restrictions that are not accounted for in the parametrisation. The impact is particularly large at small $|t|$ because excitations with large $\My$ require a minimal\footnote{ $|t| \gtrsim \dfrac{\MySq}{W_{\gp}^4}  \left(m_{\pi\pi}^2+Q^2\right)^2$~\cite{Abramowicz:1999eq}} $|t|$. For this reason, the lowest $|t|$ bin of the proton-dissociative distribution is not included in the fit. The fit yields ${\chirStat = 15.3 / 14 }$. The resulting fit parameters are presented in \tabref{tab:results_dSigmaRho_dt_pars}. The fitted curves are compared to data in \figref{fig:results_dSigmaRho_dt_fit}.
The exponential $t$ dependencies of the elastic and proton-dissociative cross sections at small $|t|$ are quantified by the parameters ${b_{\el} = 9.61\pm0.15~(\stat)\ {}^{+0.20}_{-0.15}~(\syst)~\gevSqInv}$ and ${b_{\pd} = 4.81\pm0.24~(\stat)\ {}^{+0.39}_{-0.37}~(\syst)~\gevSqInv}$, respectively. With $b_{\pd} \sim 1/2\, b_{\el}$, the elastic spectrum falls off much more steeply than the proton-dissociative spectrum. The difference between the exponential slope parameters is $b_{\el} - b_{\pd} = 4.80 \pm 0.29~(\stat)\ {}^{+0.41}_{-0.40}~(\syst)$, taking into account correlations. In the optical interpretation (or in an eikonal model approach~\cite{eikonalModel}), this difference reflects the difference in the size of the respective target, \ie proton in elastic vs parton in proton-dissociative interactions. The measured value is consistent with HERA measurements in \myrho meson electroproduction~\cite{Aaron:2009xp}. There is evidence for a deviation from the purely exponential behaviour given by the finite values for $a_{\el} = 20.4\pm3.7~(\stat)\ {}^{+6.8}_{-5.1}~(\syst)$ and $a_{\pd} = 8.5\pm1.7~(\stat)\ {}^{+2.7}_{-2.1}~(\syst)$. The smaller $a_{\pd}$ indicates a stronger deviation for the proton-dissociative than for the elastic cross section. The present measurements of the $t$ dependencies of the elastic and proton-dissociative \myrho meson photoproduction cross sections can be compared with other HERA measurements~\cite{Aid:1996bs,Breitweg:1997ed}. The measured values for the elastic slope $b_{\el}$ are somewhat lower than observed previously but seem compatible within uncertainties when taking into account the differences in energy and the thus expected shrinkage. The present value is the most precise of these measurements. The observed deviation of the elastic slope from the exponential behaviour is compatible with the ZEUS measurement \cite{Breitweg:1997ed}. Similar to the elastic case, the slopes $b_{\pd}$ in proton-dissociative scattering are compatible with earlier measurements. The present measurement is more precise and does account for deviations from the exponential form. Differences in the definition of the proton-dissociative phase space were not considered in these comparisons.

\subsection{\myrho meson photoproduction as a function of $t$ and \wgp}
\label{sec:res_dSigmaRhoOfWt}
\noindent
The two-dimensional $t$ and \wgp dependencies of the elastic and proton-dissociative \myrho meson photoproduction cross sections are obtained by unfolding three-dimensional distributions $\trec \otimes \wgprec \otimes \mpipirec$ and calculating the corresponding \pipi cross sections\footnote{The underlying response matrix has 1243 bins on detector and 882 bins on truth level.}. Numerical results are presented in \apxref{apx:sec_pipiXSecTables} and the cross section distributions are displayed in \figref{fig:results_dSigma_dmdtW_twmFit_el} and \figref{fig:results_dSigma_dmdtW_twmFit_pd} for the elastic and proton-dissociative component, respectively.
To extract the \myrho meson contributions, \eqnref{eqn:theo_rhoSoedMass} is fitted simultaneously to the \mpipi distributions in every unfolded \wgp and $t$ bin with the parameter settings described in \tabref{tab:theo_SodingPars}.
The fit yields $\chirStat = 804.0/607$. The fitted curves are also shown in figures \ref{fig:results_dSigma_dmdtW_twmFit_el} and \ref{fig:results_dSigma_dmdtW_twmFit_pd}.  

The fit results are used to calculate the elastic and proton-dissociative differential \myrho meson photoproduction cross sections $\dSigmaRhoYdt$ as a function of $t$ and \wgp.
The results are given in \tabref{tab:tab_xSec_twRho_el} and \tabref{tab:tab_xSec_twRho_pd} for the elastic and proton-dissociative component, respectively, and are shown in \figref{fig:results_dSigmaRho_dtW_twfit}.

\subsubsection{Regge fit of the $t$ and $\wgp$ dependence}
The elastic and proton-dissociative \myrho meson photoproduction cross sections are parametrised by the function
\begin{equation} 
 \dfrac{\dd \sigmarho}{\dd t}(t;\wgp) = \dfrac{\dd \sigmarho}{\dd t}(t;W_0) \left( \dfrac{\wgp}{W_0} \right)^{4 (\alpha(t)-1) } \text,
 \label{eqn:sigma_walpha_fcn}
\end{equation}
with an energy dependence predicted by Regge theory and with a single leading trajectory $\alpha(t)$. The $t$ dependence at the reference energy $W_0=40~\gev$ is parametrised according to \eqnref{eqn:results_dSigma_dt_fit_fcn}. Regge trajectories are expected to continue linearly only in the region of small $|t|$. Since the analysis extends to $|t| \leq 1.5~\gevSq$, a modified parametrisation based on a Fermi function is used:
\begin{equation}
  \alpha(t) = \alpha_0 + \beta\, \left( \left(\ee^{ - \frac{4\alpha_1 t}{\beta} }+1\right)^{-1}  -\dfrac{1}{2}\right)\text .
  \label{eqn:sigma_alpha_fermi}
\end{equation}
It approximates a linear function $\alpha_0 + \alpha_1\, t$ for small $|t|$ and approaches a constant value ${\alpha_0 - \beta/2}$ for ${t \rightarrow -\infty}$. Such a lower bound is expected for large $|t|$ from QCD calculations~\cite{BoundReggeTraj} but deviations from a linear behaviour have not been observed yet in this reaction in the $t$ range considered here. The parametrisation in \eqnref{eqn:sigma_walpha_fcn} is fitted simultaneously to the elastic and proton-dissociative cross sections assuming independent elastic and proton-dissociative fit parameters. In the $\chi^2$ calculation, the function is integrated over each $t$ bin and evaluated at the geometric \wgp bin centres to match the cross section definition. Following the argumentation discussed above, the respective lowest $|t|$ bins of the proton-dissociative cross sections are excluded from the fit. The fit yields $\chirStat =31.7/32 $. The resulting fit parameters are presented \tabref{tab:results_fPar_wtRhoFit}. The fitted curves are compared to data in \figref{fig:results_dSigmaRho_dtW_twfit}. 

\subsubsection{Shrinkage of the elastic differential cross section}
The fit parameters in \tabref{tab:results_fPar_wtRhoFit} that describe the $t$ dependencies of the differential \myrho meson photoproduction cross sections at $W_0=40~\gev$ are in agreement with the parameters from the one-dimensional fit of the $t$ dependencies that is described above.  For the elastic differential cross section, the finite slope of the measured trajectory $\alpha(t)$ (see below) results in a shrinkage of the forward peak with increasing \wgp. At small $|t|$, where the differential cross section $\dSigmaRhopdt$ falls exponentially with $t$ and the trajectory is linear, the shrinkage can be expressed in terms of a \wgp dependence of the exponential slope parameter
\begin{equation}
  b_{\el}(\wgp) = b_{\el}\left(W_0\right) + 4\alpha_{1} \log\left(\dfrac{\wgp}{W_0} \right)\text.
\end{equation}
The function is plotted in \figref{fig:results_fPar_wtRho_b} using the parameters extracted for the elastic cross section in the two-dimensional fit (cf. \tabref{tab:results_fPar_wtRhoFit}, $m_Y{=}m_p$). For comparison, $b(\wgp)$ is measured in every \wgp bin by fitting the parametrisation \eqnref{eqn:results_dSigma_dt_fit_fcn} with free fit parameters $b_t$ to the $t$ dependencies in all \wgp bins. Deviations from the exponential form are accounted for by including a single parameter $a$ common to all for \wgp bins.
The result is presented in \tabref{tab:tab_fPar_bRho} and also shown in \figref{fig:results_fPar_wtRho_b}. Further data from previous HERA~\cite{Aid:1996bs,Derrick:1995vq,Breitweg:1997ed} and fixed-target~\cite{Ballam:1971wq,Park:1971ts,Ballam:1972eq,Struczinski:1975ik,Egloff:1979mg,Aston:1982hr} measurements are included in the figure, as well. The present slope values are somewhat lower than those from previous HERA measurements but are much more precise. The present data alone clearly indicate that there is shrinkage of the elastic peak with increasing energy.

\subsubsection{Effective leading Regge trajectory}
For visualisation of the Regge trajectory parameters, $\alpha(t)$ is measured separately in each $t$ bin by fitting a simple power law $\propto \wgp^{4(\alpha_t - 1)}$ with free fit parameters $\alpha_t$ to the \wgp distributions in all $t$ bins. The resulting parameters $\alpha_t$ are presented in \tabref{tab:tab_fPar_alphaRho} and displayed as a function of the $t$ bin centres $t_{bc}$\footnote{
  The bin centres $t_{bc}$ are defined to satisfy $\frac{\dd\sigma}{\dd t}\left(t_{bc}\right) = \frac{1}{\left(t_\mathrm{max}-t_\mathrm{min}\right)} \int_{t_\mathrm{min}}^{t_\mathrm{max}} \frac{\dd \sigma}{\dd t}\left(t\right)\, \dd t$ and calculated using \eqnref{eqn:results_dSigma_dt_fit_fcn} and the fit parameters for the elastic and proton-dissociative cross-sections given in \tabref{tab:results_dSigmaRho_dt_pars}.
  } in \figref{fig:results_fPar_wtRho_alpha}. The data are compared to the parametrisations $\alpha(t)$ obtained in the two-dimensional fit and to other curves as discussed below. A direct fit of a parametrisation $\alpha(t)$ to the so extracted $\alpha_t$ is ambiguous because of the definition of the bin centres $t_{bc}$.

For the elastic cross section, the measured trajectory is linear at small $|t|$. The intercept \linebreak and slope parameters are measured at ${\alpha_{0}=1.0654 \pm 0.0044~(\stat)\ {}^{+0.0088}_{-0.0050}~(\syst)}$ and \linebreak ${\alpha_{1} =  0.233 \pm 0.064~(\stat)\ {}^{+0.020 }_{-0.038 }~(\syst) ~\gevSqInv}$, respectively. Potential non-linearities start occurring for $t\lesssim -0.2~\gevSq$ and the trajectory is compatible with approaching a constant asymptotic value of approximately $0.98 \pm 0.04~(\tot)$ for $t \rightarrow -\infty$. However, a linear trajectory over the full considered $t$ range is also able to describe the elastic cross section. The resulting intercept and slope parameters are ${\alpha_{0,\mathrm{lin}} = 1.0624 \pm 0.0033~(\stat)~{}^{+0.0083}_{-0.0052}~(\syst)}$ and $\alpha_{1,\mathrm{lin}} = 0.175 \pm 0.026~(\stat)~{}^{+0.021}_{-0.027}~(\syst)~\gevSqInv$, respectively. For comparison, the linear trajectory is also included in \figref{fig:results_fPar_wtRho_alpha}. 
In contrast, the proton-dissociative cross section is not compatible with a linear trajectory $\alpha(t)$. The fitted parametrisation falls off steeply at low $|t|$ but is constant with a value of approximately ${0.93 \pm 0.03~(\tot)}$ for $t \lesssim -0.2~\gevSq$. However, significant shaping effects are expected from the phase space restrictions, as discussed earlier. No attempt is made here to relate these fit parameters with the underlying amplitude.

In the high energy limit, the elastic cross section is expected to be governed by the soft pomeron trajectory. The canonical pomeron trajectory has been determined by Donnachie and Landshoff from $pp$ and $p\bar{p}$ scattering data as $\alpha_\mathrm{DL}(t) = 1.0808 + 0.25\,t/\gevSq$~\cite{Donnachie:1983hf}. The trajectory parameters have been investigated in various vector meson photo- and electroproduction measurements. A recent analysis of HERA data with a two tensor pomeron model furthermore has lead to a precise measurement of the soft pomeron intercept using inclusive DIS cross sections at low Bjorken-$x$ and total photoproduction cross sections~\cite{Britzger:2019lvc}. In \myrho meson photoproduction, the trajectory has previously been directly extracted~\cite{Breitweg:1999jy} in an analysis of the energy dependence of the differential cross section \dSigmaRhopdt in the range $8.2 < \wgp < 94~\gev$ as measured by various experiments~\cite{Aston:1982hr,Aid:1996bs,Breitweg:1997ed,Derrick:1996vw,Breitweg:1999jy}. The present analysis has the advantage that it measures the leading trajectory from a single dataset. In \figref{fig:results_alphaPom_comp}, the measured intercept and slope at $t=0$ are compared to the canonical pomeron trajectory parameters and those measured by the cited works~\cite{Britzger:2019lvc,Breitweg:1999jy}. The measured intercept seems to be comparatively low. The analysis of the one-dimensional energy dependence (\secref{sec:res_sigmaRhoOfW}) suggests the presence of a subleading reggeon contribution that indeed may have lowered the measured effective trajectory by about $0.01$ units. For the measured slope, good agreement is observed with the canonical DL value. However, there is some tension with previous HERA measurements, such as the cited ZEUS measurement. A potential explanation could arise from possible deviations from the linear form of the trajectory, which are taken into account in the present analysis. For proton-dissociative \myrho meson photoproduction, a constant trajectory has been previously measured at HERA for higher $|t|$ than are considered here~\cite{Chekanov:2002rm}.

\section{Summary}
\label{sec:Summary}
Elastic and proton-dissociative \pipi and \myrho meson photoproduction cross sections are measured as a function of the invariant \pipi mass \mpipi, the scattering energy \wgp, and the squared momentum transfer at the proton vertex $t$. The measurement covers a phase space of $0.5 < \mpipi < 2.2~\gev$, $20 < \wgp < 80~\gev$, and $|t|<1.5~\gevSq$. The cross sections are obtained by unfolding up to three-dimensional \pipi distributions. In the procedure, the elastic and proton-dissociative components are extracted simultaneously. The results are more precise than previous measurements, in particular the measurement of the proton-dissociative cross sections.

The \pipi mass spectra are analysed with a model in which the skewing of the \myrho meson resonance peak is attributed to interference with non-resonant \pipi production, as originally proposed by S\"oding in 1966.
A fit of the model to the data yields ${m_\rho = 770.8 \pm 1.3~(\stat)\ {}^{+2.3}_{-2.4}~(\syst) ~\mev}$ and ${\Gamma_\rho = 151.3 \pm 2.2~(\stat)\ {}^{+1.6}_{-2.8}~(\syst) ~\mev}$ for the mass and width of the \myrho meson, respectively. For the first time at HERA, the sensitivity of the data is sufficient to constrain the $\omega$ meson contribution to \pipi photoproduction and measure the $\omega$ meson mass at $m_\omega = 777.9 \pm 2.2~(\stat)\ {}^{+4.3}_{-2.2}~(\syst) ~\mev$. 

The fit is used to extract the \myrho meson contributions to the elastic and proton-dissociative \pipi photoproduction cross sections.
One- and two-dimensional \myrho meson photoproduction cross sections as functions of \wgp and $t$ are presented and subsequently interpreted with phenomenological models in the context of Regge theory. Precise parameters describing the $t$ and \wgp dependencies are measured. The slope parameters describing the exponential drop of the cross sections with $t$ are measured at an average energy of $\langle \wgp \rangle = 43~\gev$ as ${b_{\el} = 9.61\pm0.15~(\stat)\ {}^{+0.20}_{-0.15}~(\syst)~\gevSqInv}$ and 
$b_{\pd} = 4.81\pm0.24~(\stat)\ {}^{+0.39}_{-0.37}~(\syst)~\gevSqInv$ for the elastic and proton-dissociative components, respectively. From the analysis of the elastic cross section \dSigmaRhopdt as a function of \wgp and $t$ the effective leading Regge trajectory is extracted. Allowing for deviations from the linear form at large $|t|$, an intercept of $\alpha(t{=}0) = 1.0654 \pm 0.0044~(\stat)\ {}^{+0.0088}_{-0.0050}~(\syst)$ and a slope of $\alpha^\prime(t{=}0) = 0.233 \pm 0.064~(\stat)\ {}^{+0.020 }_{-0.038 }~(\syst) ~\gevSqInv$ are measured. Since the probed energies are low, the intercept measurement may be obscured somewhat by sub-leading reggeon contributions that cannot be constrained by the present data alone. There are some indications that the trajectory deviates from a linear behaviour for $t < -0.2~\gevSq$. Within uncertainties, the data are compatible with the trajectory becoming equal to unity for large $|t|$. 

\section*{Acknowledgements}

We are grateful to the HERA machine group whose outstanding efforts have made this experiment possible. We thank the engineers and technicians for their work in constructing and maintaining the H1 detector, our funding agencies for financial support, the DESY technical staff for continual assistance and the DESY directorate for support and for the hospitality which they extend to the non DESY members of the collaboration. We would like to give credit to all partners contributing to the EGI computing infrastructure for their support for the H1 Collaboration.

We express our thanks to all those involved in securing not only the H1 data but also the software and working environment for long term use allowing the unique H1 dataset to continue to be explored in the coming years. The transfer from experiment specific to central resources with long term support, including both storage and batch systems has also been crucial to this enterprise. We therefore also acknowledge the role played by DESY-IT and all people involved during this transition and their future role in the years to come.


\newpage
{
\begin{table}[!htp]
\footnotesize \centering
\renewcommand{\arraystretch}{1.3}
\setlength{\tabcolsep}{5pt}
\begin{tabularx}{\textwidth}{|c|c|}
\hline
\begin{tabular}[t]{ c  r | d{3.3} d{3.3} l}  
\multicolumn{5}{l}{$m_Y = m_p$} \\ 
\midrule 
$W_{\gamma p} $ range & \multicolumn{1}{c|}{$\Phi_{\gamma/e}$}& \multicolumn{1}{c}{$\sigma_{\rho} $}          & \multicolumn{1}{c}{$\Delta_{\stat}$} & \multicolumn{1}{c}{$\Delta_{\syst}$}  \\ 
$[\gev]$  &      & \multicolumn{1}{c}{$[\mub]$} & \multicolumn{1}{c}{$[\mub]$} & \multicolumn{1}{c}{$[\mub]$} \\ 
\midrule 
$20.0 - 25.0$ & 0.0248  & 9.93 & 0.11 & $ ^{+0.80}_{-0.94} $\\ 
$25.0 - 30.0$ & 0.0195  & 10.33 & 0.10 & $ ^{+0.75}_{-0.77} $\\ 
$30.0 - 35.0$ & 0.0160  & 10.57 & 0.10 & $ ^{+0.66}_{-0.68} $\\ 
$35.0 - 40.0$ & 0.0134  & 10.71 & 0.11 & $ ^{+0.61}_{-0.64} $\\ 
$40.0 - 45.0$ & 0.0115  & 11.13 & 0.11 & $ ^{+0.64}_{-0.67} $\\ 
$45.0 - 50.0$ & 0.0101  & 11.53 & 0.12 & $ ^{+0.72}_{-0.73} $\\ 
$50.0 - 56.0$ & 0.0105  & 11.82 & 0.13 & $ ^{+0.82}_{-0.82} $\\ 
$56.0 - 66.0$ & 0.0147  & 11.80 & 0.12 & $ ^{+0.87}_{-0.86} $\\ 
$66.0 - 80.0$ & 0.0163  & 11.73 & 0.16 & $ ^{+0.93}_{-0.92} $\\ 
 \end{tabular} 

\begin{tabular}[t]{ c  r | d{3.3} d{3.3} l}  
\multicolumn{5}{l}{$m_p < m_Y < 10~\gev$} \\ 
\midrule 
$W_{\gamma p} $ range & \multicolumn{1}{c|}{$\Phi_{\gamma/e}$}& \multicolumn{1}{c}{$\sigma_{\rho} $}          & \multicolumn{1}{c}{$\Delta_{\stat}$} & \multicolumn{1}{c}{$\Delta_{\syst}$}  \\ 
$[\gev]$  &      & \multicolumn{1}{c}{$[\mub]$} & \multicolumn{1}{c}{$[\mub]$} & \multicolumn{1}{c}{$[\mub]$} \\ 
\midrule 
$20.0 - 26.0$ & 0.0290  & 4.86 & 0.13 & $ ^{+0.71}_{-0.69} $\\ 
$26.0 - 32.0$ & 0.0220  & 4.89 & 0.11 & $ ^{+0.65}_{-0.63} $\\ 
$32.0 - 38.0$ & 0.0175  & 4.85 & 0.11 & $ ^{+0.60}_{-0.58} $\\ 
$38.0 - 46.0$ & 0.0188  & 4.69 & 0.10 & $ ^{+0.57}_{-0.56} $\\ 
$46.0 - 56.0$ & 0.0185  & 4.46 & 0.11 & $ ^{+0.61}_{-0.58} $\\ 
$56.0 - 80.0$ & 0.0311  & 4.15 & 0.11 & $ ^{+0.55}_{-0.55} $\\ 
&& \\
&& \\
&& \\
\end{tabular} 
 \\ 
 \hline 
\end{tabularx}
\vspace*{2ex}
\addtocounter{table}{-1}
\caption{Elastic ($\My{=}m_p$) and proton-dissociative ($m_p{<}\My{<}10~\gev$) \myrho meson photoproduction cross sections \sigmaRhoY in bins of \wgp.
The cross sections are obtained by analysing the \pipi photoproduction cross sections given in \tabref{tab:tab_xSec_wm_el} and \tabref{tab:tab_xSec_wm_pd} with the fit model defined in \eqnref{eqn:theo_rhoSoedMass}.
  The effective photon flux factors \phiWW for each \wgp bin are given. The statistical and full systematic uncertainties including normalisation uncertainties are given by $\Delta_{\stat}$ and $\Delta_{\syst}$, respectively. A full uncertainty breakdown and statistical correlations are provided~\cite{H1_data}. The data are shown in \figref{fig:results_sigmaRho_w_fit}.}
\label{tab:tab_xSec_wRho}
\end{table}
}

\begin{table}[!htp]\centering
\small
\renewcommand{\arraystretch}{1.3}
\begin{tabular}{@{}l@{\hspace{4em}}   d{2.3} d{1.3} l  c@{\hspace{4em}}  d{2.3} d{2.3} l@{}}
  \toprule
  & \multicolumn{3}{l}{$\My = m_p$} && \multicolumn{3}{l@{}}{$m_p < \My < 10~{\gev}$} \\ 
                                              \cmidrule{2-4} \cmidrule{6-8}
Parameter & \multicolumn{1}{l}{Value} & \multicolumn{1}{l}{$\Delta_{\stat}$} & \multicolumn{1}{l}{$\Delta_{\syst}$} && \multicolumn{1}{l}{Value} & \multicolumn{1}{l}{$\Delta_{\stat}$} & \multicolumn{1}{l}{$\Delta_{\syst}$}\\ 
    \midrule
    $\sigmarho(W_0{=}40~\gev)~[\mub]$  & 10.98    &      0.07   & $^{+0.72 }_{-0.74}$  &&   4.62     &      0.06   & $^{+0.59 }_{-0.57 }$\\
    $\delta$                            & 0.171   &      0.009  & $^{+0.039}_{-0.026}$ &&  -0.156   &      0.026  & $^{+0.081}_{-0.079}$\\
    \bottomrule
\end{tabular}
  \caption{Fit parameters for the fit of the energy dependence of the elastic and proton-dissociative \myrho meson photoproduction cross sections \sigmaRhoY in the range ${20<\wgp < 80~\gev}$. The nominal fit values and statistical and systematic uncertainties are given. A full uncertainty breakdown and statistical correlations are provided~\cite{H1_data}. The corresponding fit is shown in \figref{fig:results_sigmaRho_w_fit}.}
  \label{tab:results_sigmaRho_w_pars}
 \end{table}

\begin{table}[!ttt]\centering
  \small
\renewcommand{\arraystretch}{1.3}
\begin{tabular}{@{}l@{\hspace{4em}}   d{2.3} d{1.3} l }
  \toprule
  & \multicolumn{3}{l}{$\My = m_p$} \\ 
                                              \cmidrule{2-4}
Parameter & \multicolumn{1}{l}{Value} & \multicolumn{1}{l}{$\Delta_{\stat}$} & \multicolumn{1}{l}{$\Delta_{\syst}$} \\
    \midrule
    $\sigmarho(W_0{=}40~\gev)~[\mub]$  & 10.66    &      0.11   & $^{+0.75 }_{-1.00}$  \\
    $f_\reg$                             & 0.020   &      0.007  & $^{+0.029}_{-0.013}$ \\
    $\delta_\reg$                           & -1.45   &      0.12  & $^{+0.35}_{-0.21}$ \\
    $\delta_\pom$                           & 0.207   &      0.015  & $^{+0.053}_{-0.033}$ \\
    \bottomrule
\end{tabular}
  \caption{Fit parameters for the fit of the energy dependence of the elastic \myrho meson photoproduction cross section in the range ${2<\wgp < 200~\gev}$. The nominal fit values and statistical and systematic uncertainties are given. A full uncertainty breakdown and statistical correlations are provided~\cite{H1_data}. The corresponding fit is shown in \figref{fig:results_sigmaRho_w_hist}.}
  \label{tab:results_world_w_pars}
 \end{table}

\newpage
{
\begin{table}[!thp]
\footnotesize \centering
\renewcommand{\arraystretch}{1.3}
\setlength{\tabcolsep}{3pt}
\begin{tabularx}{\textwidth}{|c|c|}
\hline
\begin{tabular}[t]{ c  c | d{3.3} d{3.3} l}  
\multicolumn{5}{l}{$m_Y = m_p$} \\ 
\midrule 
$|t| $ range & $|t_\text{bc}|$ & \multicolumn{1}{c}{$\frac{\dd\sigma_{\rho}}{\dd t} (|t_\text{bc}|)$} & \multicolumn{1}{c}{$\Delta_{\stat}$} & \multicolumn{1}{c}{$\Delta_{\syst}$}  \\ 
$[\gev^{2}]$  & $[\gev^{2}]$ & \multicolumn{1}{c}{$\left[\frac{\mub}{\gev^{2}}\right]$} & \multicolumn{1}{c}{$\left[\frac{\mub}{\gev^{2}}\right]$} & \multicolumn{1}{c}{$\left[\frac{\mub}{\gev^{2}}\right]$} \\ 
\midrule 
$0.000 - 0.008$ & 0.004  & 96.7 & 2.0 & $ ^{+6.3}_{-6.3} $\\ 
$0.008 - 0.018$ & 0.013  & 86.0 & 1.7 & $ ^{+5.5}_{-5.5} $\\ 
$0.018 - 0.030$ & 0.024  & 77.0 & 1.4 & $ ^{+5.0}_{-4.9} $\\ 
$0.030 - 0.044$ & 0.037  & 69.7 & 1.3 & $ ^{+4.4}_{-4.5} $\\ 
$0.044 - 0.060$ & 0.052  & 60.1 & 1.1 & $ ^{+3.8}_{-3.8} $\\ 
$0.060 - 0.078$ & 0.069  & 49.26 & 0.91 & $ ^{+3.18}_{-3.16} $\\ 
$0.078 - 0.100$ & 0.089  & 41.43 & 0.74 & $ ^{+2.63}_{-2.69} $\\ 
$0.100 - 0.126$ & 0.113  & 33.37 & 0.61 & $ ^{+2.13}_{-2.19} $\\ 
$0.126 - 0.156$ & 0.141  & 26.53 & 0.47 & $ ^{+1.74}_{-1.72} $\\ 
$0.156 - 0.200$ & 0.177  & 19.12 & 0.32 & $ ^{+1.25}_{-1.28} $\\ 
$0.200 - 0.280$ & 0.238  & 11.36 & 0.18 & $ ^{+0.79}_{-0.83} $\\ 
$0.280 - 1.500$ & 0.565  & 0.771 & 0.020 & $ ^{+0.086}_{-0.087} $\\ 
 \end{tabular} 
&
\begin{tabular}[t]{ c  c | d{3.3} d{3.3} l}  
\multicolumn{5}{l}{$m_p < m_Y < 10~\gev$} \\ 
\midrule 
$|t| $ range & $|t_\text{bc}|$ & \multicolumn{1}{c}{$\frac{\dd\sigma_{\rho}}{\dd t} (|t_\text{bc}|)$} & \multicolumn{1}{c}{$\Delta_{\stat}$} & \multicolumn{1}{c}{$\Delta_{\syst}$}  \\ 
$[\gev^{2}]$  & $[\gev^{2}]$ & \multicolumn{1}{c}{$\left[\frac{\mub}{\gev^{2}}\right]$} & \multicolumn{1}{c}{$\left[\frac{\mub}{\gev^{2}}\right]$} & \multicolumn{1}{c}{$\left[\frac{\mub}{\gev^{2}}\right]$} \\ 
\midrule 
$0.000 - 0.030$ & 0.015  & 13.96 & 0.71 & $ ^{+2.37}_{-2.23} $\\ 
$0.030 - 0.060$ & 0.045  & 15.33 & 0.70 & $ ^{+2.38}_{-2.20} $\\ 
$0.060 - 0.096$ & 0.078  & 14.15 & 0.57 & $ ^{+1.95}_{-1.85} $\\ 
$0.096 - 0.140$ & 0.118  & 11.21 & 0.45 & $ ^{+1.46}_{-1.49} $\\ 
$0.140 - 0.200$ & 0.169  & 8.88 & 0.33 & $ ^{+1.21}_{-1.18} $\\ 
$0.200 - 0.280$ & 0.239  & 6.87 & 0.24 & $ ^{+0.91}_{-0.86} $\\ 
$0.280 - 0.390$ & 0.333  & 4.55 & 0.16 & $ ^{+0.60}_{-0.59} $\\ 
$0.390 - 0.600$ & 0.487  & 2.418 & 0.073 & $ ^{+0.300}_{-0.302} $\\ 
$0.600 - 1.500$ & 0.940  & 0.524 & 0.019 & $ ^{+0.082}_{-0.087} $\\ 
&&\\
&&\\
&&\\
 \end{tabular} 
\\ 
 \hline 
\end{tabularx} 
\vspace*{2ex}
\addtocounter{table}{-1}
\caption{Elastic ($\My{=}m_p$) and proton-dissociative ($m_p{<}\My{<}10~\gev$) differential \myrho meson photoproduction cross sections \dSigmaRhoYdt in bins of $t$.
  The cross sections are obtained by analysing the \pipi photoproduction cross sections given in \tabref{tab:tab_xSec_mt_el} and \tabref{tab:tab_xSec_mt_pd}  with the fit model defined in \eqnref{eqn:theo_rhoSoedMass}.
  The $t$ bin centres $t_{bc}$ are calculated as described in the text. The statistical and full systematic uncertainties including normalisation uncertainties are given by $\Delta_{\stat}$ and $\Delta_{\syst}$, respectively. A full uncertainty breakdown and statistical correlations are provided~\cite{H1_data}. The data are shown in \figref{fig:results_dSigmaRho_dt_fit}.}
\label{tab:tab_xSec_tRho}
\end{table}
}

\begin{table}[!ttt]\centering
\small
\renewcommand{\arraystretch}{1.3}
\begin{tabular}{@{}l@{\hspace{4em}}   d{2.3} d{1.3} l    c@{\hspace{4.0em}}  d{2.3} d{2.3} l@{}}
\toprule
    & \multicolumn{3}{l}{$\My = m_p$} && \multicolumn{3}{l@{}}{$m_p < \My < 10~{\gev}$} \\ 
                                              \cmidrule{2-4} \cmidrule{6-8}
  Parameter & \multicolumn{1}{l}{Value} & \multicolumn{1}{l}{$\Delta_{\stat}$} & \multicolumn{1}{l}{$\Delta_{\syst}$} && \multicolumn{1}{l}{Value} & \multicolumn{1}{l}{$\Delta_{\stat}$} & \multicolumn{1}{l}{$\Delta_{\syst}$}\\ 
    \midrule
  $ {\dd\sigmarho}/{\dd t}(t=0)~[\mub/\gevSq]    $ &   97.3 &  1.2   & $^{+6.3}_{-6.3}$   &&   19.5  &   0.7    & $^{+3.0}_{-2.9}$    \\
  $  b~[\gevSqInv]                                      $ &   9.61 &  0.15  & $^{+0.20}_{-0.15}$ &&    4.81 &   0.24   & $^{+0.39}_{-0.37}$  \\
  $  a                                                  $ &   20.4 &  3.7   & $^{+6.8}_{-5.1}$   &&    8.5  &   1.7    & $^{+2.7}_{-2.1}$    \\
  \bottomrule
 \end{tabular} 
\\ 
  \caption{Fit parameters for the fit of the $t$ dependencies of the elastic and proton-dissociative \myrho meson photoproduction cross sections \dSigmaRhoYdt. The nominal fit values and statistical and systematic uncertainties are given. A full uncertainty breakdown and statistical correlations are provided~\cite{H1_data}. The corresponding fit is shown in \figref{fig:results_dSigmaRho_dt_fit}. }
  \label{tab:results_dSigmaRho_dt_pars}
\end{table}

\newpage
{
\begin{table}[!thp]
\scriptsize \centering
\renewcommand{\arraystretch}{1.3}
\setlength{\tabcolsep}{5pt}
\begin{tabularx}{\textwidth}{|c|c|}
\hline
\begin{tabular}[t]{ c  c | d{3.3} d{3.3} l}  
\multicolumn{5}{l}{$m_Y = m_p$} \\ 
\multicolumn{5}{l}{$20.0 < W_{\gamma p}  < 28.0~\gev$} \\ 
\multicolumn{5}{l}{$\Phi_{\gamma/e} = 0.0370$} \\ 
\midrule 
$|t| $ range & $|t_\text{bc}|$ & \multicolumn{1}{c}{$\frac{\dd\sigma_{\rho}}{\dd t} (|t_\text{bc}|)$} & \multicolumn{1}{c}{$\Delta_{\stat}$} & \multicolumn{1}{c}{$\Delta_{\syst}$}  \\ 
$[\gev^{2}]$  & $[\gev^{2}]$ & \multicolumn{1}{c}{$\left[\frac{\mub}{\gev^{2}}\right]$} & \multicolumn{1}{c}{$\left[\frac{\mub}{\gev^{2}}\right]$} & \multicolumn{1}{c}{$\left[\frac{\mub}{\gev^{2}}\right]$} \\ 
\midrule 
$0.000 - 0.016$ & 0.008  & 80.5 & 1.8 & $ ^{+6.0}_{-6.3} $\\ 
$0.016 - 0.036$ & 0.026  & 66.9 & 1.5 & $ ^{+5.0}_{-5.4} $\\ 
$0.036 - 0.062$ & 0.049  & 55.4 & 1.2 & $ ^{+4.2}_{-4.6} $\\ 
$0.062 - 0.100$ & 0.080  & 41.40 & 0.85 & $ ^{+3.10}_{-3.33} $\\ 
$0.100 - 0.150$ & 0.124  & 28.45 & 0.64 & $ ^{+2.22}_{-2.39} $\\ 
$0.150 - 0.230$ & 0.188  & 16.83 & 0.40 & $ ^{+1.34}_{-1.48} $\\ 
$0.230 - 1.500$ & 0.517  & 1.067 & 0.035 & $ ^{+0.111}_{-0.124} $\\ 
 \end{tabular} 
&
\begin{tabular}[t]{ c  c | d{3.3} d{3.3} l}  
\multicolumn{5}{l}{$m_Y = m_p$} \\ 
\multicolumn{5}{l}{$28.0 \leq W_{\gamma p}  < 38.0~\gev$} \\ 
\multicolumn{5}{l}{$\Phi_{\gamma/e} = 0.0316$} \\ 
\midrule 
$|t| $ range & $|t_\text{bc}|$ & \multicolumn{1}{c}{$\frac{\dd\sigma_{\rho}}{\dd t} (|t_\text{bc}|)$} & \multicolumn{1}{c}{$\Delta_{\stat}$} & \multicolumn{1}{c}{$\Delta_{\syst}$}  \\ 
$[\gev^{2}]$  & $[\gev^{2}]$ & \multicolumn{1}{c}{$\left[\frac{\mub}{\gev^{2}}\right]$} & \multicolumn{1}{c}{$\left[\frac{\mub}{\gev^{2}}\right]$} & \multicolumn{1}{c}{$\left[\frac{\mub}{\gev^{2}}\right]$} \\ 
\midrule 
$0.000 - 0.016$ & 0.008  & 86.9 & 1.7 & $ ^{+5.0}_{-4.9} $\\ 
$0.016 - 0.036$ & 0.026  & 72.6 & 1.4 & $ ^{+4.1}_{-4.1} $\\ 
$0.036 - 0.062$ & 0.049  & 59.1 & 1.1 & $ ^{+3.4}_{-3.4} $\\ 
$0.062 - 0.100$ & 0.080  & 42.42 & 0.79 & $ ^{+2.48}_{-2.49} $\\ 
$0.100 - 0.150$ & 0.124  & 29.30 & 0.56 & $ ^{+1.76}_{-1.79} $\\ 
$0.150 - 0.230$ & 0.188  & 17.31 & 0.33 & $ ^{+1.10}_{-1.12} $\\ 
$0.230 - 1.500$ & 0.517  & 1.132 & 0.029 & $ ^{+0.109}_{-0.112} $\\ 
 \end{tabular} 
\\ 
 \hline 
\hline
\begin{tabular}[t]{ c  c | d{3.3} d{3.3} l}  
\multicolumn{5}{l}{$m_Y = m_p$} \\ 
\multicolumn{5}{l}{$38.0 \leq W_{\gamma p}  < 50.0~\gev$} \\ 
\multicolumn{5}{l}{$\Phi_{\gamma/e} = 0.0267$} \\ 
\midrule 
$|t| $ range & $|t_\text{bc}|$ & \multicolumn{1}{c}{$\frac{\dd\sigma_{\rho}}{\dd t} (|t_\text{bc}|)$} & \multicolumn{1}{c}{$\Delta_{\stat}$} & \multicolumn{1}{c}{$\Delta_{\syst}$}  \\ 
$[\gev^{2}]$  & $[\gev^{2}]$ & \multicolumn{1}{c}{$\left[\frac{\mub}{\gev^{2}}\right]$} & \multicolumn{1}{c}{$\left[\frac{\mub}{\gev^{2}}\right]$} & \multicolumn{1}{c}{$\left[\frac{\mub}{\gev^{2}}\right]$} \\ 
\midrule 
$0.000 - 0.016$ & 0.008  & 94.9 & 1.9 & $ ^{+5.6}_{-5.6} $\\ 
$0.016 - 0.036$ & 0.026  & 77.4 & 1.5 & $ ^{+4.4}_{-4.4} $\\ 
$0.036 - 0.062$ & 0.049  & 63.2 & 1.2 & $ ^{+3.6}_{-3.6} $\\ 
$0.062 - 0.100$ & 0.080  & 45.87 & 0.84 & $ ^{+2.59}_{-2.62} $\\ 
$0.100 - 0.150$ & 0.124  & 30.86 & 0.59 & $ ^{+1.76}_{-1.80} $\\ 
$0.150 - 0.230$ & 0.188  & 17.82 & 0.36 & $ ^{+1.03}_{-1.08} $\\ 
$0.230 - 1.500$ & 0.517  & 1.161 & 0.029 & $ ^{+0.098}_{-0.103} $\\ 
 \end{tabular} 
&
\begin{tabular}[t]{ c  c | d{3.3} d{3.3} l}  
\multicolumn{5}{l}{$m_Y = m_p$} \\ 
\multicolumn{5}{l}{$50.0 \leq W_{\gamma p}  < 80.0~\gev$} \\ 
\multicolumn{5}{l}{$\Phi_{\gamma/e} = 0.0416$} \\ 
\midrule 
$|t| $ range & $|t_\text{bc}|$ & \multicolumn{1}{c}{$\frac{\dd\sigma_{\rho}}{\dd t} (|t_\text{bc}|)$} & \multicolumn{1}{c}{$\Delta_{\stat}$} & \multicolumn{1}{c}{$\Delta_{\syst}$}  \\ 
$[\gev^{2}]$  & $[\gev^{2}]$ & \multicolumn{1}{c}{$\left[\frac{\mub}{\gev^{2}}\right]$} & \multicolumn{1}{c}{$\left[\frac{\mub}{\gev^{2}}\right]$} & \multicolumn{1}{c}{$\left[\frac{\mub}{\gev^{2}}\right]$} \\ 
\midrule 
$0.000 - 0.016$ & 0.008  & 102.7 & 2.1 & $ ^{+7.4}_{-7.4} $\\ 
$0.016 - 0.036$ & 0.026  & 85.1 & 1.7 & $ ^{+6.0}_{-5.8} $\\ 
$0.036 - 0.062$ & 0.049  & 69.2 & 1.3 & $ ^{+4.9}_{-4.7} $\\ 
$0.062 - 0.100$ & 0.080  & 49.14 & 0.89 & $ ^{+3.48}_{-3.40} $\\ 
$0.100 - 0.150$ & 0.124  & 33.08 & 0.62 & $ ^{+2.36}_{-2.29} $\\ 
$0.150 - 0.230$ & 0.188  & 18.55 & 0.37 & $ ^{+1.34}_{-1.34} $\\ 
$0.230 - 1.500$ & 0.517  & 1.109 & 0.029 & $ ^{+0.108}_{-0.109} $\\ 
 \end{tabular} 
\\ 
 \hline 
 \end{tabularx} 
\vspace*{2ex}
\addtocounter{table}{-1}
\caption{Elastic ($\My{=}m_p$) differential \myrho meson photoproduction cross section \dSigmaRhoYdt in bins of \wgp and $t$.
  The cross section is obtained by analysing the \pipi photoproduction cross section given in \tabref{tab:tab_xSec_twm_el} with the fit model defined in \eqnref{eqn:theo_rhoSoedMass}.
The effective photon flux factors \phiWW for each \wgp bin are given. The $t$ bin centres $t_{bc}$ are calculated as described in the text. The statistical and full systematic uncertainties including normalisation uncertainties are given by $\Delta_{\stat}$ and $\Delta_{\syst}$, respectively. A full uncertainty breakdown and statistical correlations are provided~\cite{H1_data}. The data are shown in  \figref{fig:results_dSigmaRho_dtW_twfit}.}
 \label{tab:tab_xSec_twRho_el}
 \end{table}
\newpage
\begin{table}[!htb]
\scriptsize \centering
\renewcommand{\arraystretch}{1.3}
\setlength{\tabcolsep}{5pt}
\begin{tabularx}{\textwidth}{|c|c|}
\endlastfoot
\hline
\begin{tabular}[t]{ c  c | d{3.3} d{3.3} l}  
\multicolumn{5}{l}{$m_p < m_Y < 10~\gev$} \\ 
\multicolumn{5}{l}{$20.0 < W_{\gamma p}  < 28.0~\gev$} \\ 
\multicolumn{5}{l}{$\Phi_{\gamma/e} = 0.0370$} \\ 
\midrule 
$|t| $ range & $|t_\text{bc}|$ & \multicolumn{1}{c}{$\frac{\dd\sigma_{\rho}}{\dd t} (|t_\text{bc}|)$} & \multicolumn{1}{c}{$\Delta_{\stat}$} & \multicolumn{1}{c}{$\Delta_{\syst}$}  \\ 
$[\gev^{2}]$  & $[\gev^{2}]$ & \multicolumn{1}{c}{$\left[\frac{\mub}{\gev^{2}}\right]$} & \multicolumn{1}{c}{$\left[\frac{\mub}{\gev^{2}}\right]$} & \multicolumn{1}{c}{$\left[\frac{\mub}{\gev^{2}}\right]$} \\ 
\midrule 
$0.000 - 0.050$ & 0.024  & 13.17 & 0.72 & $ ^{+2.31}_{-2.06} $\\ 
$0.050 - 0.110$ & 0.079  & 13.70 & 0.68 & $ ^{+2.07}_{-1.98} $\\ 
$0.110 - 0.210$ & 0.158  & 10.19 & 0.47 & $ ^{+1.51}_{-1.43} $\\ 
$0.210 - 0.400$ & 0.298  & 5.89 & 0.25 & $ ^{+0.86}_{-0.83} $\\ 
$0.400 - 1.500$ & 0.783  & 0.986 & 0.048 & $ ^{+0.155}_{-0.150} $\\ 
 \end{tabular} 
&
\begin{tabular}[t]{ c  c | d{3.3} d{3.3} l}  
\multicolumn{5}{l}{$m_p < m_Y < 10~\gev$} \\ 
\multicolumn{5}{l}{$28.0 \leq W_{\gamma p}  < 38.0~\gev$} \\ 
\multicolumn{5}{l}{$\Phi_{\gamma/e} = 0.0316$} \\ 
\midrule 
$|t| $ range & $|t_\text{bc}|$ & \multicolumn{1}{c}{$\frac{\dd\sigma_{\rho}}{\dd t} (|t_\text{bc}|)$} & \multicolumn{1}{c}{$\Delta_{\stat}$} & \multicolumn{1}{c}{$\Delta_{\syst}$}  \\ 
$[\gev^{2}]$  & $[\gev^{2}]$ & \multicolumn{1}{c}{$\left[\frac{\mub}{\gev^{2}}\right]$} & \multicolumn{1}{c}{$\left[\frac{\mub}{\gev^{2}}\right]$} & \multicolumn{1}{c}{$\left[\frac{\mub}{\gev^{2}}\right]$} \\ 
\midrule 
$0.000 - 0.050$ & 0.024  & 14.38 & 0.70 & $ ^{+2.20}_{-2.11} $\\ 
$0.050 - 0.110$ & 0.079  & 14.82 & 0.66 & $ ^{+2.03}_{-1.96} $\\ 
$0.110 - 0.210$ & 0.158  & 9.85 & 0.40 & $ ^{+1.29}_{-1.20} $\\ 
$0.210 - 0.400$ & 0.298  & 5.58 & 0.21 & $ ^{+0.71}_{-0.70} $\\ 
$0.400 - 1.500$ & 0.783  & 0.933 & 0.032 & $ ^{+0.127}_{-0.121} $\\ 
 \end{tabular} 
\\ 
 \hline 
\hline
\begin{tabular}[t]{ c  c | d{3.3} d{3.3} l}  
\multicolumn{5}{l}{$m_p < m_Y < 10~\gev$} \\ 
\multicolumn{5}{l}{$38.0 \leq W_{\gamma p}  < 50.0~\gev$} \\ 
\multicolumn{5}{l}{$\Phi_{\gamma/e} = 0.0267$} \\ 
\midrule 
$|t| $ range & $|t_\text{bc}|$ & \multicolumn{1}{c}{$\frac{\dd\sigma_{\rho}}{\dd t} (|t_\text{bc}|)$} & \multicolumn{1}{c}{$\Delta_{\stat}$} & \multicolumn{1}{c}{$\Delta_{\syst}$}  \\ 
$[\gev^{2}]$  & $[\gev^{2}]$ & \multicolumn{1}{c}{$\left[\frac{\mub}{\gev^{2}}\right]$} & \multicolumn{1}{c}{$\left[\frac{\mub}{\gev^{2}}\right]$} & \multicolumn{1}{c}{$\left[\frac{\mub}{\gev^{2}}\right]$} \\ 
\midrule 
$0.000 - 0.050$ & 0.024  & 15.61 & 0.70 & $ ^{+2.37}_{-2.30} $\\ 
$0.050 - 0.110$ & 0.079  & 13.59 & 0.62 & $ ^{+1.87}_{-1.81} $\\ 
$0.110 - 0.210$ & 0.158  & 8.89 & 0.39 & $ ^{+1.16}_{-1.08} $\\ 
$0.210 - 0.400$ & 0.298  & 5.29 & 0.20 & $ ^{+0.71}_{-0.65} $\\ 
$0.400 - 1.500$ & 0.783  & 0.798 & 0.030 & $ ^{+0.100}_{-0.099} $\\ 
 \end{tabular} 
&
\begin{tabular}[t]{ c  c | d{3.3} d{3.3} l}  
\multicolumn{5}{l}{$m_p < m_Y < 10~\gev$} \\ 
\multicolumn{5}{l}{$50.0 \leq W_{\gamma p}  < 80.0~\gev$} \\ 
\multicolumn{5}{l}{$\Phi_{\gamma/e} = 0.0416$} \\ 
\midrule 
$|t| $ range & $|t_\text{bc}|$ & \multicolumn{1}{c}{$\frac{\dd\sigma_{\rho}}{\dd t} (|t_\text{bc}|)$} & \multicolumn{1}{c}{$\Delta_{\stat}$} & \multicolumn{1}{c}{$\Delta_{\syst}$}  \\ 
$[\gev^{2}]$  & $[\gev^{2}]$ & \multicolumn{1}{c}{$\left[\frac{\mub}{\gev^{2}}\right]$} & \multicolumn{1}{c}{$\left[\frac{\mub}{\gev^{2}}\right]$} & \multicolumn{1}{c}{$\left[\frac{\mub}{\gev^{2}}\right]$} \\ 
\midrule 
$0.000 - 0.050$ & 0.024  & 16.08 & 0.79 & $ ^{+2.77}_{-2.48} $\\ 
$0.050 - 0.110$ & 0.079  & 11.95 & 0.58 & $ ^{+1.92}_{-1.83} $\\ 
$0.110 - 0.210$ & 0.158  & 7.85 & 0.36 & $ ^{+1.14}_{-1.08} $\\ 
$0.210 - 0.400$ & 0.298  & 4.37 & 0.17 & $ ^{+0.57}_{-0.57} $\\ 
$0.400 - 1.500$ & 0.783  & 0.755 & 0.028 & $ ^{+0.102}_{-0.106} $\\ 
 \end{tabular} 
\\ 
 \hline 
\end{tabularx} 
\vspace*{2ex}
\addtocounter{table}{-1}
\caption{Proton-dissociative ($m_p{<}\My{<}10~\gev$)  differential \myrho meson photoproduction cross section \dSigmaRhoYdt in bins of \wgp and $t$.
  The cross section is obtained by analysing the \pipi photoproduction cross section given in \tabref{tab:tab_xSec_twm_pd} with the fit model defined in \eqnref{eqn:theo_rhoSoedMass}.
  The effective photon flux factors \phiWW for each \wgp bin are given. The $t$ bin centres $t_{bc}$ are calculated as described in the text. The statistical and full systematic uncertainties including normalisation uncertainties are given by $\Delta_{\stat}$ and $\Delta_{\syst}$, respectively. A full uncertainty breakdown and statistical correlations are provided~\cite{H1_data}. The data are shown in  \figref{fig:results_dSigmaRho_dtW_twfit}.}
 \label{tab:tab_xSec_twRho_pd}
 \end{table}
}

\begin{table}[!htb] \centering
\small
\renewcommand{\arraystretch}{1.3}
\begin{tabular}{@{}l@{\hspace{4em}}   d{2.3} d{1.3} l c@{\hspace{4em}}  d{2.3} d{2.3} l@{}}
\toprule
& \multicolumn{3}{l}{$\My = m_p$} && \multicolumn{3}{l@{}}{$m_p < \My < 10~{\gev}$} \\ 
                                              \cmidrule{2-4} \cmidrule{6-8}
Parameter & \multicolumn{1}{l}{Value} & \multicolumn{1}{l}{$\Delta_{\stat}$} & \multicolumn{1}{l}{$\Delta_{\syst}$} && \multicolumn{1}{l}{Value} & \multicolumn{1}{l}{$\Delta_{\stat}$} & \multicolumn{1}{l}{$\Delta_{\syst}$}\\ 
 \midrule
$\dd \sigmarho/\dd t(0;W_0)[\mub/\gevSq] $ & 97.3     &  1.1     & $^{+6.2}_{-6.1}      $ && 18.7   &  0.9   & $^{+3.0}_{-2.8}$\\ 
$                        b[\gevSqInv]    $ &  9.59    &  0.14    & $^{+0.14}_{-0.12}    $ &&  4.58  &  0.33  & $^{+0.46}_{-0.44}$\\ 
$                                   a    $ & 21.6     &  4.0     & $^{+6.4}_{-4.8}      $ && 10.5   &  4.2   & $^{+5.4}_{-3.7}$\\ 
$                          \alpha_{0}    $ &  1.0654  &  0.0044  & $^{+0.0088}_{-0.0050}$ &&  1.06  &  0.25  & $^{+0.12}_{-0.09}$\\ 
$               \alpha_{1}[\gevSqInv]    $ &  0.233   &  0.064   & $^{+0.020}_{-0.038}  $ &&  1.6   &  5.0   & $^{+2.7}_{-1.7}$\\ 
$                               \beta    $ &  0.164   &  0.068   & $^{+0.051}_{-0.045}  $ &&  0.27  &  0.49  & $^{+0.24}_{-0.16}$\\ 
\bottomrule
 \end{tabular} 
\\ 
  \caption{Fit parameters of the Regge fit of the $t$ and \wgp dependencies of the \myrho meson photoproduction cross sections in the range $20 < \wgp < 80~\gev$ and $|t|<1.5~\gevSq$. The nominal fit values and statistical and systematic uncertainties are given. A full uncertainty breakdown and statistical correlations are provided~\cite{H1_data}. The corresponding fit is shown in \figref{fig:results_dSigmaRho_dtW_twfit}.}
  \label{tab:results_fPar_wtRhoFit}
\end{table}

{
  \begin{table}[!htb]
\footnotesize \centering
\renewcommand{\arraystretch}{1.3}
\setlength{\tabcolsep}{4pt}
\begin{tabularx}{\textwidth}{|c|}
\hline
\begin{tabular}[t]{ c | d{3.3} d{3.3} l}  
\multicolumn{4}{l}{$m_Y = m_p$} \\ 
\hline 
   $\wgp$ range & \multicolumn{1}{c}{$b_W$} & \multicolumn{1}{c}{$\Delta_{\stat}$} & \multicolumn{1}{c}{$\Delta_{\syst}$}  \\ 
   $[\gev]$   & \multicolumn{3}{c}{$[\gevSqInv]$} \\
  \midrule
  $20.0 - 28.0$ &  9.24 & 0.16 & $^{+0.16}_{-0.12}$ \\ 
  $28.0 - 38.0$ &  9.40 & 0.15 & $^{+0.15}_{-0.12}$ \\ 
  $38.0 - 50.0$ &  9.60 & 0.16 & $^{+0.17}_{-0.12}$ \\ 
  $50.0 - 80.0$ & 10.05 & 0.16 & $^{+0.15}_{-0.15}$ \\ 
  \end{tabular} 
\\
\hline
\end{tabularx}
\vspace*{2ex}
\addtocounter{table}{-1}
    \caption{Fit parameters $b_W$ for the fit of $t$ dependencies of the elastic differential \myrho photoproduction cross sections $\dSigmaRhopdt$ in bins of $\wgp$ (cf. \tabref{tab:tab_xSec_twRho_el}). The statistical and full systematic uncertainties are given by $\Delta_{\stat}$ and $\Delta_{\syst}$, respectively. A full uncertainty breakdown and statistical correlations are provided~\cite{H1_data}. The data are shown in \figref{fig:results_fPar_wtRho_b}.}
  \label{tab:tab_fPar_bRho}
\end{table} 
}

{
  \begin{table}[!htb]
\footnotesize \centering
\renewcommand{\arraystretch}{1.3}
\setlength{\tabcolsep}{4pt}
\begin{tabularx}{\textwidth}{|c|c|}
\hline
\begin{tabular}[t]{ c c | d{3.3} d{3.3} l}  
\multicolumn{5}{l}{$m_Y = m_p$} \\ 
\hline 
   $|t|$ range & $|t_\text{bc}|$ & \multicolumn{1}{c}{$\alpha_t$} & \multicolumn{1}{c}{$\Delta_{\stat}$} & \multicolumn{1}{c}{$\Delta_{\syst}$}  \\ 
   $[\gevSq]$  & $[\gevSq]$ & \multicolumn{3}{c}{$ \ $} \\
  \midrule
  $0.000 - 0.016$ & 0.008 & 1.062 & 0.006 & $^{+0.008}_{-0.007}$ \\ 
  $0.016 - 0.036$ & 0.026 & 1.060 & 0.006 & $^{+0.009}_{-0.005}$ \\ 
  $0.036 - 0.062$ & 0.049 & 1.056 & 0.006 & $^{+0.010}_{-0.005}$ \\ 
  $0.062 - 0.100$ & 0.080 & 1.046 & 0.006 & $^{+0.009}_{-0.005}$ \\ 
  $0.100 - 0.150$ & 0.124 & 1.040 & 0.006 & $^{+0.009}_{-0.006}$ \\ 
  $0.150 - 0.230$ & 0.188 & 1.025 & 0.007 & $^{+0.010}_{-0.007}$ \\ 
  $0.230 - 1.500$ & 0.517 & 1.005 & 0.009 & $^{+0.015}_{-0.009}$ \\ 
  \end{tabular} 
& 
\begin{tabular}[t]{ c c | d{3.3} d{3.3} l}  
\multicolumn{5}{l}{$m_p < m_Y < 10~\gev$} \\ 
\hline 
   $|t|$ range & $|t_\text{bc}|$ & \multicolumn{1}{c}{$\alpha_t$} & \multicolumn{1}{c}{$\Delta_{\stat}$} & \multicolumn{1}{c}{$\Delta_{\syst}$}  \\ 
   $[\gevSq]$  & $[\gevSq]$ & \multicolumn{3}{c}{$ \ $} \\
  \midrule
  $0.000 - 0.050$ & 0.024 & 1.051 & 0.016 & $^{+0.024}_{-0.023}$ \\ 
  $0.050 - 0.110$ & 0.079 & 0.961 & 0.015 & $^{+0.022}_{-0.021}$ \\ 
  $0.110 - 0.210$ & 0.158 & 0.931 & 0.015 & $^{+0.022}_{-0.020}$ \\ 
  $0.210 - 0.400$ & 0.298 & 0.925 & 0.013 & $^{+0.022}_{-0.020}$ \\ 
  $0.400 - 1.500$ & 0.783 & 0.927 & 0.013 & $^{+0.025}_{-0.027}$ \\ 
  &&& \\
  &&&
  \end{tabular} 
\\
\hline
\end{tabularx}
\vspace*{2ex}
\addtocounter{table}{-1}
  \caption{Fit parameters $\alpha_t$ for the fit of simple power-law functions to the energy dependencies of the elastic and proton-dissociative differential \myrho cross sections $\dSigmaRhoYdt$ in bins of $t$ (cf. \tabref{tab:tab_xSec_twRho_el} and \tabref{tab:tab_xSec_twRho_pd}). The $t$ bin centres $t_{bc}$ are calculated as described in the text. The statistical and full systematic uncertainties are given by $\Delta_{\stat}$ and $\Delta_{\syst}$, respectively. A full uncertainty breakdown and statistical correlations are provided~\cite{H1_data}. The data are shown in \figref{fig:results_fPar_wtRho_alpha}.}
  \label{tab:tab_fPar_alphaRho}
\end{table} 
}

\FloatBarrier
\newpage
\appendix
\section{\pipi Photoproduction Cross Section Tables}
\label{apx:sec_pipiXSecTables}
\raggedright
Unfolded \pipi photoproduction cross sections are presented in tabular form. A full breakdown of systematic uncertainties and all statistical correlations is available~\cite{H1_data}.
\begin{itemize}
\item The differential cross sections \dSigmaPiPiYdm in bins of \mpipi and for elastic and proton-dissociative production are given in \tabref{tab:tab_xSec_m}.
\item The differential cross sections \dSigmaPiPiYdm in bins of \wgp and \mpipi are given in \tabref{tab:tab_xSec_wm_el} and \tabref{tab:tab_xSec_wm_pd} for elastic and proton-dissociative production, respectively.
\item The double-differential cross sections \ddSigmaPiPiYdmdt in bins of $t$ and \mpipi are given in \tabref{tab:tab_xSec_mt_el} and \tabref{tab:tab_xSec_mt_pd} for elastic and proton-dissociative production, respectively. 
\item The double-differential cross sections \ddSigmaPiPiYdmdt in bins of $t$, \wgp, and \mpipi are given in \tabref{tab:tab_xSec_twm_el} and \tabref{tab:tab_xSec_twm_pd} for elastic and proton-dissociative production, respectively.
\end{itemize}

\newpage
{
\footnotesize \centering
\renewcommand{\arraystretch}{1.2}
\setlength{\tabcolsep}{5pt}

  \caption{Control distributions of the reconstructed \mpipirec on a linear (a) and logarithmic $y$-scale (b), of \wgprec (c), and of \trec (d). The black data points are compared to the MC simulation including various signal and background components as given in the legends. The dotted red bands denote the total uncertainty of the simulation including normalisation uncertainties. }
  \label{fig:exp_ctrl}
\end{figure}

\begin{figure}[htb]\raggedleft
  \setlength{\tabcolsep}{4pt}
  \begin{tabular}{@{}c c c@{}}
   $\quad$ (a) & $\quad$ (b) & $\quad$ (c) \\[1ex]
  \includegraphics[scale=0.30]{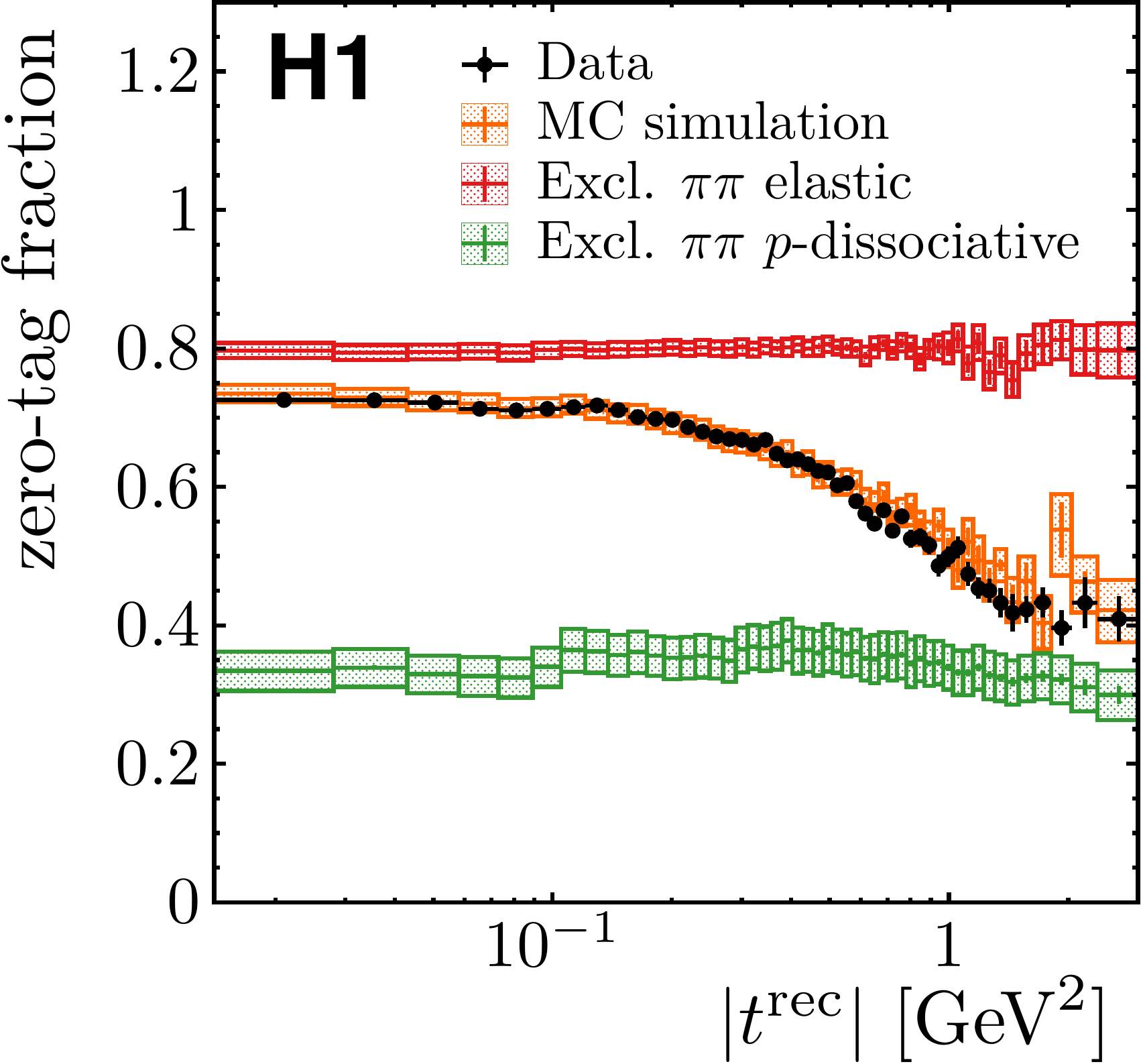}&
  \includegraphics[scale=0.30]{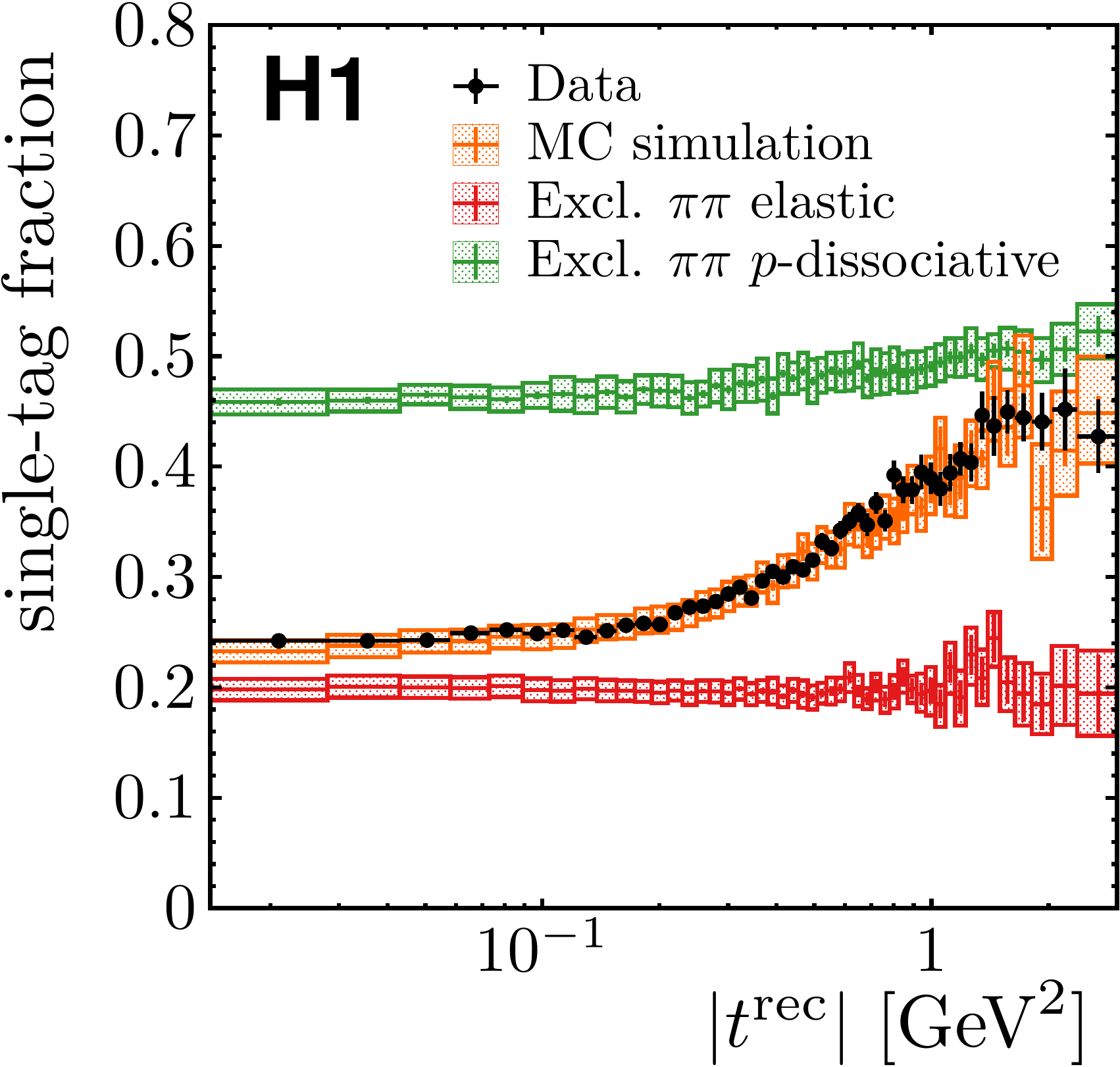} &
  \includegraphics[scale=0.30]{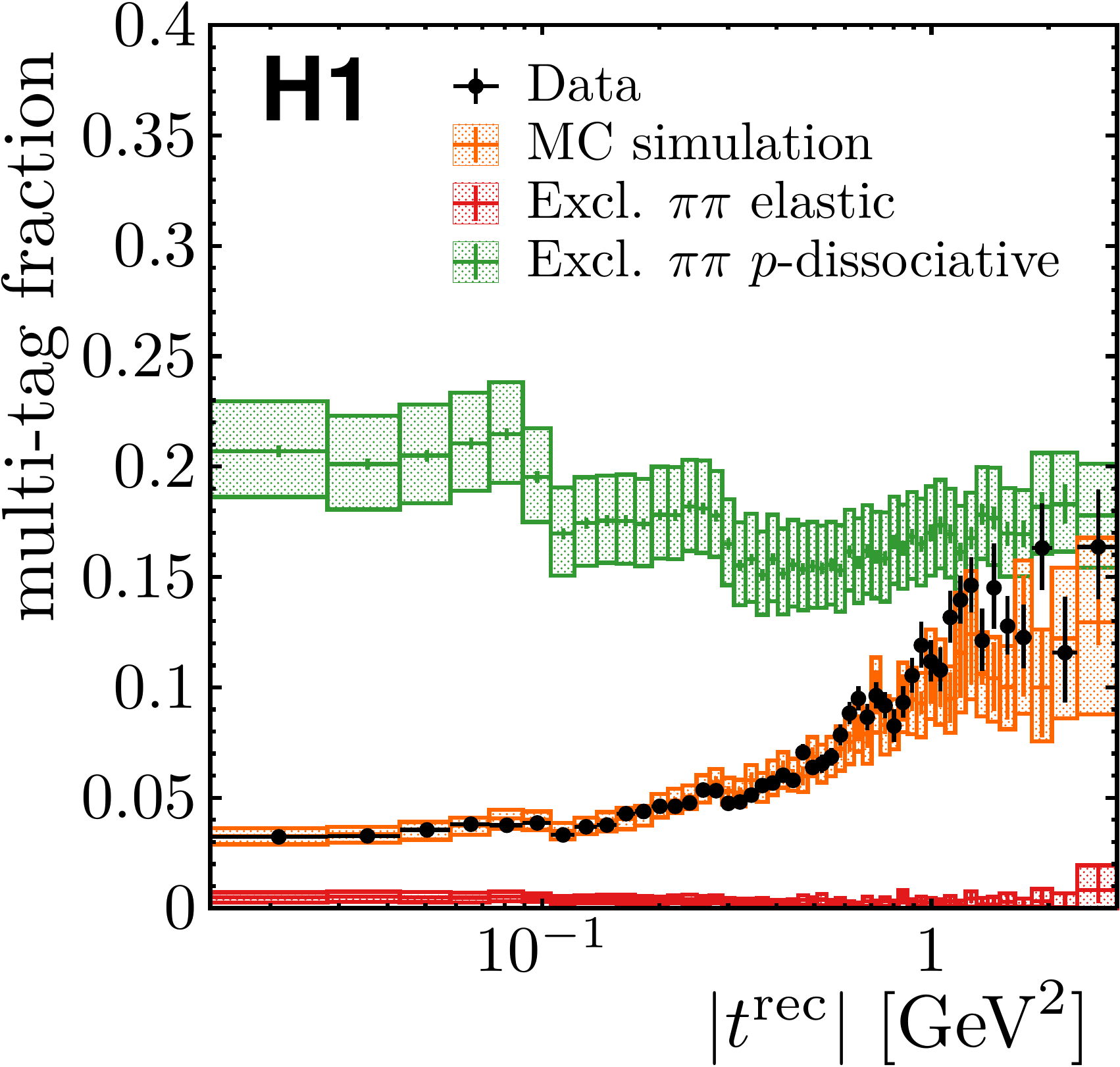} 
 \end{tabular}
  \caption{Zero- (a), single- (b), and multi-tag fraction (c) using tagging information from the FMD, the FTS, and the PLUG as a function of $|\trec|$. The fraction measured in data is given by the black points. The orange bands show the corresponding fraction predicted by the MC simulation. The tagging fractions of the simulated elastic and proton-dissociative exclusive \pipi signal channels are also shown. The dotted bands denote the total uncertainty of the simulation.}
  \label{fig:exp_ctrl_tagEff_ptsq}
\end{figure}

\begin{figure}[htb]
\centering
 \includegraphics[width=.66\textwidth]{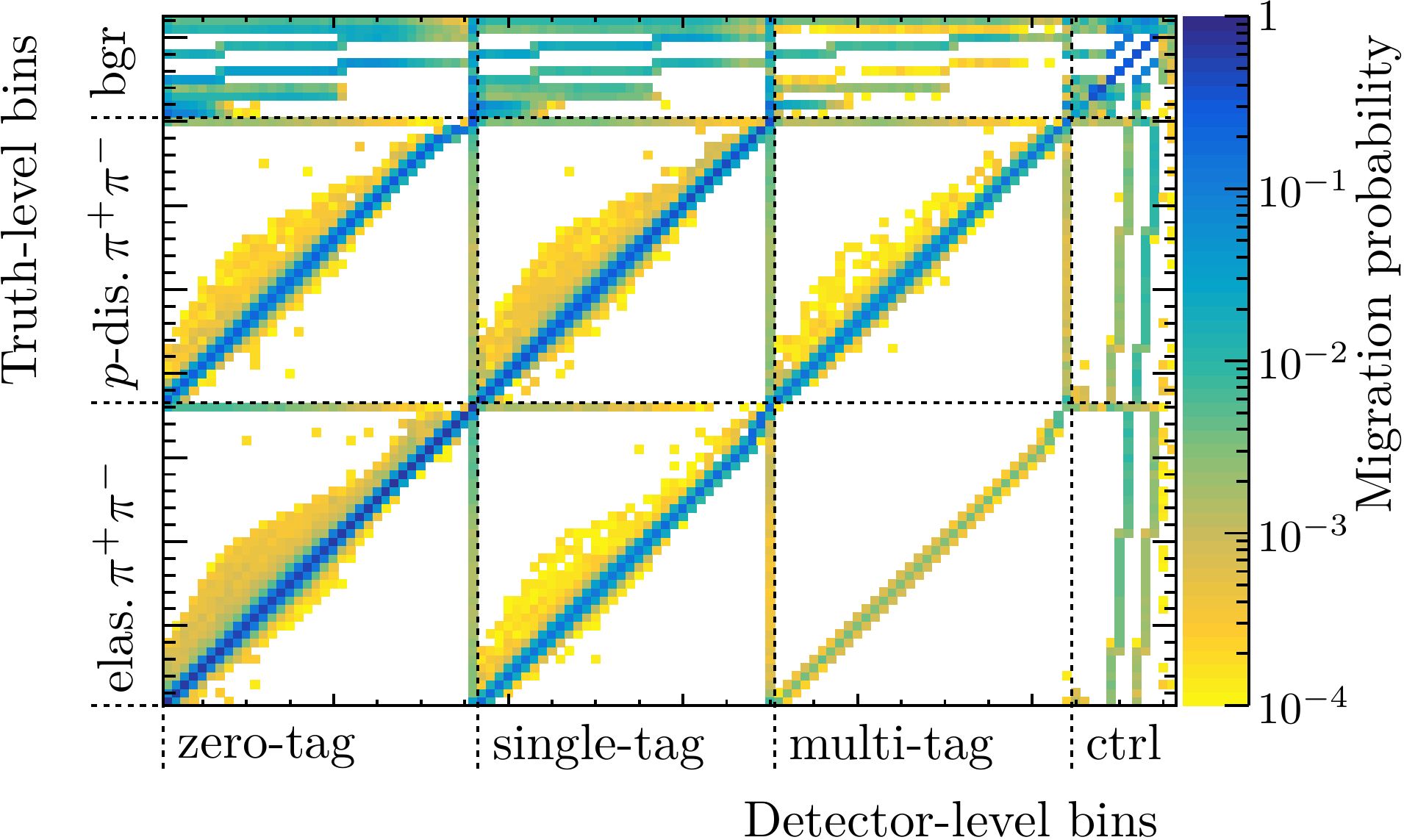
}
  \caption{Response matrix for unfolding the one-dimensional \mpipirec distributions. The matrix depicts migration probabilities from bins of the truth-level distributions along the $y$ axis into bins of the detector-level distributions along the $x$ axis. On detector-level, \mpipirec distributions in the forward tagging categories are considered, and on truth-level, \mpipi distributions from the elastic and proton-dissociative \pipi signal procesess are considered, as indicated. Migrations from outside the fiducial phasespace are accounted for in dedicated bins in each region. Background processes are considered in separate truth-level bins to be normalised in corresponding detector-level control regions. }
  \label{fig:exp_respMatrix_mpipi}
\end{figure}

\begin{figure}[htb]\centering
  \includegraphics[scale=0.5]{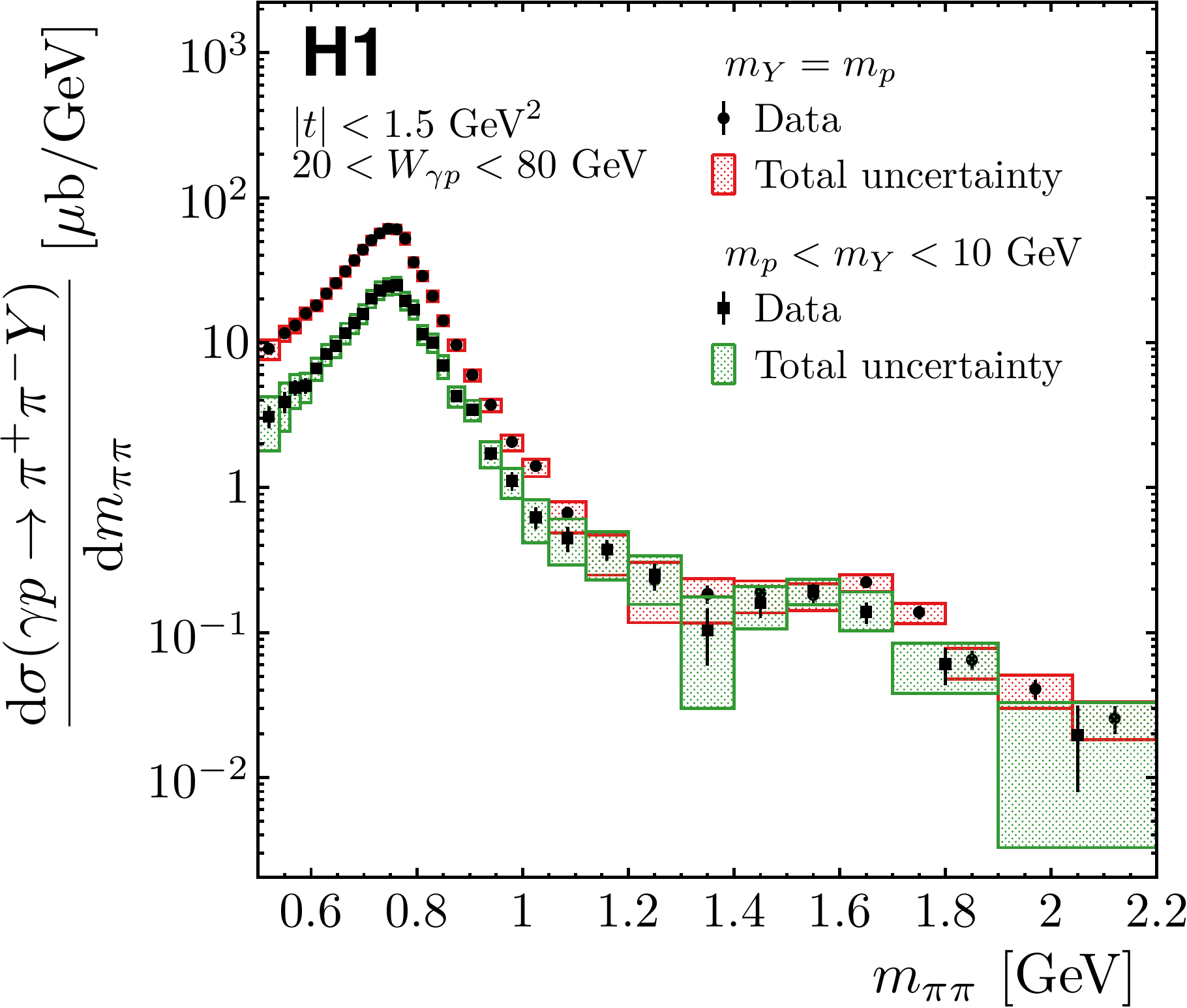
}
  \caption{Elastic ($\My{=}m_p$) and proton-dissociative ($m_p{<}\My{<}10~\gev$) differential \pipi photoproduction cross sections $\dSigmaPiPiYdm$ as functions of \mpipi. The black points and error bars show the measured cross section values with statistical uncertainties. The red and green bands show the total uncertainties of the elastic and proton-dissociative components, respectively.}
  \label{fig:results_dSigma_dm}
\end{figure}

\begin{figure}[htb]\centering
  \setlength{\tabcolsep}{5pt}
  \begin{tabular}{@{}c c@{}}
   $\qquad$  (a) & $\qquad$  (b) \\[1ex]
   \includegraphics[scale=0.45]{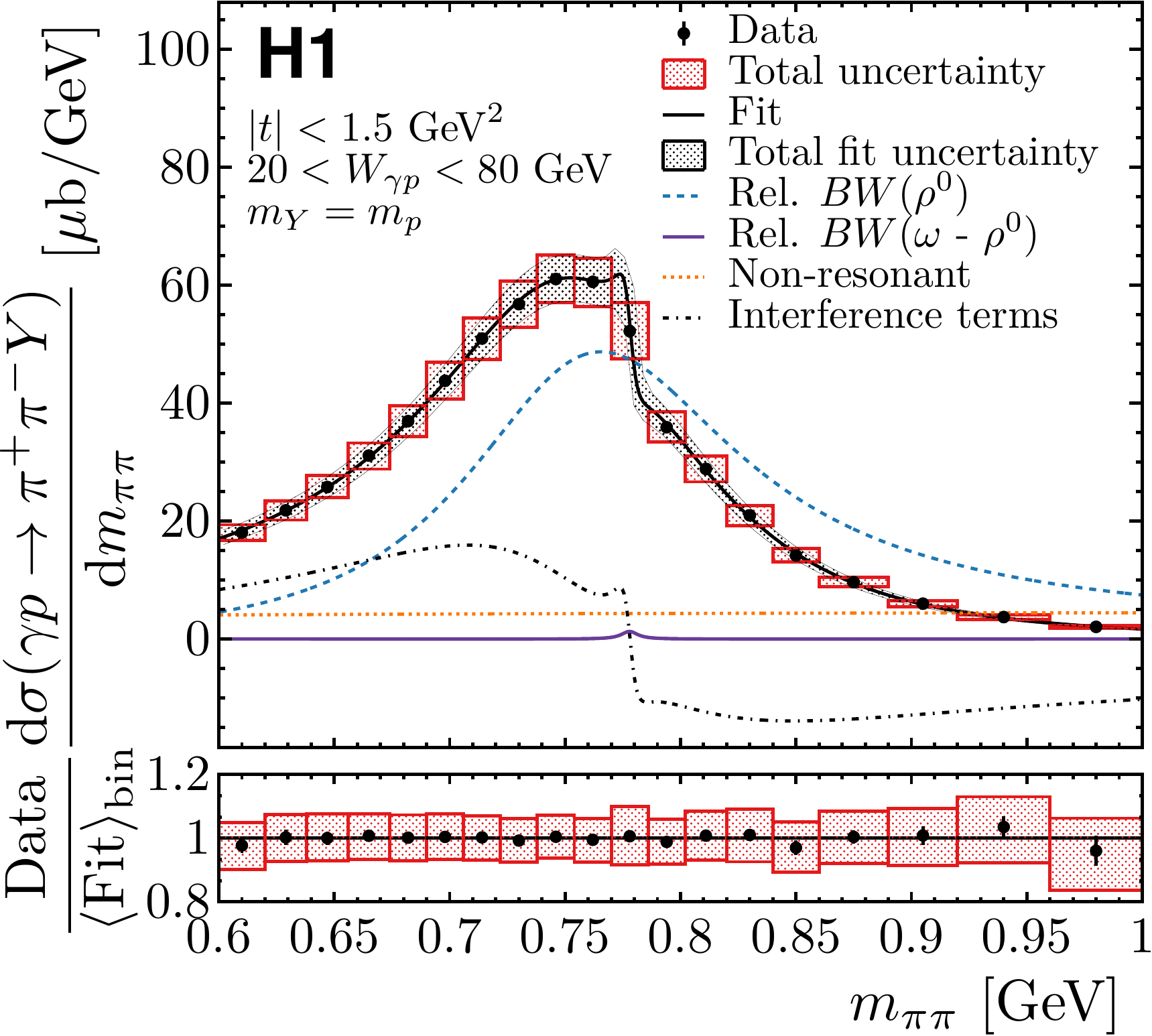}&
  \includegraphics[scale=0.45]{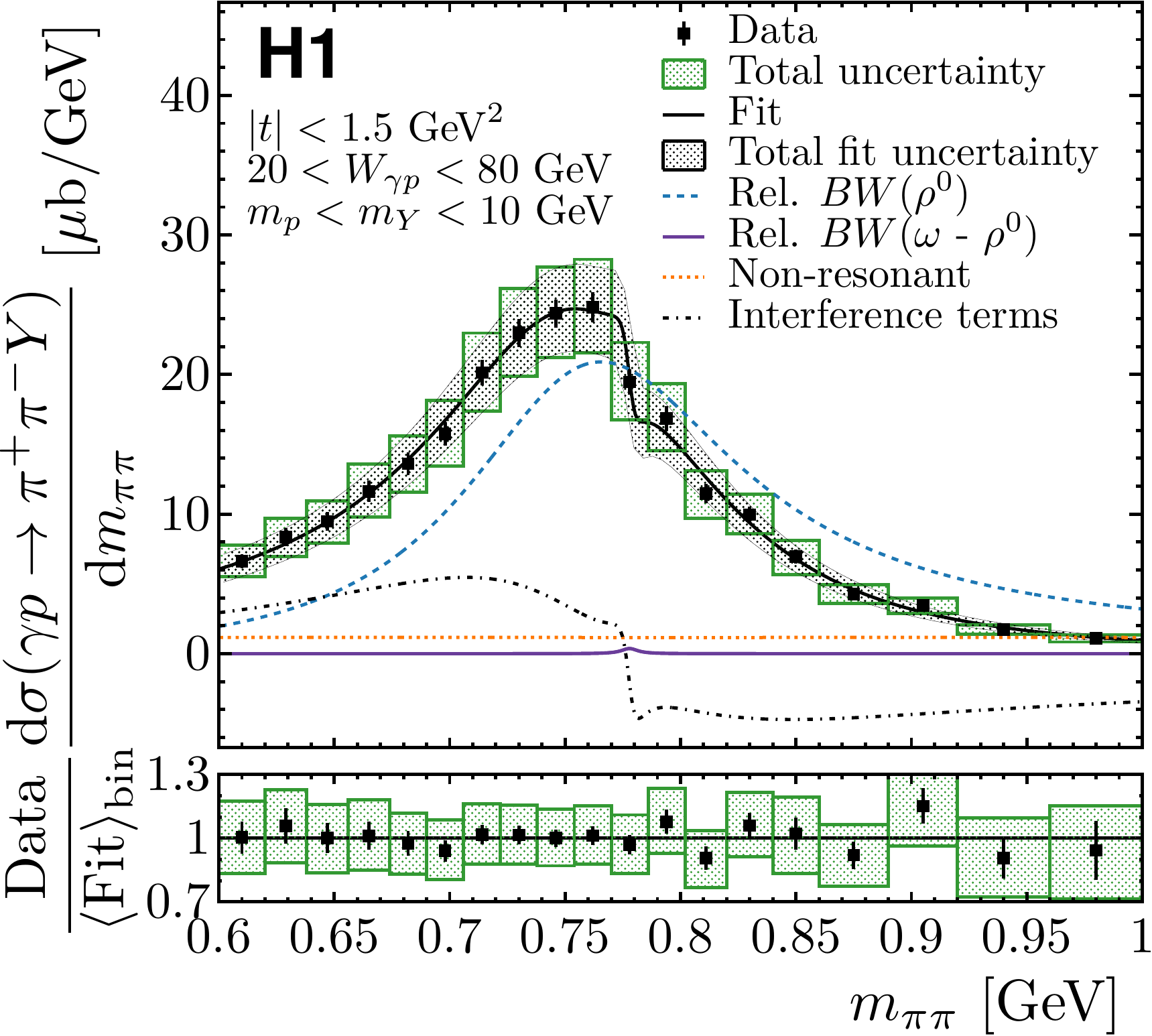} 
 \end{tabular}
  \caption{Elastic ($\My{=}m_p$) (a) and proton-dissociative ($m_p{<}\My{<}10~\gev$) (b) differential \pipi photoproduction cross sections \dSigmaPiPiYdm as functions of \mpipi.
    \eqnref{eqn:theo_rhoSoedMass} is fitted to the distributions in the mass region $0.6~\gev\leq \mpipi \leq 1.0~\gev$ as described in the text.
    The model and its components are displayed as indicated in the legend. In the ratio panel, the data are compared to the bin-averaged fit function values as they entered in the $\chi^2$ calculation for the fit. }
  \label{fig:results_dSigma_dm_fit}
\end{figure}

\begin{figure}[htb]\centering
  \includegraphics[scale=0.90]{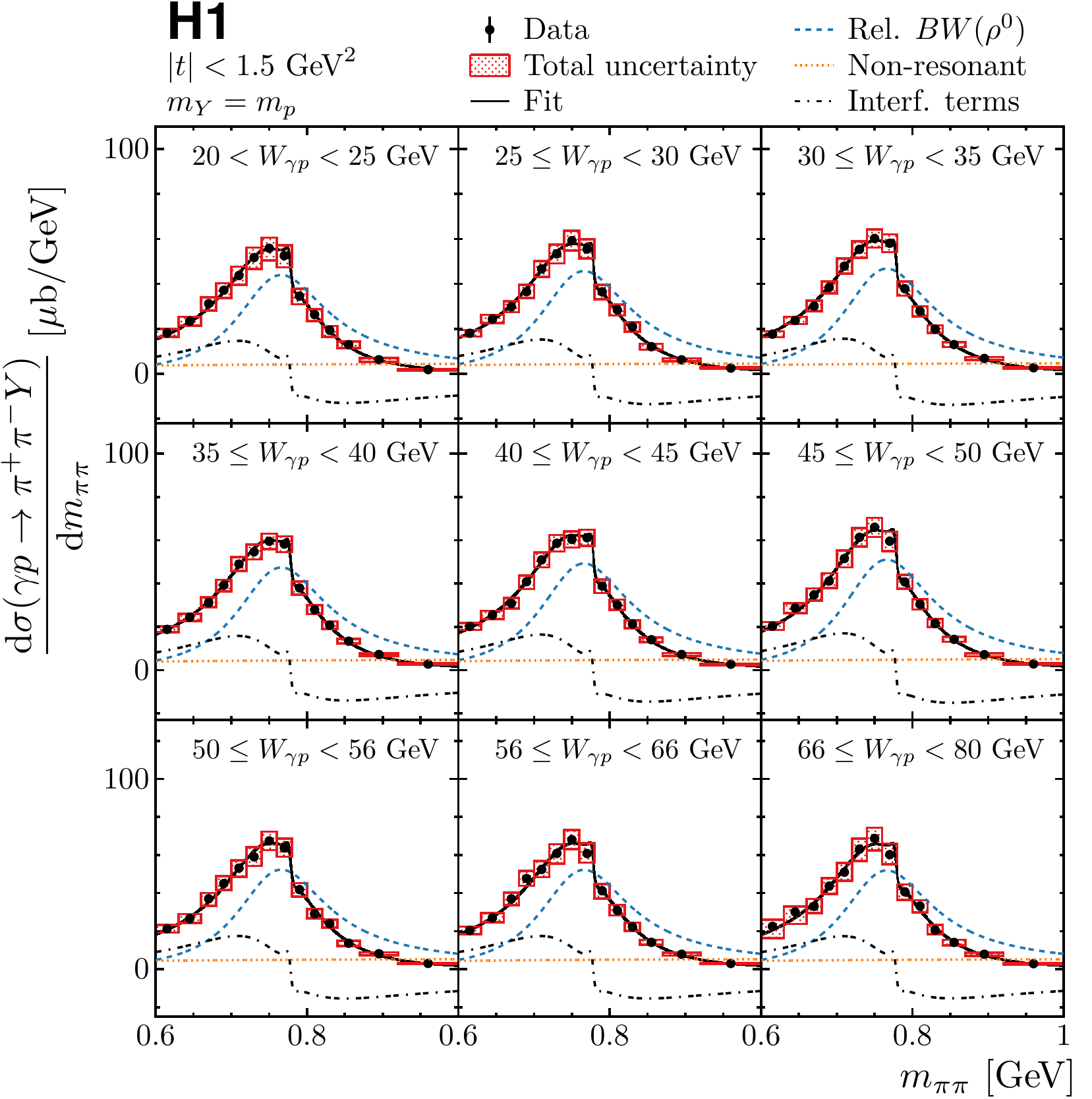
}
  \caption{Elastic ($\My{=}m_p$) differential \pipi photoproduction cross sections \dSigmaPiPipdm in bins of \wgp and as functions of \mpipi.
    The model that is fitted to the data as described in the text and its components are also shown.%
}
  \label{fig:results_dSigma_dmW_wmFit_el}
\end{figure}
\begin{figure}[htb]\centering
  \includegraphics[scale=0.81]{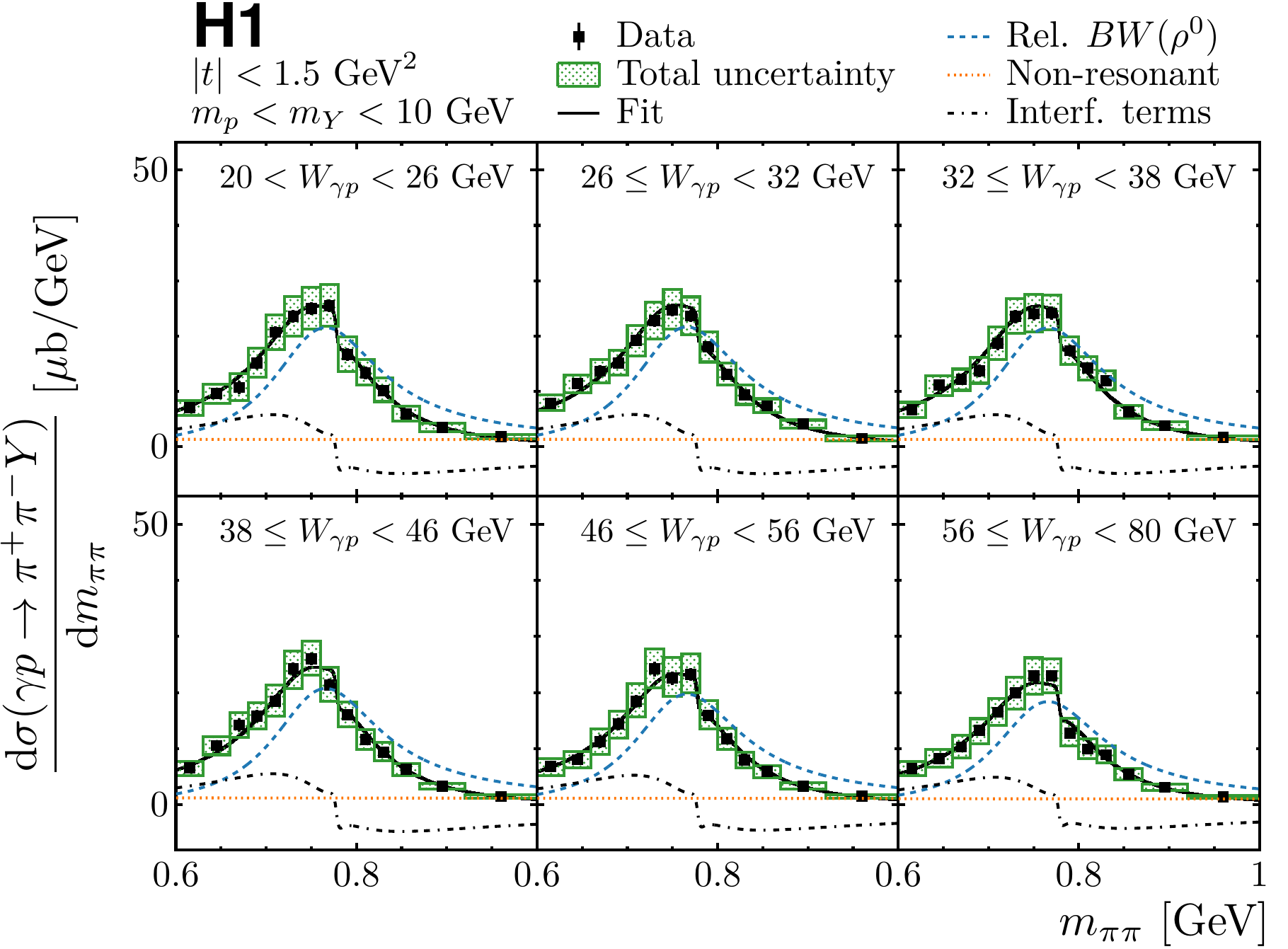
}
  \caption{Proton-dissociative ($m_p{<}\My{<}10~\gev$) differential \pipi photoproduction cross sections \dSigmaPiPiYdm in bins of \wgp and as functions of \mpipi.
    The model that is fitted to the data as described in the text and its components are also shown.%
}
  \label{fig:results_dSigma_dmW_wmFit_pd}
\end{figure}

\begin{figure}[htb]\centering
  \setlength{\tabcolsep}{5pt}
  \begin{tabular}{@{}c c@{}}
   $\qquad$  (a) & $\qquad$  (b) \\[1ex]
    \includegraphics[scale=0.45]{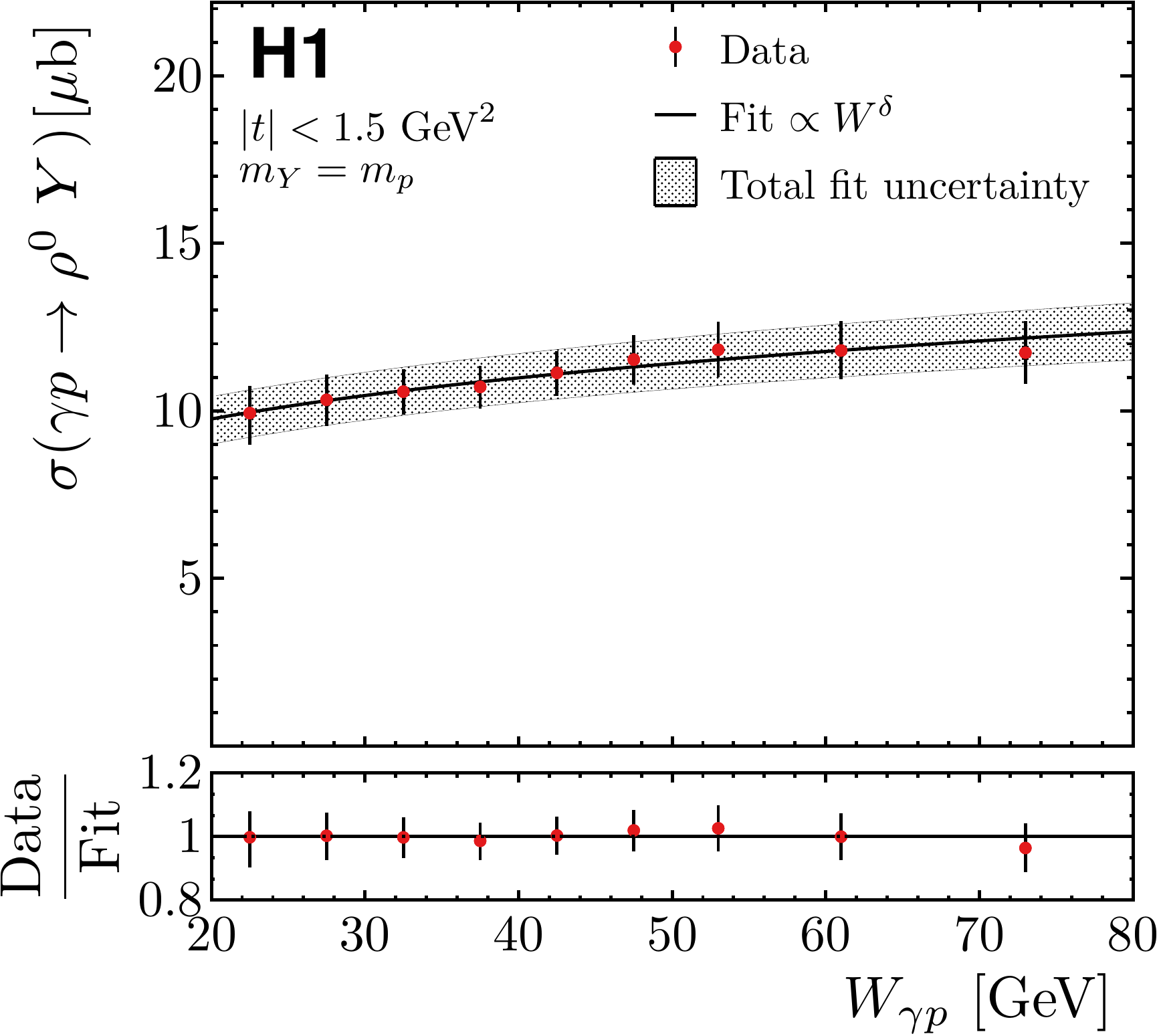}& 
    \includegraphics[scale=0.45]{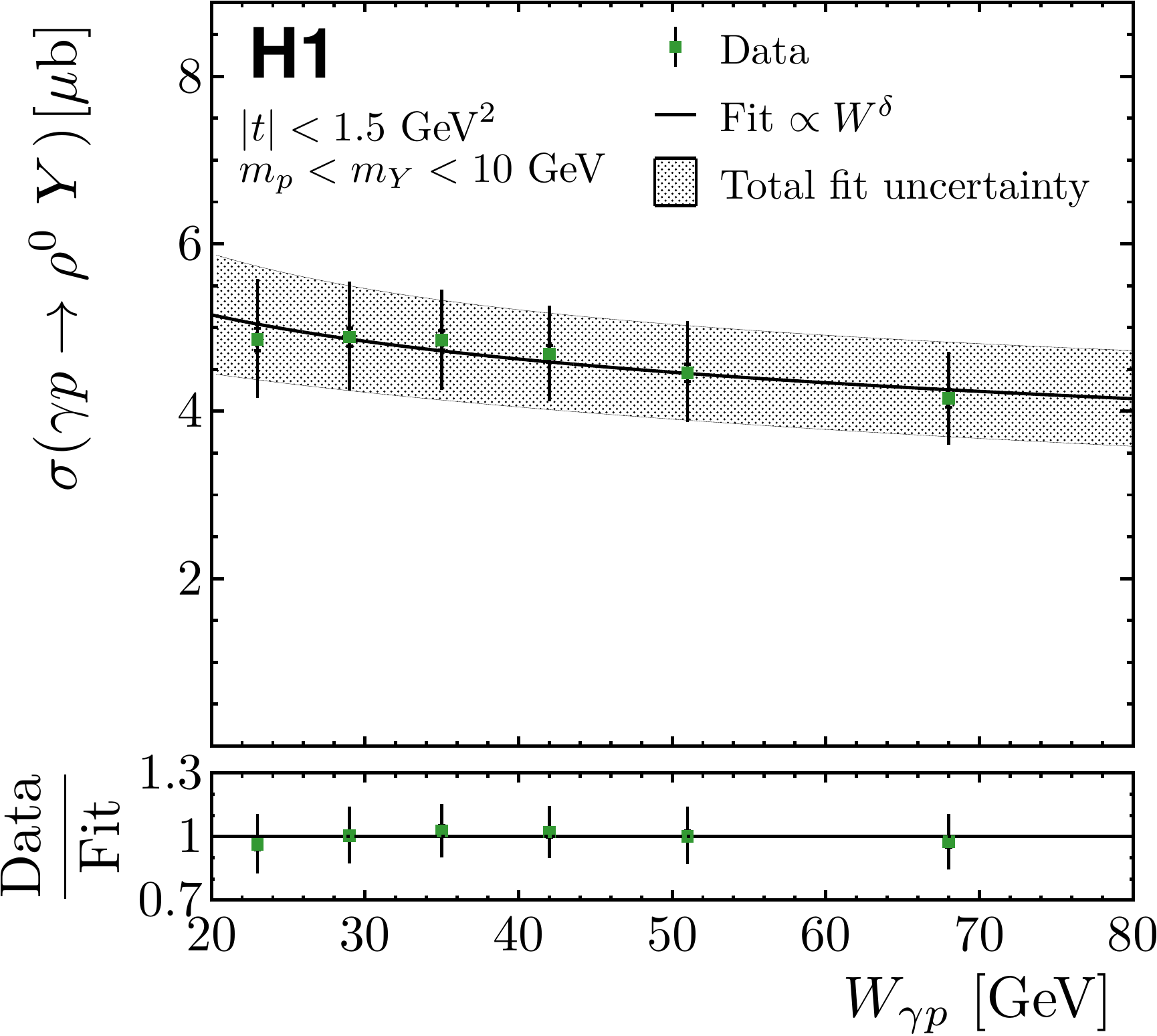} 
  \end{tabular}
  \caption{Elastic ($\My{=}m_p$) (a) and proton-dissociative ($m_p{<}\My{<}10~\gev$) (b) \myrho meson photoproduction cross sections \sigmaRhoY as functions of $\wgp$. The vertical lines indicate the total uncertainties. The statistical uncertainties are indicated by the inner error bars, which however are mostly covered by the data markers. The distributions are parametrised with fits as described in the text, and the fitted curves are also shown. In the ratio panels, the data are compared to the fit function values at the geometric bin centres as they entered the $\chi^2$ calculation in the fit.}
  \label{fig:results_sigmaRho_w_fit}
\end{figure}


\begin{figure}[htb]\centering
  \includegraphics[scale=0.5]{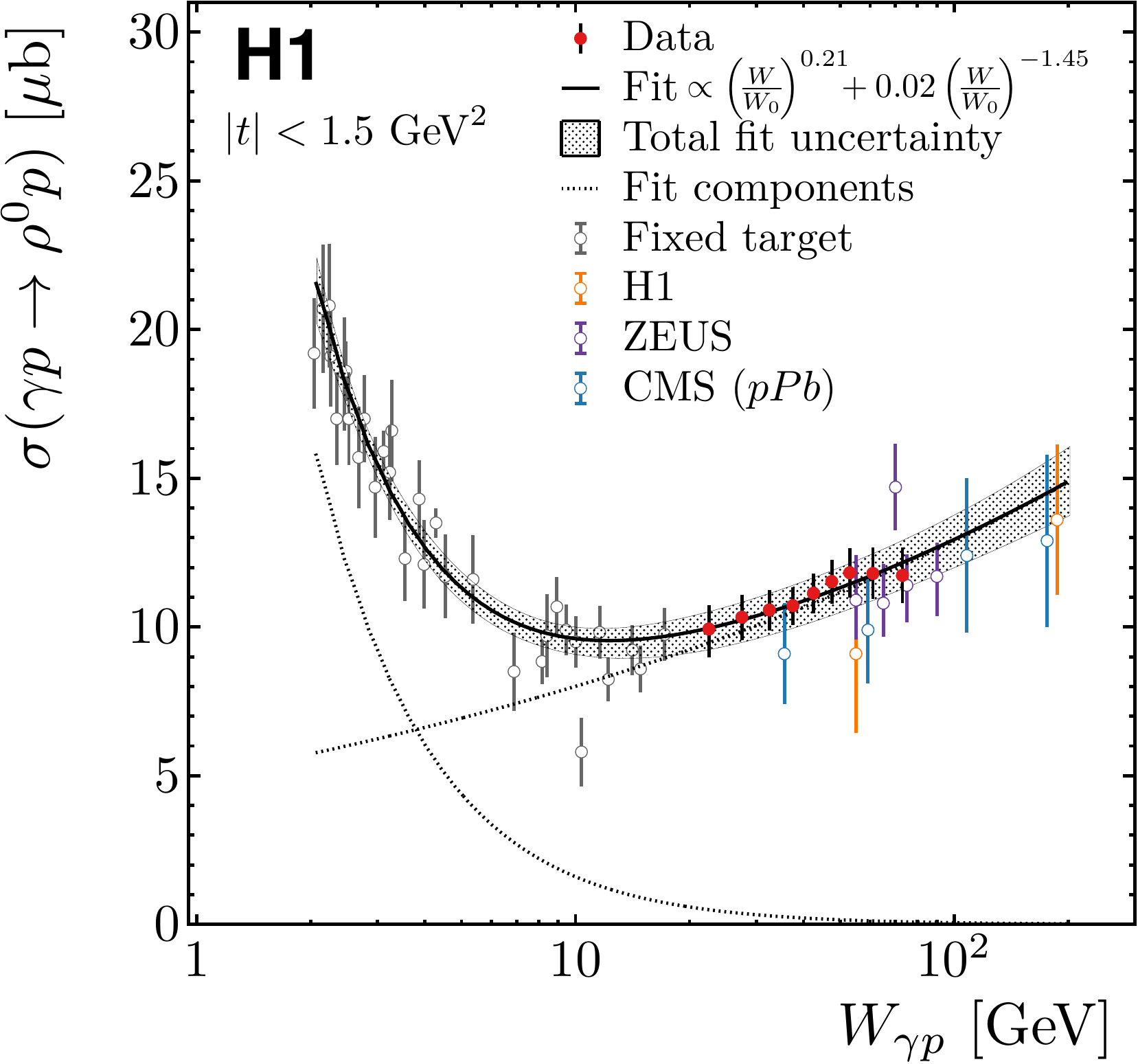
}   
  \caption{Elastic ($\My{=}m_p$) \myrho meson photoproduction cross section \sigmaRhoY as a function of \wgp. The present data are compared to measurements by fixed-target~\cite{Ballam:1971wq,Park:1971ts,Ballam:1972eq,Struczinski:1975ik,Egloff:1979mg,Aston:1982hr}, HERA~\cite{Aid:1996bs,Derrick:1995vq,Breitweg:1997ed}, and LHC~\cite{Sirunyan:2019nog} experiments as indicated in the legend. Only total uncertainties are shown. The solid line shows the fit of a sum of two power-law functions to the fixed-target and HERA data. The respective contributions are shown as dotted lines. The fit uncertainty is indicated by a band.}
  \label{fig:results_sigmaRho_w_hist}
\end{figure}

\begin{figure}[htb]\centering
  \includegraphics[scale=0.90]{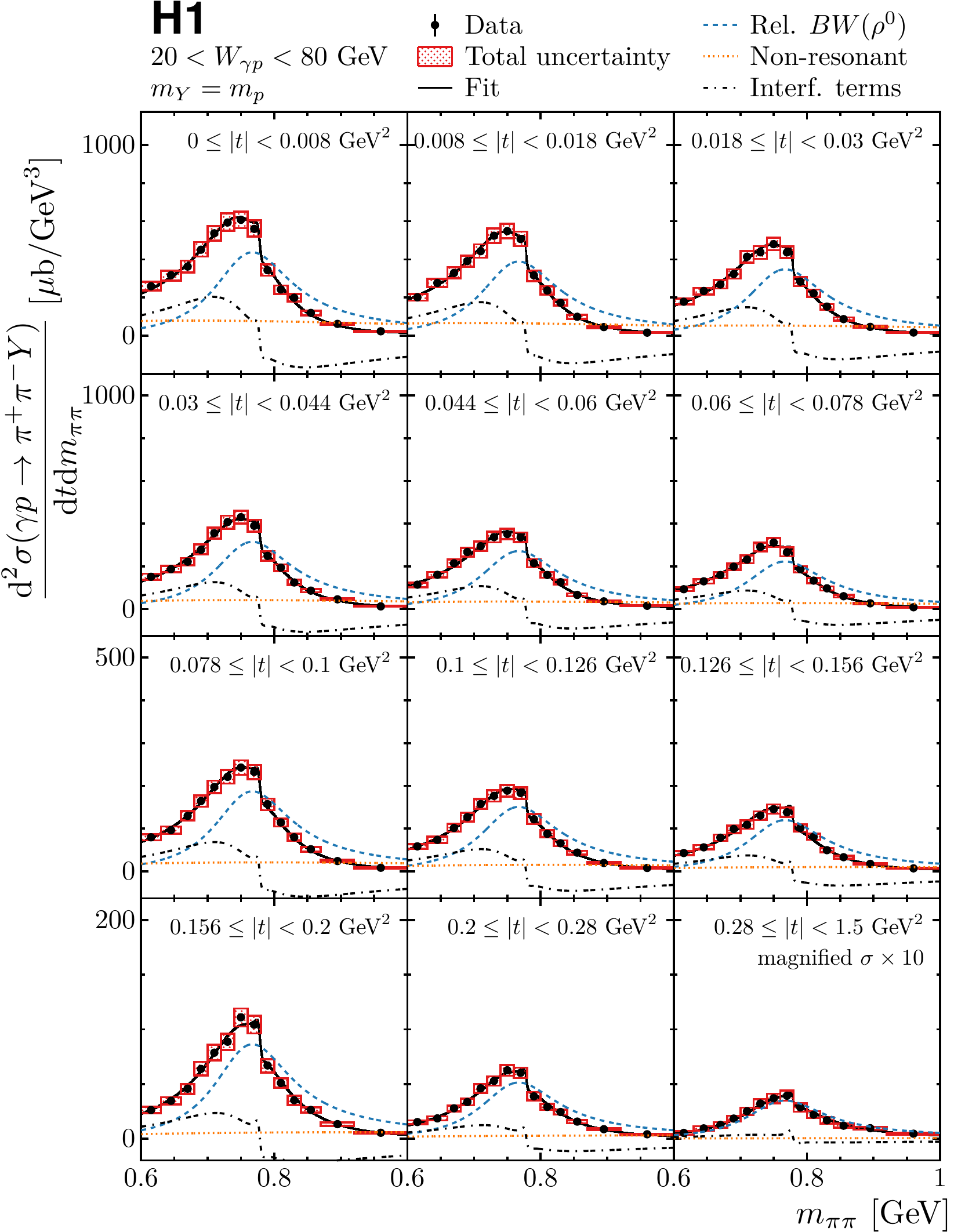
}
  \caption{Elastic ($\My{=}m_p$) double-differential \pipi photoproduction cross sections \ddSigmaPiPipdmdt in bins of $t$ and as functions of \mpipi.
    The model that is fitted to the data as described in the text and its components are also shown.%
}
  \label{fig:results_dSigma_dmdt_mtFit_el}
\end{figure}
\begin{figure}[htb]\centering
  \includegraphics[scale=0.90]{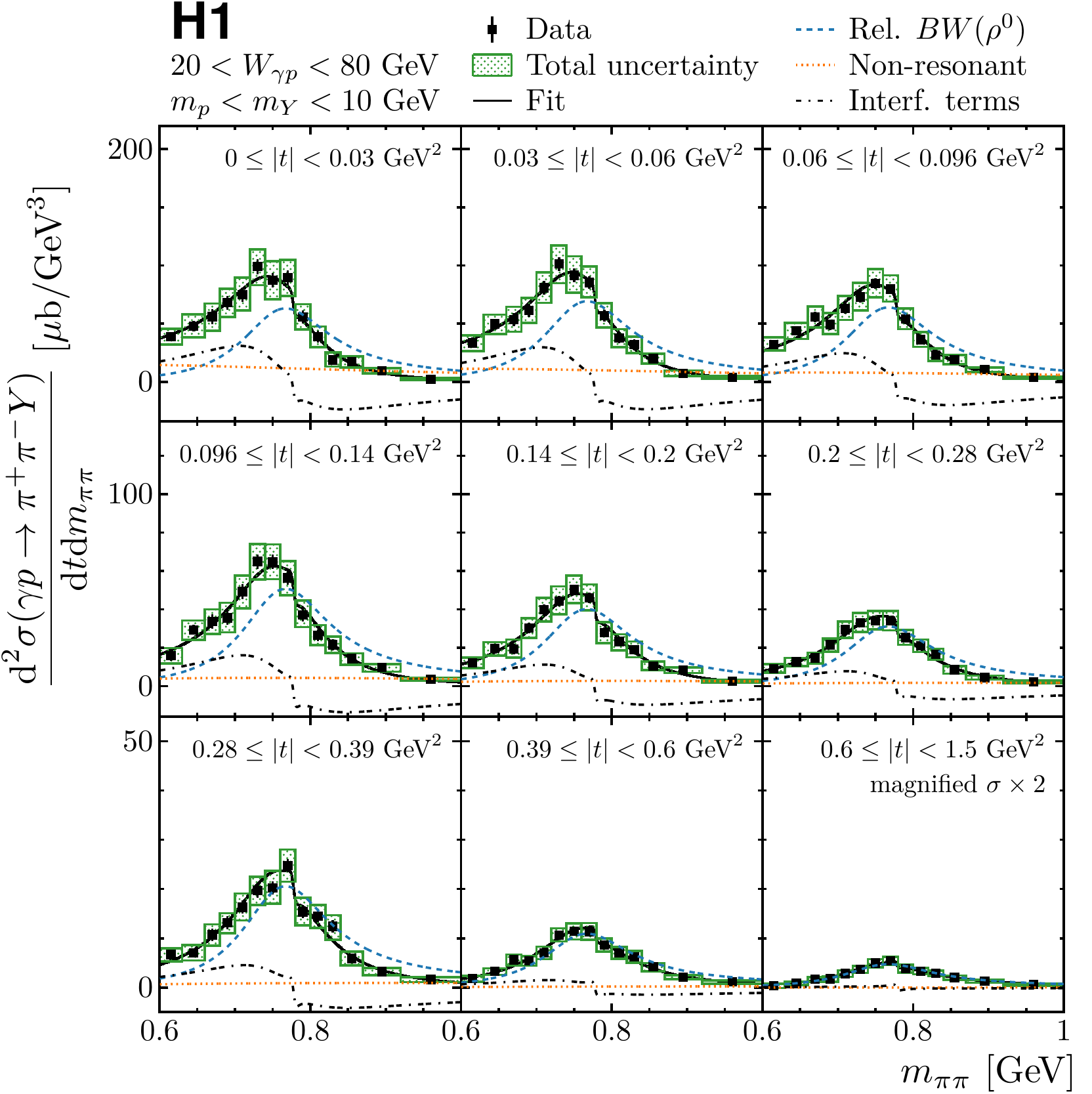
}
  \caption{Proton-dissociative ($m_p{<}\My{<}10~\gev$) double-differential \pipi photoproduction cross sections \ddSigmaPiPiYdmdt in bins of $t$ and as functions of \mpipi.
    The model that is fitted to the data as described in the text and its components are also shown.%
}
  \label{fig:results_dSigma_dmdt_mtFit_pd}
\end{figure}

\begin{figure}[htb]\centering
  \setlength{\tabcolsep}{5pt}
  \begin{tabular}{@{}c c@{}}
    $\qquad$ (a) & $\qquad$  (b) \\[1ex]
    \includegraphics[scale=0.45]{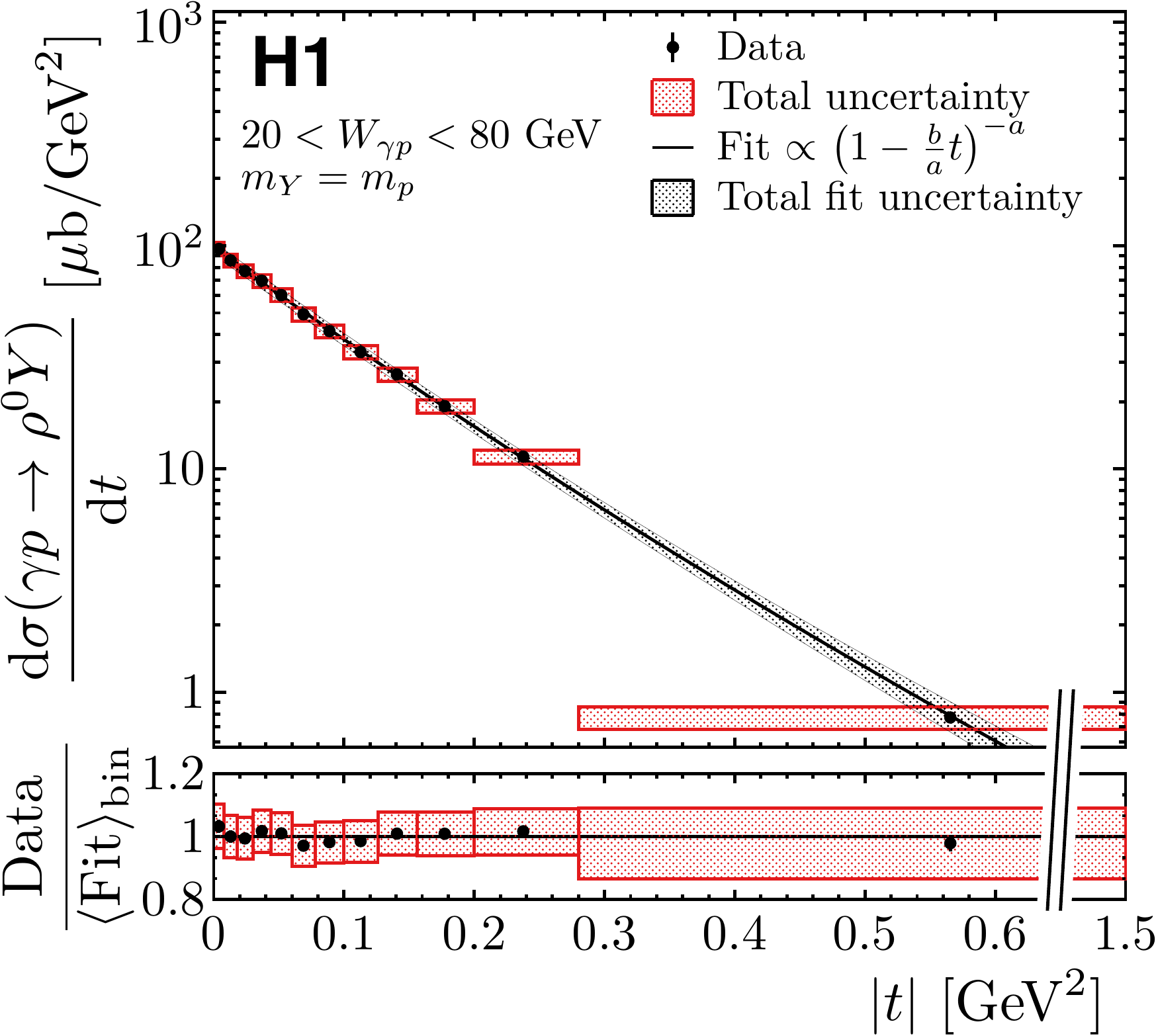} & 
    \includegraphics[scale=0.45]{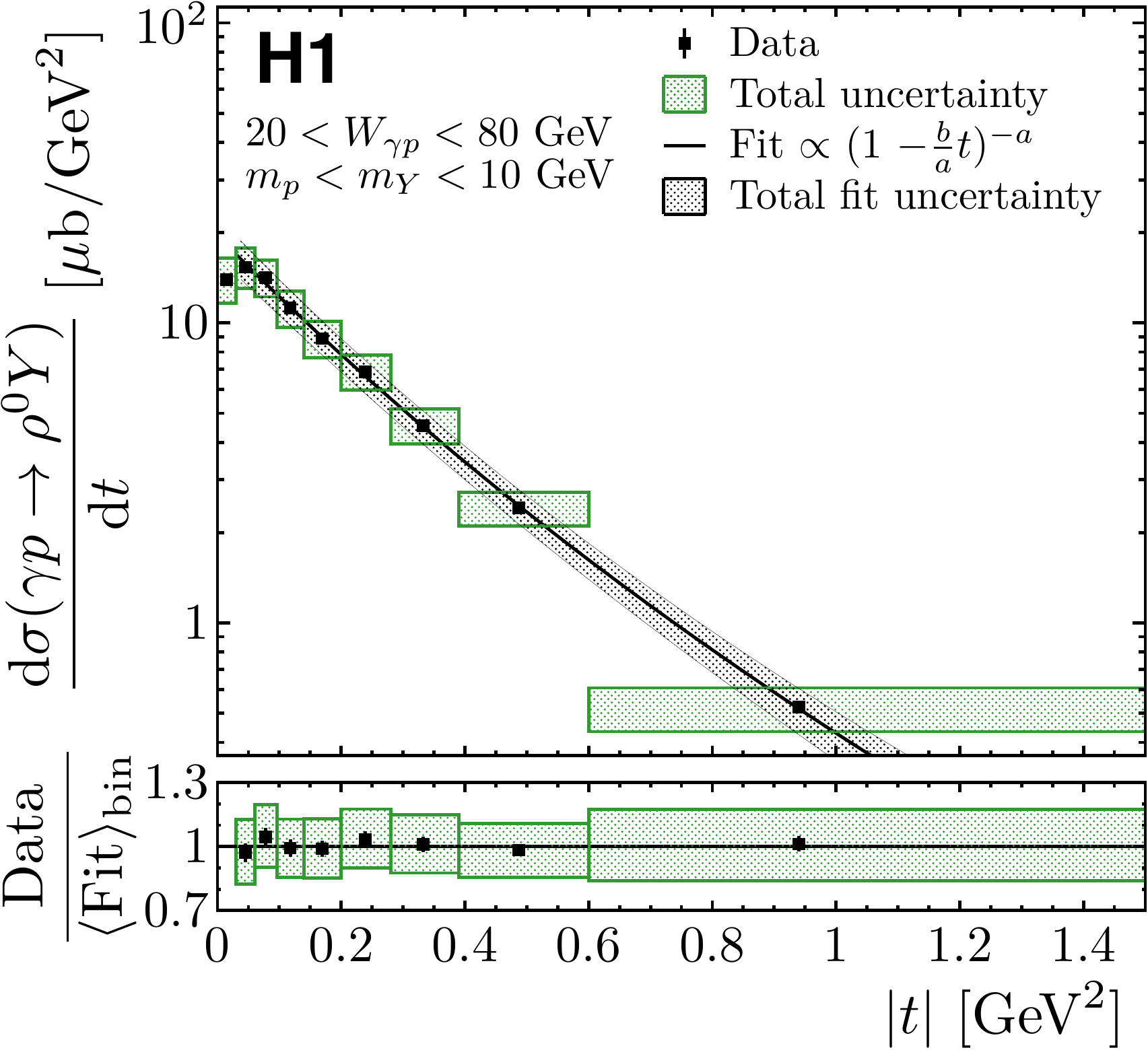} 
  \end{tabular}
  \caption{Elastic  ($\My{=}m_p$) (a) and proton-dissociative ($m_p{<}\My{<}10~\gev$) (b) differential \myrho meson photoproduction cross sections \dSigmaRhoYdt as functions of $t$. The distributions are parametrised with fits as described in the text,  and the fitted curves are also shown. In the ratio panels, the data are compared to the bin-averaged fit function values as they entered in the $\chi^2$ calculation for the fit. The depicted bin centres are evaluated using the fit functions as described in the text.}
  \label{fig:results_dSigmaRho_dt_fit}
\end{figure}

\begin{figure}[htb]\centering
  \includegraphics[scale=0.85]{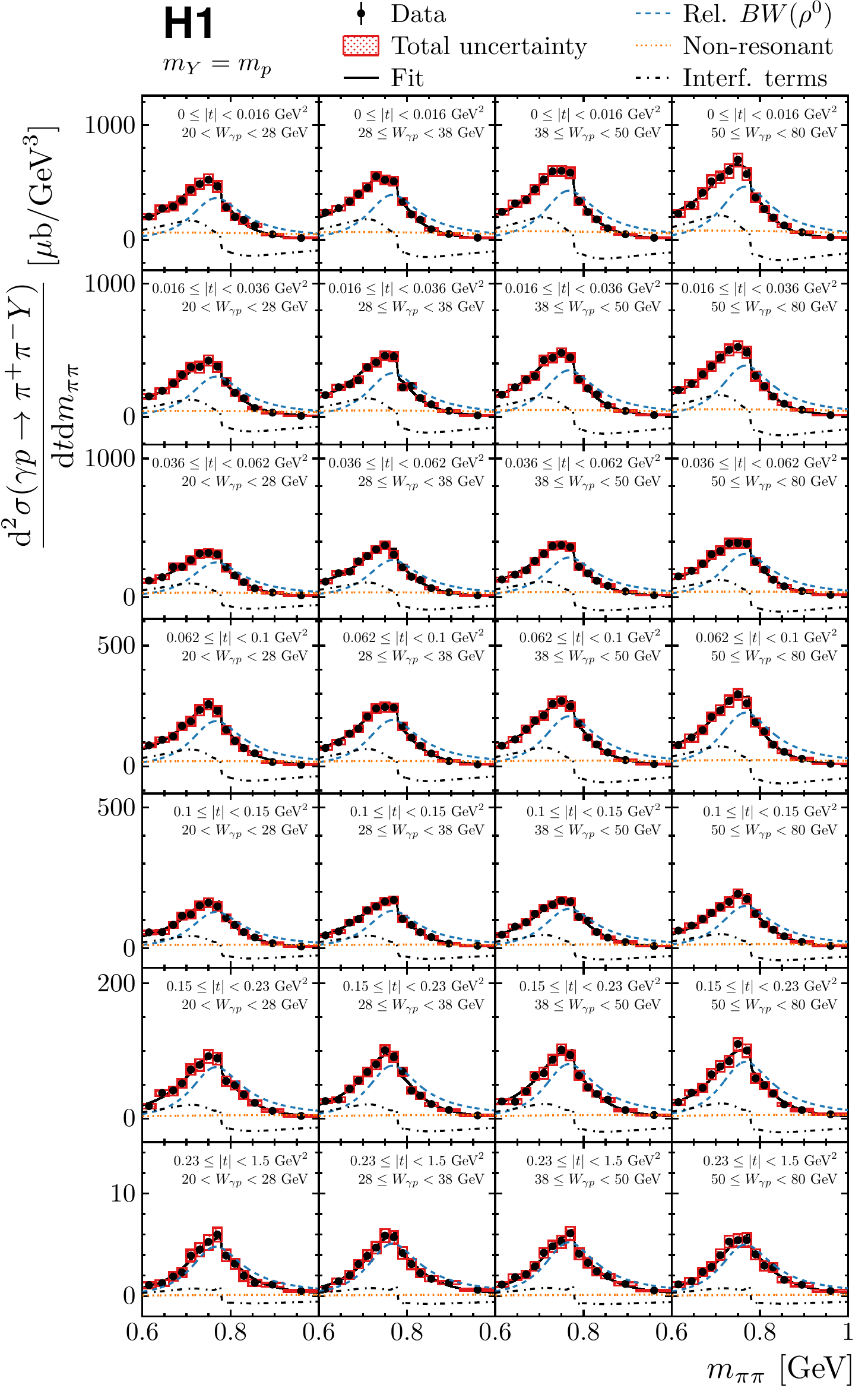
}
  \caption{Elastic ($\My{=}m_p$) double-differential \pipi photoproduction cross sections \ddSigmaPiPipdmdt in bins of \wgp and $t$ and as functions of \mpipi.
    The model that is fitted to the data as described in the text and its components are also shown.%
}
  \label{fig:results_dSigma_dmdtW_twmFit_el}
\end{figure}
\begin{figure}[htb]\centering
  \includegraphics[scale=0.85]{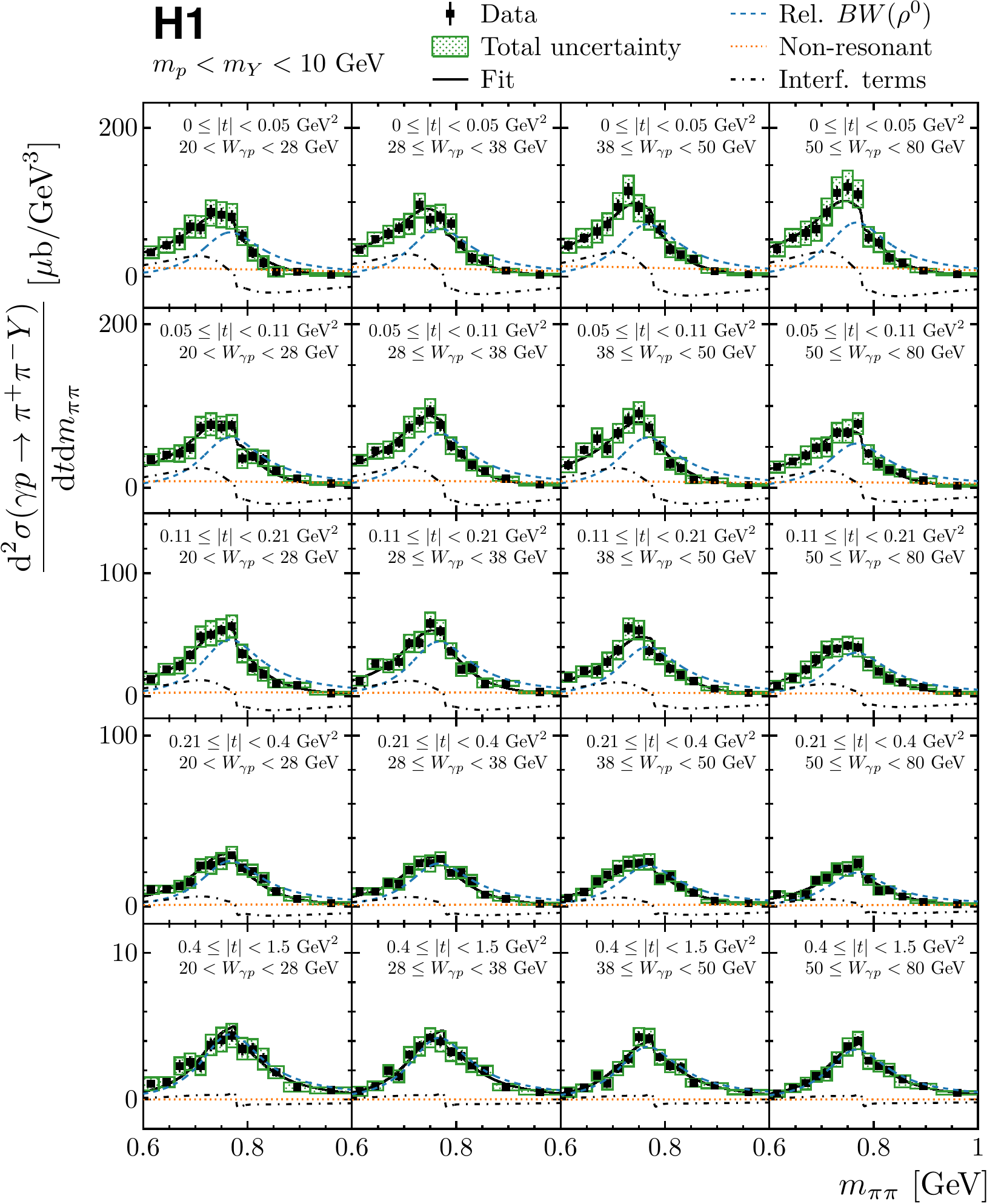
}
  \caption{Proton-dissociative ($m_p{<}\My{<}10~\gev$) double-differential \pipi photoproduction cross sections \ddSigmaPiPiYdmdt in bins of \wgp and $t$ and as functions of \mpipi.
    The model that is fitted to the data as described in the text and its components are also shown.%
}
  \label{fig:results_dSigma_dmdtW_twmFit_pd}
\end{figure}

\begin{figure}[htb]\centering
  \setlength{\tabcolsep}{5pt}
  \begin{tabular}{@{}c c@{}}
    $\qquad$ (a) & $\qquad$  (b) \\[1ex]
    \includegraphics[scale=0.45]{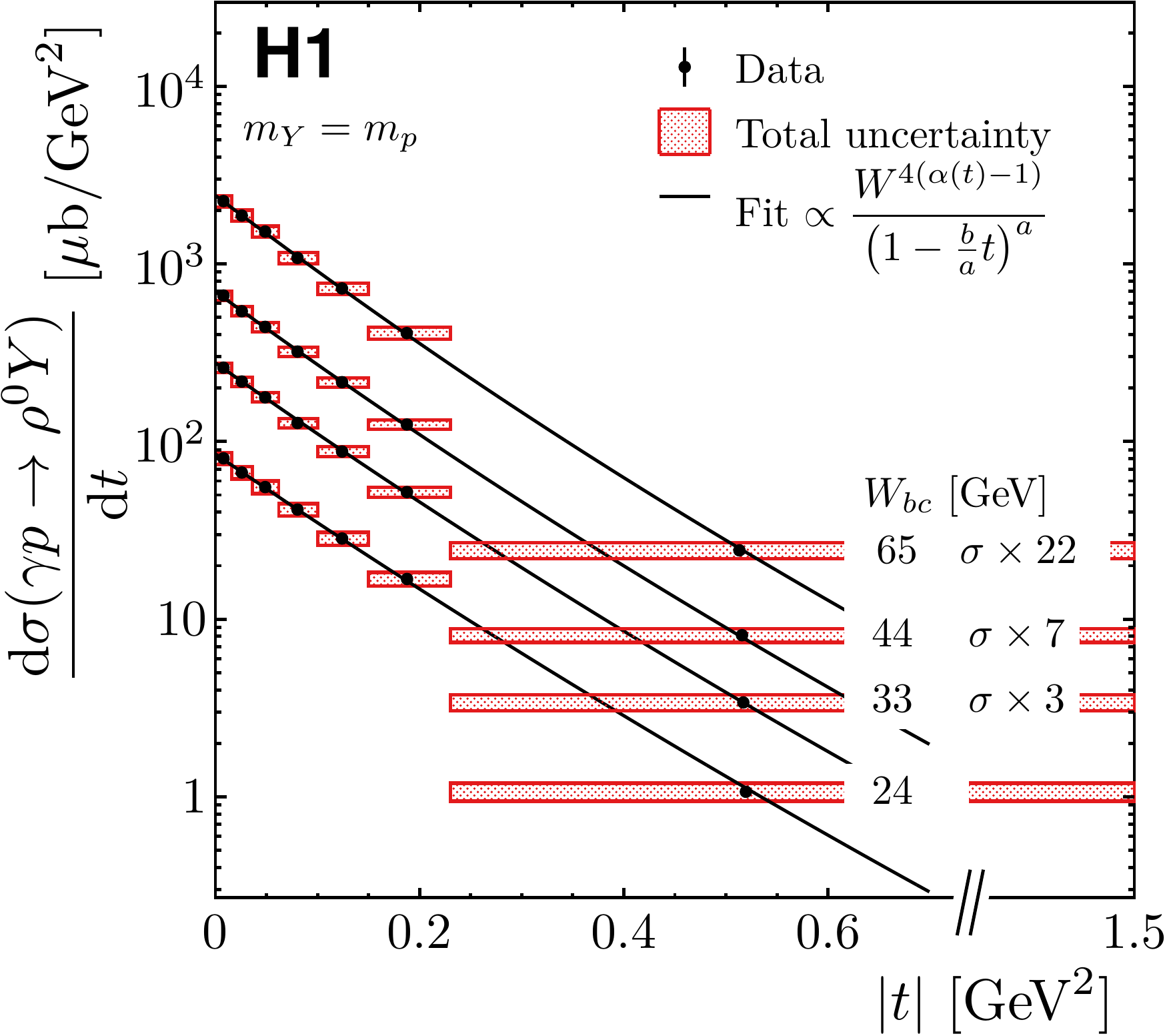}&
    \includegraphics[scale=0.45]{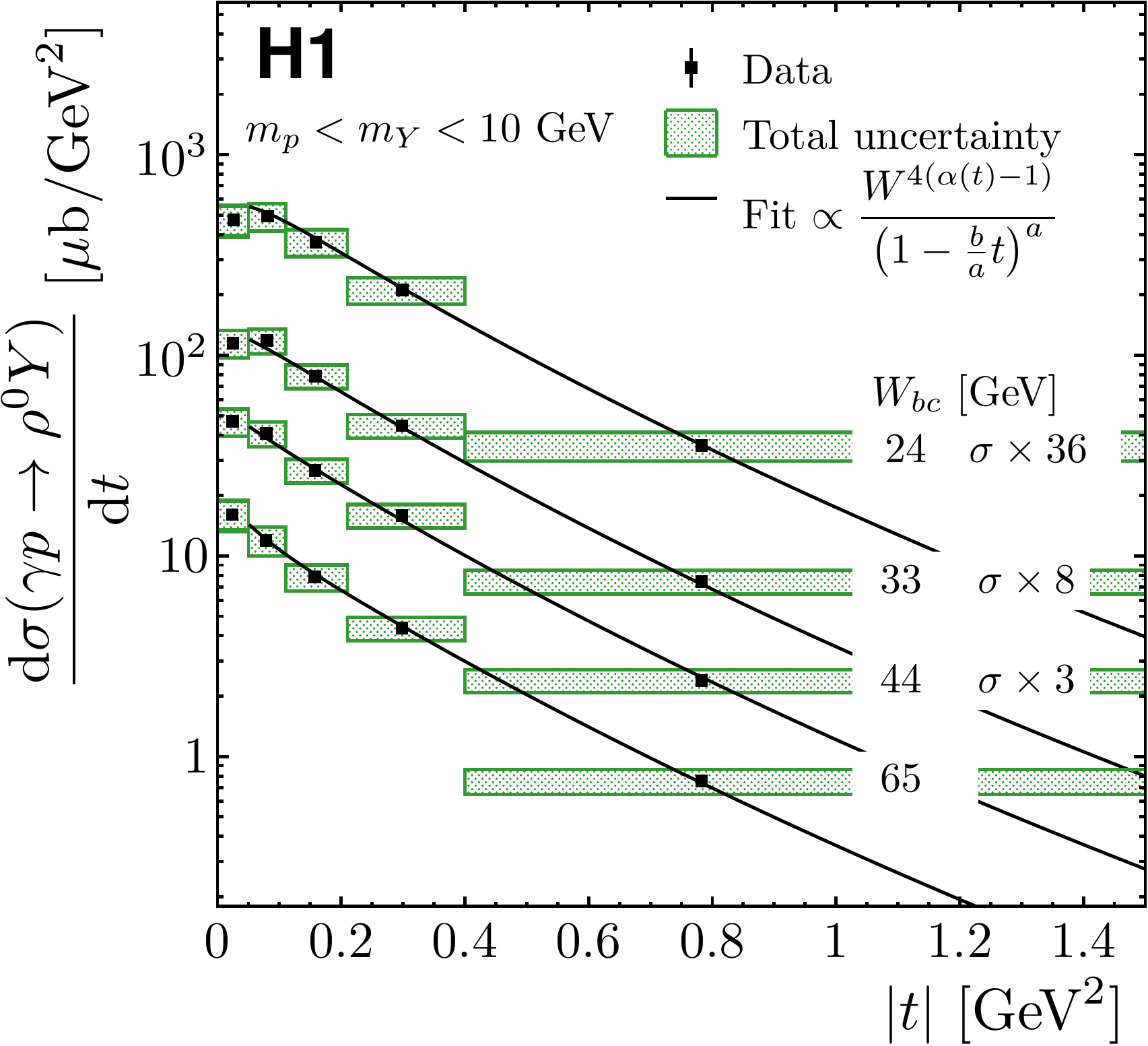}  
  \end{tabular}
  \caption{Elastic ($\My{=}m_p$) (a) and proton-dissociative ($m_p{<}\My{<}10~\gev$) (b) differential \myrho meson photoproduction cross sections \dSigmaRhoYdt in bins of \wgp and as functions of $t$. Individual distributions are scaled for visual separation, as indicated. The depicted $t$ bin centres are evaluated as described in the text. The quoted average \wgp correspond to the geometrical bin centres. The cross sections are parametrised and fitted as described in the text, and the fitted curves are also shown.}
  \label{fig:results_dSigmaRho_dtW_twfit}
\end{figure}

\begin{figure}[htb]\centering
    \includegraphics[scale=0.5]{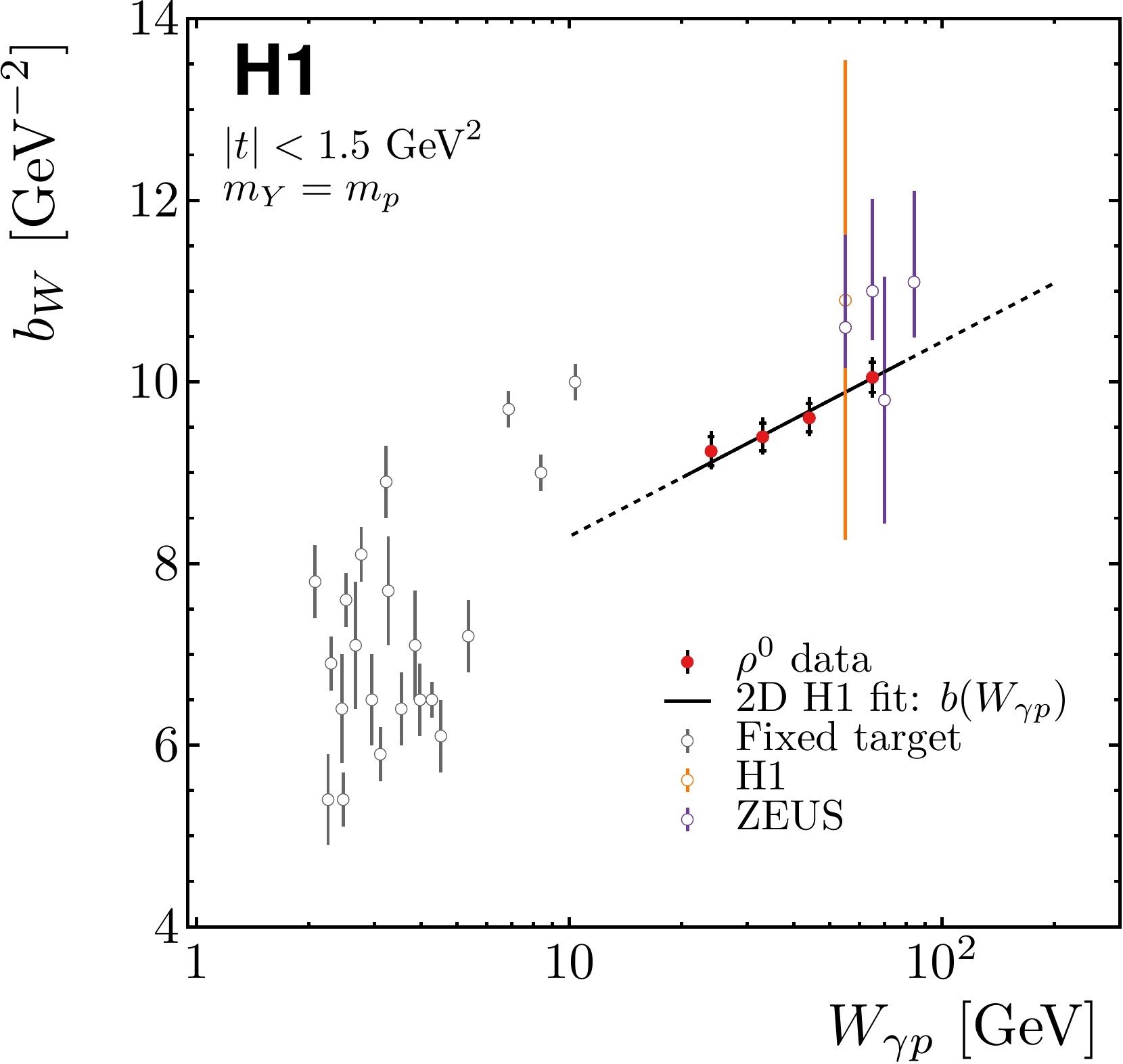
} 
  \caption{Fit parameters $b_W$ as a function of $\wgp$ for the elastic ($\My{=}m_p$) differential \myrho meson photoproduction cross section $\dSigmaRhoYdt(t;\wgp)$. The parameters are obtained from fits of modified exponential functions to the $t$ distributions in all $\wgp$ bins, as described in the text. Corresponding measurements by HERA~\cite{Aid:1996bs,Derrick:1995vq,Breitweg:1997ed} and fixed-target~\cite{Ballam:1971wq,Park:1971ts,Ballam:1972eq,Struczinski:1975ik,Egloff:1979mg,Aston:1982hr} experiments are also shown. The plotted curve depicts the dependence $b(\wgp)$ expected from the leading Regge trajectory parameters that are extracted from the present data as described in the text. } 
  \label{fig:results_fPar_wtRho_b}
\end{figure}

\begin{figure}[htb]\centering
  \setlength{\tabcolsep}{10pt}
  \begin{tabular}{@{}c c@{}}
    $\qquad$ (a) & $\qquad$  (b) \\[1ex]
    \includegraphics[scale=0.45]{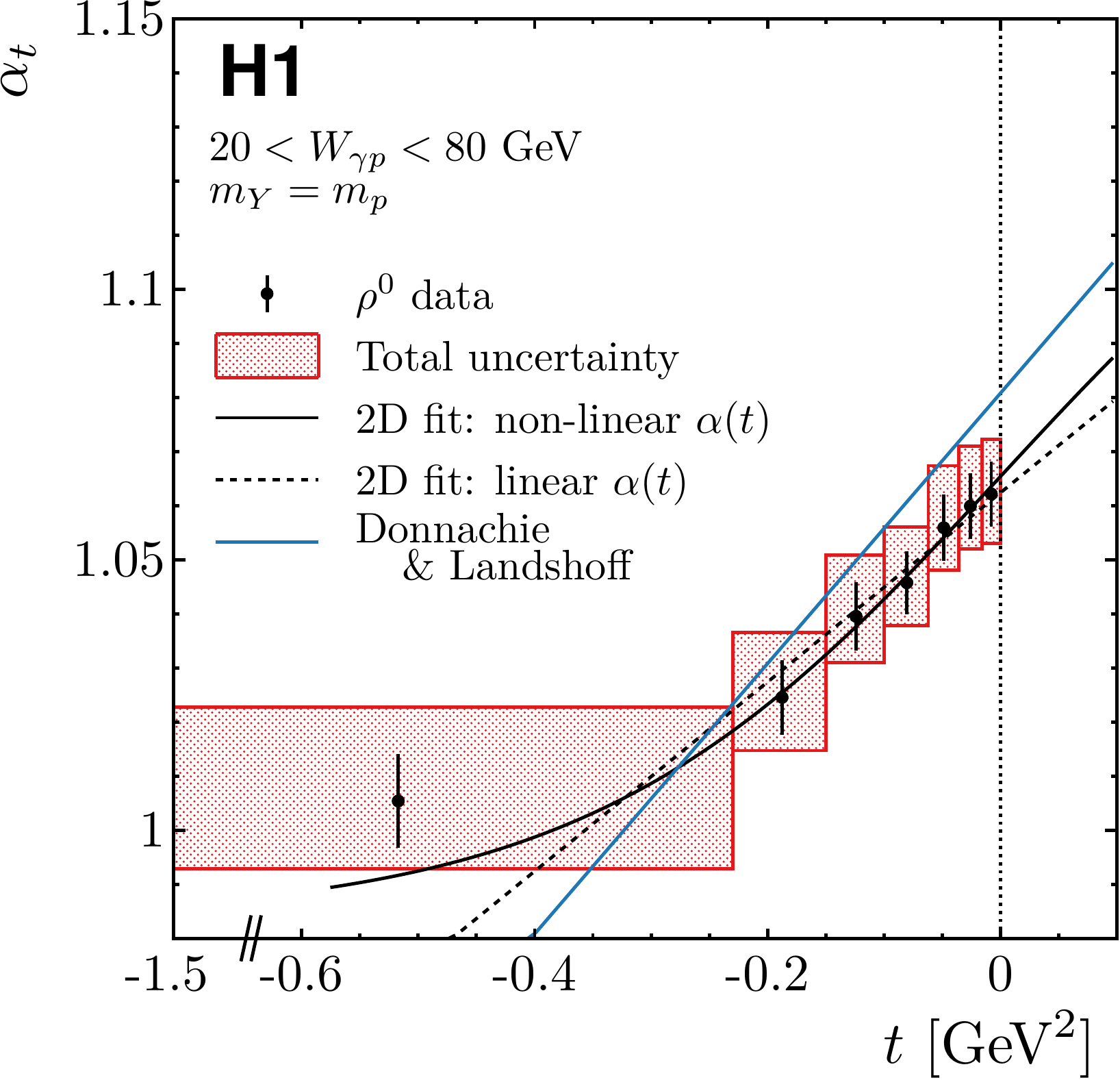}&
    \includegraphics[scale=0.45]{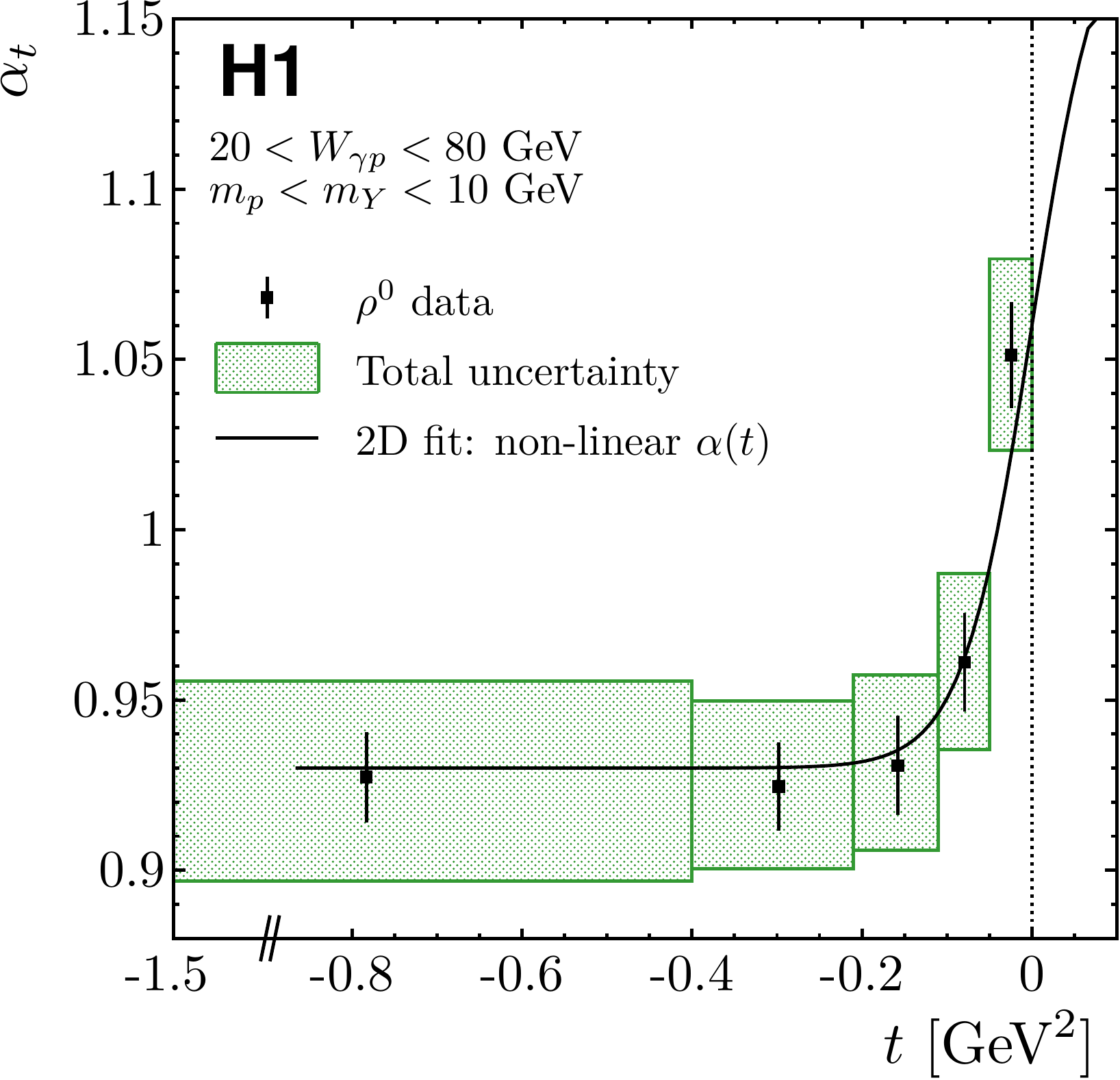}   
  \end{tabular}
  \caption{Fit parameters $\alpha_t$ as functions of $t$ for the elastic ($\My{=}m_p$) (a) and proton-dissociative ($m_p{<}\My{<}10~\gev$) (b) differential \myrho meson photoproduction cross sections $\dSigmaRhoYdt(t;\wgp)$. The parameters are obtained from fits of a power law to the energy dependencies of the cross sections in all $t$ bins, as described in the text. The trajectories extracted from a Regge model fit to the \wgp and $t$ dependencies are shown as solid curved lines (non-linear ansatz) and as a dashed line (linear ansatz). In pannel (a), the Donnachie-Landshoff trajectory is also shown. } 
  \label{fig:results_fPar_wtRho_alpha}
\end{figure}

\begin{figure}[htb]\centering
  \setlength{\tabcolsep}{10pt}
  \begin{tabular}[b]{@{}c c@{}}
    \begin{tabular}[b]{@{}r@{}}
      (a) \hspace*{13ex}\\[1ex]
    \includegraphics[scale=0.45,valign=b]{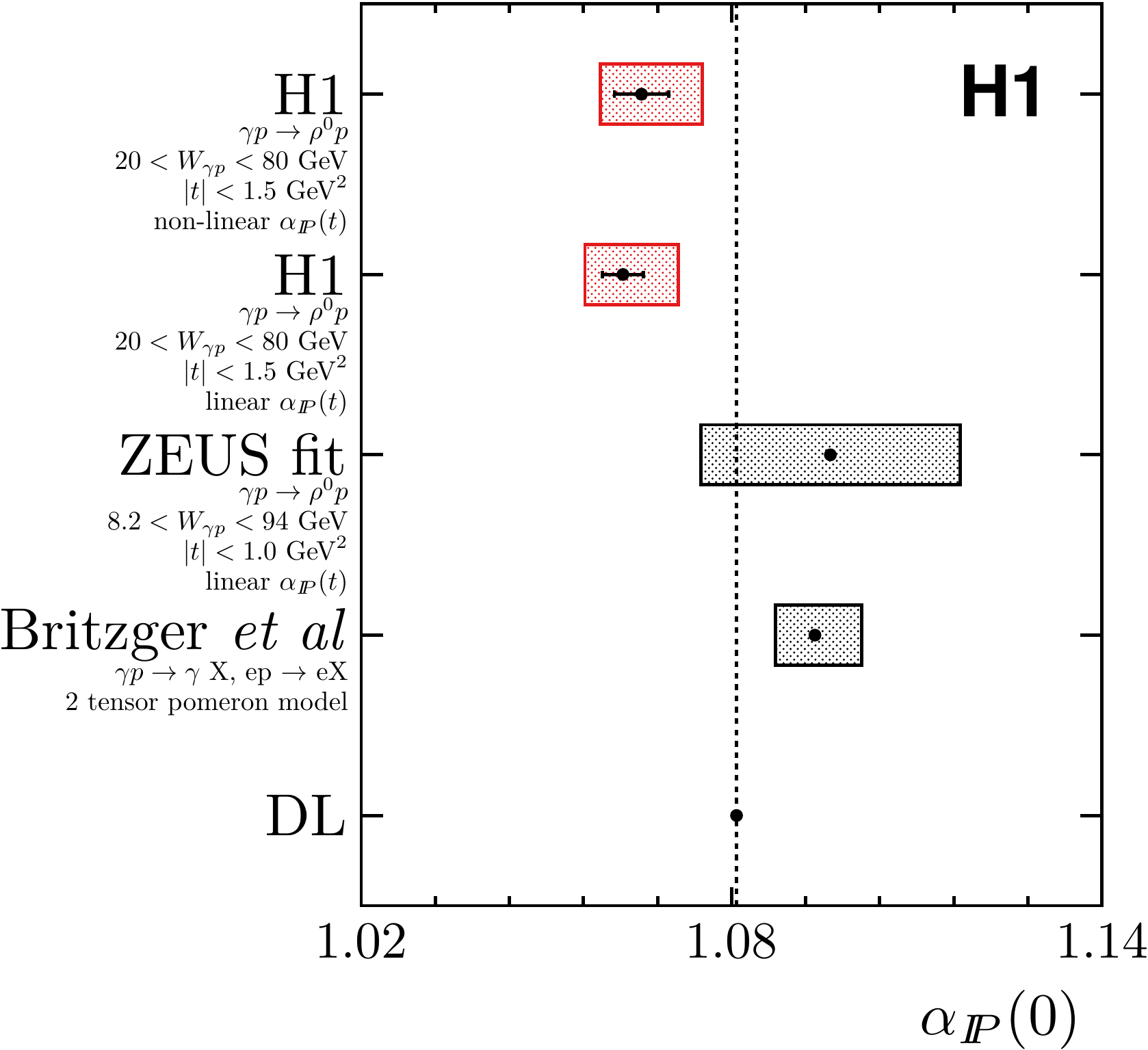}
    \end{tabular}
    &
    \begin{tabular}[b]{@{}r@{}}
      (b) \hspace*{13ex}\\[1ex]
    \includegraphics[scale=0.45,valign=b]{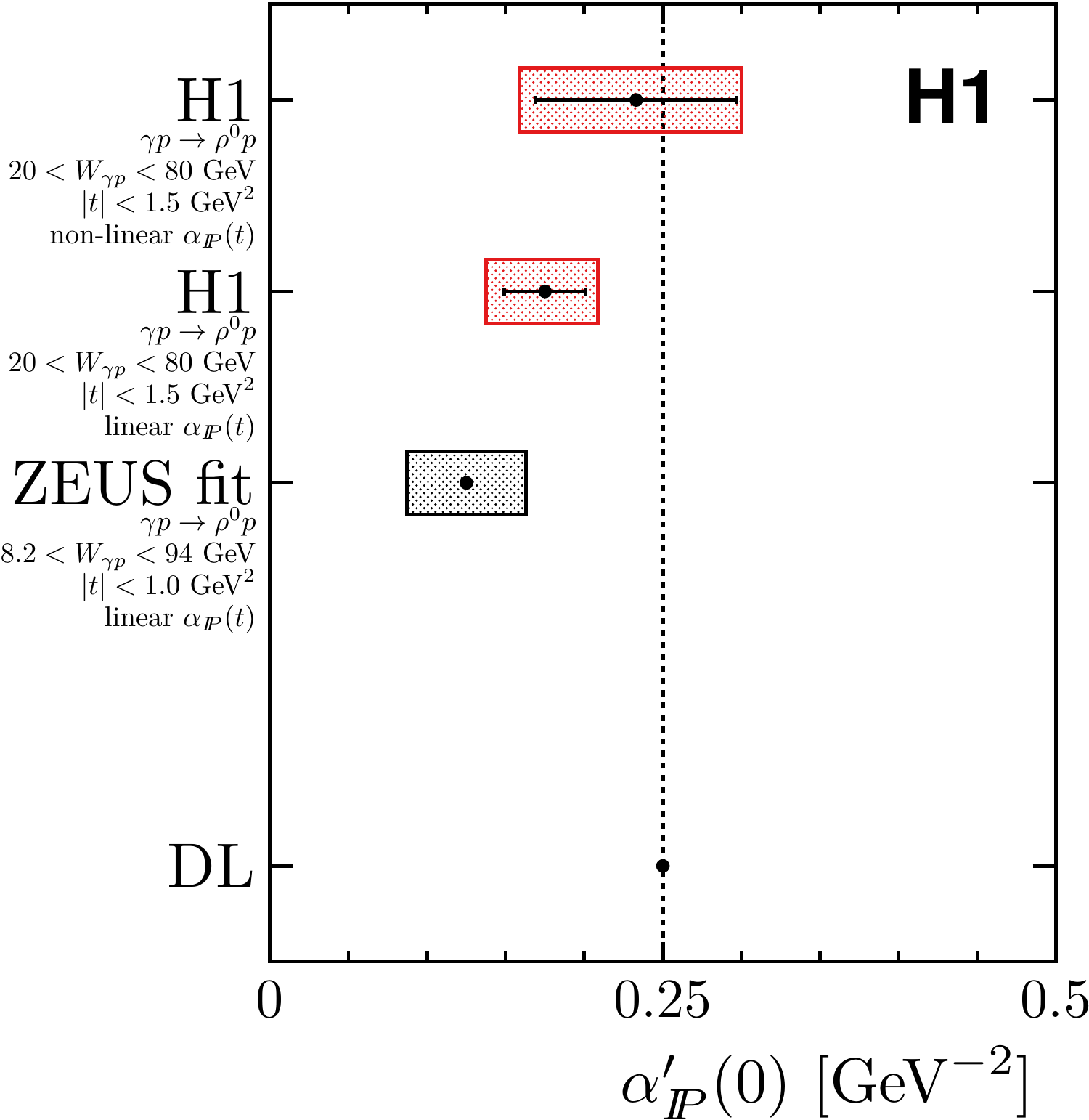}   
    \end{tabular}
  \end{tabular}
    \caption{Comparison of the soft pomeron intercept (a) and slope (b) according to Donnachie and Landshoff (DL)~\cite{Donnachie:1983hf} with measurements of the leading trajectory in \myrho meson photoproduction in the present paper and from an analysis of previous data~\cite{Breitweg:1999jy}, as well as with a measurement of the intercept from inclusive DIS and photoproduction data~\cite{Britzger:2019lvc}. The shaded bands show the total statistical and systematic uncertainties of the respective measurements. For the present data points, the statistical uncertainties alone are indicated by the error bars.}
  \label{fig:results_alphaPom_comp}
\end{figure}

\end{document}